\documentclass{article}

\usepackage{PRIMEarxiv}   
\usepackage{fancyhdr}      
\usepackage{amsmath,amssymb,amsfonts,mathtools}
\usepackage{amsthm}
\usepackage[utf8]{inputenc}
\usepackage[T1]{fontenc}
\usepackage{courier}
\usepackage{hyperref}
\usepackage{url}
\usepackage{booktabs}
\usepackage{nicefrac}
\usepackage{microtype}
\usepackage{graphicx}
\graphicspath{{./images/}{./assets/}}
\usepackage[numbers]{natbib}
\usepackage{doi}
\usepackage{tabularx,array,threeparttable}
\usepackage{enumitem}
\usepackage{caption}
\usepackage{float}
\usepackage{tikz}
\usepackage{verbatim}
\usetikzlibrary{arrows.meta,positioning,backgrounds}
\usepackage{placeins}
\usepackage{needspace}
\usepackage{xcolor}
\usepackage{comment}

\definecolor{CAMPSBlue}{HTML}{1769AA}
\definecolor{CAMPSOrange}{HTML}{C45A16}
\definecolor{AuthorRed}{HTML}{A61B1B}

\newcommand{\shorttitle}{CAMPS: Composable Mixed-Precision Adapters}

\hypersetup{
  colorlinks=true,
  linkcolor=CAMPSBlue,
  citecolor=CAMPSBlue,
  urlcolor=CAMPSBlue,
  pdftitle={CAMPS: Composable Adapters for Mixed-Precision Solvers},
  pdfsubject={Numerical Analysis; Mixed-Precision Linear Solvers},
  pdfauthor={Kevin Li, Chandrajit Bajaj},
  pdfkeywords={mixed precision, iterative refinement, FGMRES, linear systems, preconditioning, composable solvers}
}

\setlist[itemize]{leftmargin=1.6em,itemsep=2pt,topsep=4pt}
\setlist[enumerate]{leftmargin=1.8em,itemsep=2pt,topsep=4pt}
\newcommand{\IR}[2]{\operatorname{IR}_{#1\leftarrow #2}}
\newcommand{\FGMRES}{\operatorname{FGMRES}}
\newcommand{\ILU}{\operatorname{ILU}}
\newcommand{\fl}{\operatorname{fl}}
\newcommand{\norm}[1]{\left\lVert #1\right\rVert}
\newcommand{\abs}[1]{\left\lvert #1\right\rvert}

\theoremstyle{plain}
\newtheorem{theorem}{Theorem}
\newtheorem{lemma}{Lemma}
\newtheorem{proposition}{Proposition}
\newtheorem{corollary}{Corollary}
\theoremstyle{definition}
\newtheorem{definition}{Definition}

\newcounter{campsalg}
\newenvironment{algorithmblock}[1]
  {\begin{center}\begin{minipage}{0.96\linewidth}\hrule\vspace{5pt}
   \refstepcounter{campsalg}%
   \noindent\textbf{#1}\par\vspace{2pt}\small}
  {\vspace{3pt}\hrule\end{minipage}\end{center}}

\title{CAMPS: Composable Adapters for\\
Mixed-Precision Solvers}

\author{
  Kevin Li \\
  Department of Electrical and Computer Engineering \\
  University of Texas at Austin \\
  \texttt{kevinlisstx@utexas.edu} \\
\And
  Chandrajit Bajaj \\
  Department of Computer Science \\
  University of Texas at Austin \\
  \texttt{bajaj@cs.utexas.edu} \\
}

\date{November 24, 2026} 

\begin{document}
\maketitle

\begin{abstract}
Mixed-precision solvers usually assign a fixed precision to each algorithmic component. This makes it hard to reuse a lower-precision solver at different depths inside a higher-precision solver. We introduce CAMPS (Composable Adapters for Mixed-Precision Solvers), a framework that uses iterative refinement to connect solvers across precisions. An adapter computes residuals and updates in the precision of its parent and passes each correction solve to a child solver in a lower precision. The child can be a direct solver, an iterative solver, or another adapter. The adapter checks every correction and keeps it if the residual does not increase. We prove three results. Adapters can be placed at any compatible set of precision boundaries in a solver tree without breaking the solver interface. Every output that the root solver reports as successful satisfies a backward-error bound set by the stopping tolerance and the root precision. Convergence follows from one sufficient condition per boundary, computed from the leaves of the tree to its root. We test five configurations on dense and sparse systems in 64-bit, 32-bit, and 16-bit floating-point arithmetic (FP64, FP32, and FP16). On dense well-conditioned systems, CAMPS is up to $9.6\times$ faster than an FP64 direct solver for a single solve that includes the factorization. On sparse nonsymmetric systems, a three-precision hierarchy is the fastest method we tested. Compared with our implementation of the nested solver F3R, CAMPS is as fast and uses about $3.4\times$ less peak memory. CAMPS does not always win. Plain conjugate gradient is faster on sparse symmetric positive definite systems. On very ill-conditioned systems, FP32 corrections fail, and the adapter detects the failure and switches to an FP64 fallback.
\end{abstract}

\section{Introduction}
\label{intro}
Numerical linear solvers face two bottlenecks: data movement and arithmetic. Sparse iterative solvers, with their matrix-vector products, preconditioner applications, and Krylov-basis accesses, are largely memory-bandwidth bound. Dense direct solvers, on the other hand, have high arithmetic intensity and are largely compute-bound~\cite{abdelfattah2021survey,higham2022survey}. Mixed precision attacks both bottlenecks: sparse iterative solvers end up moving fewer bytes because each value occupies less memory, and dense factorizations can exploit faster lower-precision arithmetic units~\cite{langou2006exploiting,haidar2018harnessing}. However, mixed precision can introduce numerical cost. Rounding, a
compressed dynamic range, and inexact lower-precision correction solves can each prevent a lower-precision
method from ever attaining the requested accuracy~\cite{higham2002asna,carson2018three}.

Mixed-precision iterative refinement (IR) is the standard response to this problem. It relies on the observation that obtaining an accurate final solution and obtaining a useful approximate correction are different requirements with different precision needs. Starting from an approximation
$x_k$, refinement forms the residual $r_k=b-Ax_k$ in a higher (parent) precision, solves the correction equation $A d_k = r_k$ only approximately in a
lower (child) precision, and applies the update $x_{k+1}=x_k+d_k$ in the parent
precision. This way, the residual and update carry the accuracy of the parent precision, while the expensive solve is delegated to a cheaper child precision. Moler analyzed iterative refinement in floating-point arithmetic in 1967~\cite{moler1967}. He proved that a system factorized in a lower precision can still be solved to full working accuracy, provided the residual is formed in a higher precision and the factorization is reused to solve for the correction. In particular, he proved that the error converges geometrically whenever the factorization is accurate enough and that the attainable accuracy is set by the precision of the residual rather than by the precision of the factorization. The two precisions therefore play separate roles: the low one pays for the expensive work and controls the rate of convergence, while the high one sets the accuracy the final answer can reach. The modern analysis by Carson and Higham~\cite{carson2017analysis} reformulated iterative refinement around a general correction solver rather than reused factorization, and they demonstrated that convergence depends on the accuracy of the computed correction, not on the factorization itself. Their GMRES-IR variant uses the low-precision factors only as a preconditioner for GMRES on the correction equation, which keeps refinement converging on matrices where the factorization alone is too inaccurate. In three precisions~\cite{carson2018three}, one each for the factorization, the working arithmetic, and the residual, they gave the exact conditions required for the variant to reach working-precision accuracy. With a half-precision factorization, GMRES-IR extends the usable condition-number range by about eight orders of magnitude over plain refinement. Amestoy et al.~\cite{amestoy2024five} went further and let the matrix-vector products and the preconditioner applications inside GMRES-IR run in their own precisions, with five precisions in total. They proved error bounds for every combination, so the cost of the inner iteration can be weighed against the range of condition numbers it tolerates. Earlier, Buttari et al.~\cite{buttari2008sparse} had demonstrated for sparse systems that a reduced-precision iterative solver can take the place of the factorization as the correction solver and still deliver 64-bit accuracy. The five-precision analysis explains when this works and when it fails.

This paper offers a different interpretation of the mixed-precision iterative refinement algorithm. Instead of treating the precision transition as a loop, we treat it as a first-class solver object. Given a solver $S_q$ that accepts and returns vectors in a lower precision~$q$, we define an adapter $\IR{p}{q}[S_q]$. This adapter accepts and returns vectors in the parent precision~$q$, stores its state and forms residuals and updates in~$p$, internally converts a safely scaled residual to~$q$, invokes~$S_q$, and promotes the returned correction back to~$p$. Very importantly, the parent precision does not just round the problem down and cast the answer up. It continually measures and corrects the error the child introduces.

The central property of this abstraction is that it is closed under composition. Because an adapter takes and returns vectors in the parent precision, the parent cannot tell it apart from any other solver in that precision. It can therefore be used anywhere a solver is expected, including as the child of another adapter. For example, an FP32 solver may wrap an FP16 solver, and the result may in turn be wrapped by an FP64 solver. One such example hierarchy is 
\[
\FGMRES{64}\bigl[\IR{64}{32}[\FGMRES{32}[\IR{32}{16}[\ILU(0)_{16}]]]\bigr].
\]
Each adapter forms its residuals and updates in its own parent precision, and only the innermost preconditioner runs in FP16. Additionally, lowering the precision of a correction solve does not lower the precision of the problem around it, as the parent's matrix-vector products, residual norms, updates, and Arnoldi process all stay in the parent precision.

The main contribution of this paper is to precisely define how refinement could appear at several levels. Theorem~\ref{thm:closure} states that, under a residual contract at the parent level, a refinement adapter can be inserted at any compatible set of edges of a finite flexible solver tree. The tree is well defined at every depth, and the residual at each replaced edge satisfies a geometric bound after $J$ corrections whose constants depend only on that edge, not on where the edge sits in the tree. We call this framework CAMPS, Composable Adapters for Mixed-Precision Solvers. The theorem is stated in terms of the residual that is actually achieved, so it still applies when the child solver is nonlinear or changes from call to call, which is the case that requires a flexible outer Krylov method~\cite{saad1993fgmres}. We prove two further theorems, which say what a tree built this way guarantees. Theorem~\ref{thm:stability} states that any output the root reports as successful satisfies a backward-error bound set by its tolerance, regardless of what lies below it. Theorem~\ref{thm:recursive} states that convergence needs one checkable condition at each edge, with the constants built up from the leaves to the root and the order of the limiting accuracy set by the root precision.

CAMPS is a generalized framework, not a single algorithm, so it can be applied to any matrix class. Specifically, we claim that for each class of matrix, there is a tree in the CAMPS grammar that composes correctly and either beats a uniform-precision solver of the same type or demonstrates where it fails. We therefore test five named instances: \textsf{CAMPS-Direct} the bare adapter, \textsf{CAMPS-Krylov} the adapter inside FGMRES, \textsf{CAMPS-Cascade} the full FP64/FP32/FP16 hierarchy with refinement at both precision edges, \textsf{CAMPS-Apex} an adapter at the top edge only with plain casts below it, and \textsf{CAMPS-Ladder} which degrades precision when the contract fails. We run them on dense well-conditioned, dense ill-conditioned, sparse SPD, and sparse nonsymmetric matrices, and we compare them against uniform-FP64 direct, Krylov, and CG solvers. We also test against our owned tuned reimplementation of the F3R algorithm of Suzuki and Iwashita~\cite{suzuki2025nested}. All methods use the same relative-residual stopping rule with residuals recomputed in FP64, median timings after warmup, both analytic and measured memory, and matrix-vector counts per precision. The experiments also test CAMPS' own assumptions. The per-correction contraction that Lemma~\ref{lem:recurrence} bounds is measured on every instance, and the ill-conditioned sweep demonstrates where the instance crosses one.

We therefore bring the following contributions:
\begin{itemize}
    \item A new interpretation of mixed-precision iterative refinement as a two-precision adapter $\IR{p}{q}[S_q]$. Its residuals and updates are allocated in the parent precision, even when the child itself is an adapter.
    \item A recursive solver grammar. Adapters may compose across any finite solver tree, and any correction solver, whether it is direct, iterative, stationary, or another adapter, can fit behind the same interface. Additionally, a residual contract at the parent level describes each edge on its own, independent of the rest of the tree.
    \item A theorem for insertion at arbitrary level. Refinement may be inserted at any compatible set of edges while the interfaces stay closed and each edge keeps a local residual bound which does not depend on depth. The propagation bound expresses the parent's constants through conversion error, child solve quality, representation mismatch, and correction size.
    \item Guarantees for arbitrary hierarchies. Every output that the root reports as successful satisfies a backward-error bound set by the tolerance and the root precision, whatever the subtree is, convergence follows from one checked condition per edge with explicit rounding constants, and a cost rule demonstrates when lowering precision is worse.
    \item A family of named instances. We implement the bare adapter, the Krylov-wrapped adapter, the fully recursive two-edge hierarchy, the adapter-topped hierarchy, and the contract-fallback ladder, with each defined separately as an algorithm.
    \item An evaluation across four matrix classes, including the recursive FP64/FP32/FP16 hierarchy, a condition-number sweep with measured contraction, an edge-placement ablation, a multi-right-hand-side amortization study, and a comparison with our tuned F3R reimplementation.
    \item Two findings about our contract. We discover that past a computatable conditioning threshold, the sufficient condition for contraction no longer holds and low-precision children we tested stop producing verifiable corrections. Additionally, verifying each application does not itself control the quality of the outer Krylov basis.
\end{itemize}
\section{Related Work}\label{related-work}
CAMPS closely aligns with several areas of research: iterative refinement and its mixed-precision analysis,
mixed-precision solver implementations on accelerators, flexible, inexact, and nested Krylov methods, and composable
solver software. This section surveys each area and states what CAMPS builds upon and what CAMPS contributes. Table~\ref{tab:positioning} in the appendix
summarizes the comparison against the closest prior work.

\paragraph{Iterative refinement and its analysis.} Let
$A\in\mathbb{R}^{n\times n}$ be nonsingular and $b\in\mathbb{R}^n$. A step in iterative refinement forms $r_k=b-Ax_k$, computes an approximate correction $d_k$
to $Ad_k=r_k$, and updates $x_{k+1}=x_k+d_k$. Wilkinson analyzed this loop for Gaussian elimination with residuals computed in extended precision~\cite{wilkinson1963rounding}. Moler's floating-point analysis distinguished accurate residual computation from a lower-accuracy correction solve, creating the basis for convergence conditions of the form $\kappa(A)u\ll1$~\cite{moler1967}. Jankowski and Woźniakowski demonstrated that refinement can make even an unstable solver backward stable if its corrections are sufficiently accurate~\cite{jankowski1977iterative}. Skeel established a component-wise result for Gaussian elimination~\cite{skeel1980}, and Higham developed a unified analysis for general correction solvers~\cite{higham1997iterative,higham2002asna}.

Later work addressed practical reliability of refinement. Demmel et al. studied the convergence rate of extra-precise refinement in LAPACK and reported failure when that rate indicated unreliable convergence~\cite{demmel2006error}. Arioli, Demmel, and Duff examined how sparsity affects backward stability, demonstrating why only analyzing relative residuals may not sufficiently determine solution quality\cite{arioli1989}.

CAMPS uses the same residual correction update procedure and allows general correction solvers, like Higham’s analysis. Its main difference is how it implements correction accuracy. Prior analysis assumes sufficient accuracy in advance for a fixed refinement loop, but CAMPS instead checks the corresponding residual condition for every correction in every refinement loop in its solver tree. If a correction fails this check, CAMPS discards it and switches to another child solver. The failure flag of Demmel et al. is the closest precedent for this runtime check, but it monitors a single loop and does not replace the solver when a failure occurs.

\paragraph{Mixed-precision refinement and the GMRES-IR family.} Early mixed-precision methods combined single-precision factorization with double-precision refinement on CPUs and the Cell processor~\cite{langou2006exploiting,buttari2007mixed,kurzak2007cell}. This approach was later extended to sparse direct solvers and inner-outer Krylov methods~\cite{buttari2008sparse} and surveyed in~\cite{baboulin2009mixed}. Arioli and Duff demonstrated that FGMRES preconditioned by a single-precision LU factorization can achieve double-precision backward stability in the case where standard refinement fails~\cite{arioli2009fgmres}. Carson and Higham introduced GMRES-IR and analyzed refinement with a general correction solver. Their three-precision framework allows different precisions for factorization, working arithmetic, and residual computation~\cite{carson2017analysis,carson2018three}.

Later research extended this framework to five precisions~\cite{amestoy2024five} and to least-squares problems~\cite{carson2020three,carson2025comparison,carson2025sketching,carson2025weighted}. Other extensions use approximate sparse factorizations~\cite{amestoy2023combining}, sparse approximate inverse and adaptive-precision preconditioners~\cite{carson2023spai,khan2025adaptive}, and low-rank corrections to factorization errors~\cite{higham2019preconditioner}. Related methods address symmetric positive definite systems using shifted low-precision Cholesky factorization~\cite{higham2021spd}, split-preconditioned FGMRES with four precisions~\cite{carson2024split}, and Newton's method~\cite{kelley2022newton}. 

Additional analyses explain how to scale matrices to fit within the range of half precision~\cite{higham2019squeezing} and how to bound rounding errors in block fused multiply-add units~\cite{blanchard2020block}. These results provide the representation and rounding-error models needed for half-precision child solvers. Kelley describes conversions between precisions within a refinement loop~\cite{kelley2024interprecision}. Abdelfattah et al.~\cite{abdelfattah2021survey} go through mixed-precision methods across dense and sparse linear algebra, including factorizations, iterative solvers, preconditioners, and multigrid, and report the speedups measured on GPUs with low-precision units. Higham and Mary~\cite{higham2022survey} take the theory side and collect the rounding-error results that say when each scheme is stable and how accurate it can get. Buttari et al. also developed a modular backward-error framework that derives error bounds for a GMRES variant from bounds on its individual components~\cite{buttari2026modular}.

These methods implement a fixed two-layer structure. An outer refinement loop calls an inner correction solver, which can then use several precisions internally. While their convergence guarantees depend on conditions imposed in advance on $\kappa(A)$ and the chosen precisions, CAMPS extends this structure in three ways. First, it represents a precision transition explicitly as $\IR{p}{q}[S_q]$. This allows the correction solver to contain another refinement loop, so refinement can be nested to any depth (Theorem~\ref{thm:closure}). Second, it supports arbitrary correction solvers, including methods beyond factorization and factorization-preconditioned GMRES. Third, it checks an accuracy condition at runtime for each correction, rather than only relying on an accuracy assumption made in advance.

CAMPS does not establish a new error bound for two-level refinement. For that situation, the five-precision analysis gives sharper bounds. Instead, CAMPS tackles how error bounds combine across nested levels. Theorem~\ref{thm:propagation} describes how the constants at one level affect the next, and Theorem~\ref{thm:recursive} extends this relation through the solver tree. For an LU leaf solver, the resulting condition reduces to the requirement $\kappa u_f\lesssim1$. 

The framework of Buttari et al.~\cite{buttari2026modular} offers a complementary approach. A CAMPS adapter’s contract specifies the component accuracy that their framework requires. Combining the two frameworks would be future work.

\paragraph{Adaptivity, fallback, and escalation.} Several methods already incorporate responses to low-precision solver failures at runtime. Buttari et al. limit the number of refinement iterations and refactorize in double precision if needed~\cite{buttari2007mixed}. Hogg and Scott’s sparse symmetric solver first factorizes in single precision. It then refines the solution using IR or FGMRES in double precision. If refinement fails to converge, it falls back to a double-precision factorization~\cite{hogg2010mixed}. NVIDIA’s cuSOLVER provides standard and GMRES-based refinement, with an optional fallback to a full-precision solver~\cite{nvidia2026cusolver}.

Oktay and Carson’s multistage refinement starts with the cheapest variant and observes its convergence rate. When progress stalls, it escalates through SGMRES-IR to GMRES-IR~\cite{oktay2022multistage}. Wu et al. add a line search to each refinement step and prove that this prevents divergence even when the inner solves are very inaccurate~\cite{wu2026stable}. Carson and Chen formulate runtime precision selection as a contextual bandit problem~\cite{carson2026autotuning}. The QUDA lattice-QCD solvers use reliable updates within a Krylov solve. They perform iterations in single or half precision and periodically recompute the true residual in double precision~\cite{clark2010lattice}.

CAMPS does not introduce fallback, residual recomputation, or escalation as a novelty. Its contribution has to do with where and when these mechanisms operate. CAMPS checks each correction at every adapter in the solver tree, and it rolls back rejected corrections instead of just counting them. Its grammar also includes fallback as a composition rule (Algorithm~\ref{alg:ladder}). This rule can work at any edge of the tree and wrap any child solver. 

Multistage refinement is the closest precedent for \textsf{CAMPS-Ladder}. It uses the convergence rate to decide when to switch between a fixed sequence of refinement variants. \textsf{CAMPS-Ladder} can escalate after a single rejected correction. Its child solvers can also use methods other than refinement.

Section~\ref{exp-dense-ic} evaluates residual-minimizing damping, following the approach of~\cite{wu2026stable}. In these experiments, damping allows corrections to pass the acceptance test without restoring progress. It therefore prevents rejection from triggering escalation. To address this, damping must be paired with a threshold that requires sufficient progress.

\paragraph{Mixed precision on accelerators.} Early research evaluated half-precision factorization for iterative refinement on GPUs without tensor cores~\cite{haidar2017half}. Later studies used tensor cores with FP32 accumulation. These studies showed that FP16 LU within GMRES-IR can achieve FP64 accuracy several times faster than FP64 LU~\cite{haidar2018design,haidar2018harnessing,haidar2020tensor}. The HPL-AI and HPL-MxP benchmarks established this two-level configuration as a community standard~\cite{kudo2020prompt,dongarra2026hplmxp}. Lopez and Mary store parts of the factors in FP16 to reduce data movement~\cite{lopez2023lu}. Batched half-precision kernels provide building blocks for these methods~\cite{abdelfattah2019batched}, and MAGMA provides the library foundation for vendor routines~\cite{tomov2010magma}. Error-free transformations and integer-unit emulation can produce products with FP32 or FP64 accuracy using low-precision hardware~\cite{ozaki2012errorfree,ootomo2022recovering,ootomo2024dgemm,ozaki2026ozakiII}. Thus, low-precision hardware can still support high-accuracy computation.

The performance benefits vary. Zounon et al. found that single precision often speeds up sparse solvers by much less than the ideal factor of two~\cite{zounon2022performance}. Energy studies provide another reason to use low-precision child solvers: they can reduce energy use as well as runtime~\cite{anzt2010energy}.

CAMPS uses generic PyTorch kernels and no tensor-core-specific factorization. It treats the methods above as potential child solvers. The dense experiments in this paper first reproduce the benefits of standard two-level refinement (Section~\ref{exp-dense-wc}). They then study how solver composition changes those benefits. Specifically, they test what happens when an additional adapter wraps the child solver, when the child occupies a different edge in the solver tree, or when an iterative solver replaces it through the same interface.

\paragraph{Mixed precision inside sparse and Krylov solvers.} Another line of research reduces precision within individual solver operations. Anzt, Heuveline, and Rocker use an outer error-correction loop with a low-precision inner Krylov solver. They analyze the tolerance that the inner solver must satisfy~\cite{anzt2011errorcorrection,anzt2012convergence}. Lindquist et al. interpret restarted GMRES as refinement with a single-precision inner cycle~\cite{lindquist2020improving,lindquist2022gmres}. Loe et al. compare GMRES-IR, low-precision preconditioners, and reduced-precision orthogonalization on GPUs in Trilinos. They note that Belos requires wrappers to combine precisions because it does not support it directly~\cite{loe2021experimental,loe2021mixed}.

Other methods reduce precision at specific components. Gratton et al. progressively lower the precision of inner products in GMRES~\cite{gratton2019variable} and operator evaluations in conjugate gradients~\cite{gratton2021variable}. Compressed-basis GMRES uses Ginkgo’s memory accessor to store the Krylov basis in reduced precision~\cite{aliaga2023compressed,anzt2019modular,grutzmacher2023accessor}. Adaptive-precision block-Jacobi~\cite{anzt2019adaptive,flegar2021adaptive} and reduced-precision sparse approximate inverses~\cite{gobel2021mixed} reduce preconditioner storage precision block by block. Adaptive-precision sparse matrix–vector products partition the operator according to entry magnitude~\cite{ahmad2019datadriven,graillat2024adaptive}. Scott and
T\r{u}ma prevent breakdown during the construction of half-precision incomplete factorizations~\cite{scott2024avoiding}. Sparse direct libraries also provide modes that combine single-precision factorization with double-precision refinement~\cite{amestoy2023combining,li2023superlu}. AmgX offers mixed-precision storage for multigrid~\cite{naumov2015amgx}.

These methods change precision within a solver at a single level. While they target the basis, preconditioner storage, inner products, or operator, CAMPS controls precision across solver levels. It retains the parent-precision operator for residual computation and performs updates in the parent precision. It uses lower precision only for the correction solve. Each method above can therefore serve as a child solver within CAMPS. 

\paragraph{Mixed-precision multigrid.}
Early research used defect correction with a single-precision multigrid
solver before GPUs became common~\cite{strzodka2006pipelined,goddeke2007performance}.
Later studies placed single-precision algebraic multigrid inside a
double-precision Krylov loop. They evaluated this approach on
clusters~\cite{emans2010mixed},
accelerators~\cite{sumiyoshi2014mixed,tsai2023mixed}, and in matrix-free
finite-element codes~\cite{kronbichler2019multigrid}.

McCormick, Benzaken, and Tamstorf analyze refinement with multigrid
as the inner solver, measuring error in the energy norm. They also
introduce progressive precision, which sets each level's precision
based on the discretization~\cite{mccormick2021algebraic,tamstorf2021discretization}.

Progressive precision and CAMPS both build a precision hierarchy, but they differ in how they make decisions. Progressive precision assigns
precisions in advance using discretization theory, while CAMPS checks each
correction at runtime using the measured residual.
A progressive-precision multigrid cycle could serve as a child solver
in a CAMPS tree. We leave this combination for future work
(Section~\ref{discussion}).

\paragraph{Flexible, inexact, and nested Krylov methods.}
GMRES minimizes the residual over a Krylov subspace. Its standard
right-preconditioned form assumes a fixed linear
preconditioner~\cite{saad1986gmres,saad2003iterative}.
Flexible GMRES stores each preconditioned basis vector separately, allowing the preconditioner to change between Arnoldi
steps~\cite{saad1993fgmres}.
Flexible variants also exist for short-recurrence
methods~\cite{vuik1995insights,szyld2001fqmr,vogel2007flexible,chen2016flexible}.

For SPD systems, Axelsson and Vassilevski, Golub and Ye, Notay, and
Knyazev and Lashuk study variable
preconditioning~\cite{axelsson1991blackbox,golub1999inexact,notay2000fcg,knyazev2007variable}.
Knyazev and Lashuk show that CG with variable preconditioning may not
converge faster than steepest descent~\cite{knyazev2007variable}.
A precision adapter therefore needs a parent solver that supports
variable preconditioning. Local stopping rules, scaling, and failure
handling can change the adapter's behavior between iterations.
They can also make the adapter nonlinear.

The theory of inner-outer flexible
methods~\cite{simoncini2003flexible} and inexact Krylov
methods~\cite{simoncini2003inexact,vandeneshof2004inexact,bouras2005inexact,giraud2007relaxed,simoncini2005superlinear,simoncini2007recent}
provides conditions for relaxing the accuracy of inner solves or
operator applications as the outer residual decreases.
Van den Eshof, Sleijpen, and van Gijzen develop relaxation strategies
for nested Krylov methods~\cite{vandeneshof2005relaxation}.
Their work is the closest precedent for choosing inner tolerances
that avoid unnecessary computation.
Other work studies nested Krylov methods for extreme-scale
systems~\cite{mcinnes2014hierarchical}.

CAMPS does not introduce solver nesting or variable preconditioning.
Inexact-Krylov theory derives accuracy requirements and proves
convergence when they hold. CAMPS assigns an accuracy contract to
each edge in the solver tree and checks each correction at runtime.
Adapters could also use relaxation to request less accuracy from
a child as the outer residual decreases, but we do not explore this
extension here.

Axelsson and Vassilevski require an inner solve to reduce its residual
by a prescribed factor~\cite{axelsson1991blackbox}.
This requirement is an early form of the contraction contract used
by CAMPS: CAMPS checks that contract at each step and rolls back
corrections that violate it.

\paragraph{Residual reliability, finite-precision behavior, and recycling.}
Krylov methods often update the residual recursively. Rounding errors
can create a gap between this residual and the true residual.
This gap limits attainable
accuracy~\cite{greenbaum1997attainable,greenbaum1997book,meurant2006lanczos,liesen2013krylov}.
Residual-replacement strategies periodically recompute the residual
to reduce this
gap~\cite{sleijpen1996reliable,vandervorst2000residual,carson2014residual}.
Despite this issue, GMRES with modified Gram-Schmidt is backward
stable~\cite{drkosova1995stability,paige2006mgs}.

CAMPS checks each correction using $r-Ad$, which it computes directly
in the parent precision. It does not rely on the child's recursively
updated residual. This check costs one parent-precision matrix-vector
product per correction. It keeps the child's residual gap from
affecting the acceptance test. The parent precision determines the
accuracy of this check (Section~\ref{contract}).

Greenbaum, Pt\'ak, and Strako\v{s} show that GMRES can follow any
nonincreasing convergence curve~\cite{greenbaum1996gmres}.
Thus, eigenvalues alone cannot predict how much a Krylov child
will reduce the residual in a given call. CAMPS measures this
reduction directly.

Recycling and augmentation preserve subspace information across
restarts or related linear
systems~\cite{erhel1996deflation,chapman1997deflated,desturler1999truncation,morgan2002deflated,baker2005accelerating,parks2006recycling,giraud2010fgmresdr,gaul2013framework,soodhalter2020survey}.
Oktay and Carson show that mixed-precision GMRES-IR can use GCRO-DR
to recycle subspace information across successive correction
solves~\cite{oktay2023recycling}.

A CAMPS adapter also solves a sequence of correction equations
with the same matrix. However, the children tested here discard
their subspaces after every correction
(Section~\ref{zero-contract}).
Our SPD experiments measure the cost of this choice
(Section~\ref{exp-sparse-spd}). The SPD results suggest that a child that keeps its Krylov subspace between corrections would close most of this gap.

Block methods solve systems with multiple right-hand
sides~\cite{hestenes1952cg,oleary1980block,gutknecht2007block}.
Our study in Section~\ref{exp-dense-wc} does not include these
methods as baselines, and this remains a gap in the evaluation.

\paragraph{Composable solver software.}
Solver composition is an established software pattern.
PETSc represents solvers, preconditioners, and matrices as objects
with abstract interfaces~\cite{balay1997efficient,petsc}.
Users can assemble composite, field-split, multigrid, and
nested-Krylov preconditioners at runtime through option
strings~\cite{brown2012composable}.
Kirby and Mitchell extend this approach across the boundary
between PDE models and linear algebra~\cite{kirby2018solver}.

Trilinos connects interoperable packages through templated
interfaces~\cite{heroux2005trilinos,mayr2025trilinos}.
Belos provides pluggable status tests, while Amesos2 provides
a common interface to direct solvers~\cite{bavier2012amesos2}.
Other examples include hypre~\cite{falgout2002hypre},
AmgX~\cite{naumov2015amgx}, and the AMReX multilevel solver,
which supports interchangeable bottom
solvers~\cite{zhang2019amrex}.
Kokkos Kernels provides reusable computational
kernels~\cite{rajamanickam2021kokkos}.
Krylov.jl provides Krylov methods that support multiple numerical
types~\cite{montoison2023krylov}.

Ginkgo represents matrices, preconditioners, and solvers through
a common linear-operator interface. It also provides portable
GPU backends~\cite{anzt2022ginkgo}.
Related work proposes a modular precision ecosystem at the
kernel level~\cite{anzt2019modular}.

These systems establish the software foundations for solver
composition. Using one solver as another solver's preconditioner
is therefore not a contribution of CAMPS.
The distinction concerns how composition handles precision
boundaries. Ginkgo and Trilinos specify value types per object.
Ginkgo also documents iterative refinement in double precision around a
single-precision inner solver, with conversions inserted where value types
differ~\cite{ginkgo2026mixed}.
PETSc option strings do not specify precision transitions.
Mixed-precision studies using these libraries report that
developers must implement precision boundaries manually through
wrappers~\cite{loe2021experimental}.

CAMPS makes each precision boundary an explicit, typed edge
in the solver tree. Each edge specifies a residual contract
and defines how to handle failure. Our closure theorem establishes
when these components can be composed.
Belos status tests are the closest precedent for the contract, and
PETSc option strings are the closest precedent for the CAMPS grammar.

\paragraph{Algorithm selection and learned solver components.}
Rice formulated algorithm selection as choosing an algorithm
from problem features~\cite{rice1968polyalgorithms,rice1976}.
For sparse solvers, early work includes the self-adapting
numerical software program~\cite{dongarra2003selfadapting},
machine-learned selection among PETSc
configurations~\cite{bhowmick2006application,eijkhout2008multistage},
and the Lighthouse
framework~\cite{motter2015lighthouse,jessup2016lighthouse}.
Recent work explores parameter autotuning~\cite{petrushov2023automated},
learned matrix embeddings~\cite{liuweng2026embedding},
and benchmark datasets for solver
selection~\cite{xiong2026solverset}.
Surveys cover learned iterative methods~\cite{zou2023survey}
and automated algorithm selection more
broadly~\cite{kerschke2019automated}.

Researchers have also developed learned components for individual
solvers. These include multigrid prolongation and
coarsening~\cite{greenfeld2019learning,luz2020learning,taghibakhshi2021optimization}
and conjugate-gradient
preconditioners~\cite{li2023learning,hausner2024neural,yang2025learning}.
Other work automates precision assignment.
At the variable level, approaches include dynamic
search~\cite{rubiogonzalez2013precimonious},
rigorous static bounds~\cite{chiang2017rigorous},
and algorithmic differentiation~\cite{menon2018adapt}.
For neural networks, hardware-in-the-loop reinforcement learning
assigns precision by layer~\cite{wang2019haq}.
Carson and Chen use a contextual bandit to select solver
precisions at runtime~\cite{carson2026autotuning}.

Several existing methods could train a selector over solver trees.
Policy-gradient and actor--critic methods could learn from
solver runs~\cite{schulman2017ppo,haarnoja2018sac}.
Offline reinforcement learning could use logged
solves~\cite{kumar2020cql,kostrikov2022iql}.
Constrained and Lyapunov-based policy optimization could impose
constraints during learning~\cite{achiam2017cpo,chow2019lyapunov}.
Other options include planning with a learned world
model~\cite{hafner2023dreamerv3} and sampling diverse, high-reward
candidates with generative flow
networks~\cite{bengio2021gflownet}.
Bandit-based hyperparameter optimization provides a
non-sequential baseline for evaluating such
policies~\cite{falkner2018bohb}.
Reinforcement learning combined with correctness checks has
already discovered faster matrix-multiplication and sorting
routines~\cite{fawzi2022alphatensor,mankowitz2023alphadev}.

Control-theoretic approaches to
learning~\cite{nguyen2025differential,wang2026descent,bajaj2026physics,bhardwaj2026phast,mclennan2025learning,ellendula2026learning}
and port-Hamiltonian interconnection theory offer a conceptual
analogy for adapter composition.
Power-conserving interconnections of port-Hamiltonian subsystems
preserve the port-Hamiltonian structure~\cite{vanderschaft2014port}.
This suggests viewing adapters as ports and asking which
guarantees their composition preserves.

This paper implements neither a learned component nor a selector.
It defines a search space for future selectors: a grammar of
precision-typed solver trees with runtime accuracy checks.
These checks detect corrections that violate the residual
contract, including those produced by a poor policy choice.
They operate after each correction rather than only at the
end of a solve.
Each learned preconditioner cited above is a potential child
solver, given that it satisfies the interface in
Section~\ref{zero-contract}.

\paragraph{The closest prior art.}
Four works are especially close to CAMPS.
Buttari et al.\ interpret mixed-precision refinement as
preconditioned Richardson iteration. They also use a
reduced-precision iterative solver as the preconditioner
for an outer FGMRES method. This establishes the inner-outer
mixed-precision pattern that CAMPS
generalizes~\cite{buttari2008sparse}.
Hogg and Scott incorporate runtime fallback to a double-precision
factorization in a production sparse solver~\cite{hogg2010mixed}.
Oktay and Carson use the convergence rate to decide when to
escalate between refinement variants~\cite{oktay2022multistage}.

Suzuki and Iwashita combine FP64, FP32, and FP16 arithmetic
in a deeply nested CPU/GPU Krylov solver.
Their evaluated F3R configuration uses three FGMRES levels
and an innermost Richardson level.
Only the outermost FGMRES level checks
convergence~\cite{suzuki2025nested}.
Their general framework also allows Richardson iteration
at multiple levels in principle.
However, they do not define a mixed-precision adapter contract, and they also do not prove that independently stopping refinement
layers can be inserted at arbitrary nesting depths.

We therefore do not claim solver nesting, progressive precision
reduction, or runtime fallback as new contributions.
Nor do we claim to introduce the general idea of using
Richardson iteration at multiple levels.

\paragraph{Positioning of CAMPS.}
CAMPS makes both mathematical and architectural contributions.
It represents each precision transition as a typed edge
$\IR{p}{q}$. Each edge can use a different child solver family.
Every application must satisfy a residual contract evaluated
at the parent level.

Our arbitrary-layer theorem establishes when refinement layers
can be composed at any depth. It states explicit hypotheses
and a residual bound for each edge
(Section~\ref{arbitrary-layer}).
Our propagation bound describes how the constants at one level
affect the next (Section~\ref{propagation}).

A CAMPS hierarchy can place an independently stopping adapter
at any or every precision boundary. This contrasts the published F3R configuration, which places Richardson iteration
only at the innermost level. 
CAMPS also supports escalation after an individual correction
fails its contract. It rolls back that correction and switches
child solvers. This mechanism can operate at any edge with
any child solver that satisfies the interface.

Section~\ref{exp-f3r} evaluates these design choices.
We test adapters at one edge, at two edges, and in configurations
that combine adapters with precision casts. A cast edge connects a parent to a lower-precision child like an adapter does, but it accepts the child’s correction without checking the residual.
We also reimplement the F3R structure in the same experimental
framework and apply identical tuning.
These experiments place F3R, the fully instrumented CAMPS
hierarchy, and the intermediate configurations within a common
design space. 

The best measured configuration checks corrections at exactly
one edge. It matches or outperforms F3R while using
$3.4\times$ less peak memory.
We leave comparison with the authors' public
implementation~\cite{suzuki2025nested} at their problem scales
to future work.
\section{Grammar}
\label{grammar}
\begin{figure}[H]
\centering
\definecolor{campsfpA}{HTML}{86B6EF}
\definecolor{campsfpB}{HTML}{CDE2FB}
\definecolor{campsfpC}{HTML}{F3F8FE}
\definecolor{campsadapter}{HTML}{2A78D6}
\definecolor{campsink}{HTML}{52514E}
\definecolor{campsband}{HTML}{F4F4F1}
\begin{tikzpicture}[
  font=\footnotesize,
  nd/.style={draw=campsink, rounded corners=1.5pt, minimum height=5.6mm,
             minimum width=19mm, inner xsep=3pt, align=center},
  p64/.style={nd, fill=campsfpA},
  p32/.style={nd, fill=campsfpB},
  p16/.style={nd, fill=campsfpC},
  ad/.style={nd, draw=campsadapter, line width=1.1pt, fill=white},
  ed/.style={draw=campsink, line width=0.6pt},
  lbl/.style={font=\scriptsize, text=campsink},
  note/.style={font=\scriptsize, anchor=west, align=left},
  flow/.style={draw=campsink, line width=0.5pt, ->},
]
\fill[campsband] (0,0.55) rectangle (12.6,-1.0);
\fill[campsband] (0,-3.0) rectangle (12.6,-4.55);
\node[lbl, anchor=west] at (0.15,-0.2) {FP64};
\node[lbl, anchor=west] at (0.15,-2.0) {FP32};
\node[lbl, anchor=west] at (0.15,-3.8) {FP16};

\node[p64] (n0) at (5.0,0)    {FGMRES$_{64}$};
\node[ad]  (n1) at (5.0,-1.0) {$\mathrm{IR}_{64\leftarrow32}$};
\node[p32] (n2) at (5.0,-2.0) {FGMRES$_{32}$};
\node[ad]  (n3) at (5.0,-3.0) {$\mathrm{IR}_{32\leftarrow16}$};
\node[p16] (n4) at (5.0,-4.0) {ILU(0)$_{16}$};
\draw[ed] (n0) -- (n1) -- (n2) -- (n3) -- (n4);

\draw[flow] (2.2,0.1) -- node[lbl, left, align=right] {residuals\\go down} (2.2,-4.1);
\draw[flow] (2.6,-4.1) -- node[lbl, right, align=left] {corrections\\come up} (2.6,0.1);

\node[note] at (6.2,0)    {outer Krylov solver, with FP64 residuals and basis};
\node[note] at (6.2,-1.0) {adapter: scales and casts the residual to FP32,\\
                           and checks the returned correction in FP64};
\node[note] at (6.2,-2.0) {inner Krylov solver in FP32};
\node[note] at (6.2,-3.0) {adapter: casts the residual to FP16,\\
                           and checks the returned correction in FP32};
\node[note] at (6.2,-4.0) {base preconditioner, applied in FP16};
\end{tikzpicture}
\caption{A solver tree generated by the grammar,
$\FGMRES_{64}\bigl[\IR{64}{32}[\FGMRES_{32}[\IR{32}{16}[\ILU(0)_{16}]]]\bigr]$.
Shaded bands mark the precision of each node. Each adapter sits on the
boundary between the two precisions it joins. It sends the residual down to
the lower precision and checks the correction that comes back in the higher
one.}
\label{fig:tree}
\end{figure}
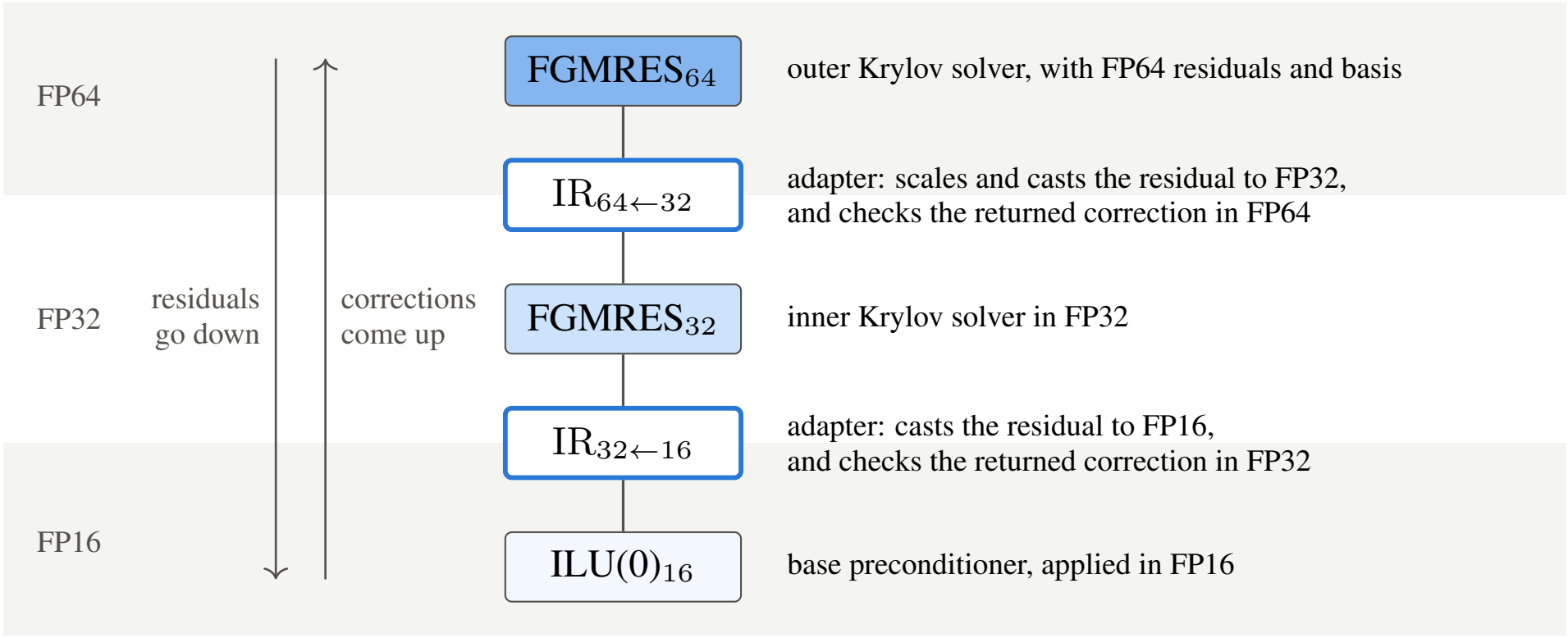

Let $F_0,F_1,\dots,F_L$ denote the supported floating-point formats, ordered
from highest to lowest precision, with unit roundoffs
$u_0<u_1<\cdots<u_L$. Let $Q_p(z)$ denote rounding or conversion to format~$p$.
After any required equilibration, $A_p=Q_p(A)$ denotes the coefficient values
stored at level~$p$. A solver node $S_p$ maps a right-hand side in
$F_p^{\,n}$ to an approximate solution in $F_p^{\,n}$. The following
conceptual grammar generates CAMPS solver trees:
\[
  S_p ::= \operatorname{Base}_p(\text{parameters})
        \;\mid\; \IR{p}{q}^{\,J,\tau}[S_q]
        \;\mid\; \FGMRES_{p}^{\,m,\tau}[S_p]
        \;\mid\; \operatorname{Cast}_{p\leftarrow q}[S_q],
        \qquad u_p<u_q .
\]
Here $\operatorname{Base}_p$ may be ILU(0), a sparse approximate inverse,
Jacobi, or any operator whose input and output formats are~$p$. The type rule
for $\IR{p}{q}[S_q]$ requires the child to accept precision~$q$ and guarantees
that the adapter returns precision~$p$. This allows recursive construction
while stopping implicit or undocumented casts between incompatible nodes.
A flexible Krylov node calls a child of its own precision, so a
lower-precision child must sit under an adapter or a cast.
An edge from $p$ to $q$ also requires $F_q\subset F_p$, so that promoting the
child's output adds no error. FP16 and BF16 are not nested in either
direction, so no edge connects them, and both sit below FP32. An adapter may
also call a fallback child in its own precision
(Algorithm~\ref{alg:adapter}), which is the one case with $q=p$.
Figure~\ref{fig:tree} shows a representative recursively typed hierarchy.

The fourth production, $\operatorname{Cast}_{p\leftarrow q}[S_q]$, is a plain
precision boundary. It converts the parent's vector to the lower precision,
invokes $S_q$, and converts the result back to the parent precision. It
performs no residual computation, scaling, or acceptance test. It is
type-correct but unchecked since it carries no residual contract.
Since a cast edge checks nothing, it should only sit below a parent that does not rely on a check, such as a flexible Krylov solver, which can absorb a poor correction, or an adapter higher in the tree, which checks the combined result. Throughout the paper, we distinguish checked
corrections (adapter edges) from unchecked directions (cast edges and
the guard substitutions of Section~\ref{strategy}).

These four productions cover the node types we use in this paper, but the
list is not complete. What decides whether a node can appear in a tree is
the contract in Section~\ref{zero-contract}, not the algorithm it runs. Any
node that meets the contract can sit at any position. For example, the
experiments in Section~\ref{exp-sparse-spd} use a conjugate gradient child,
which has no production of its own. A multigrid cycle or a sparse
triangular solve would work the same way, with no change to the adapter.
\hypertarget{adapter-approach}{%
\section{Adapter}\label{adapter-approach}}
CAMPS is a family of nested solvers, not a single algorithm. This section
describes one adapter and the condition its corrections must satisfy. It
then shows why adapters can be nested to any depth, gives a strategy for
building a solver tree, and defines the five instances we test in
Section~\ref{experiments}.

\subsection{Adapter algorithm}\label{adapter-algorithm}
\begin{figure}[H]
\centering
\includegraphics[width=0.85\linewidth]{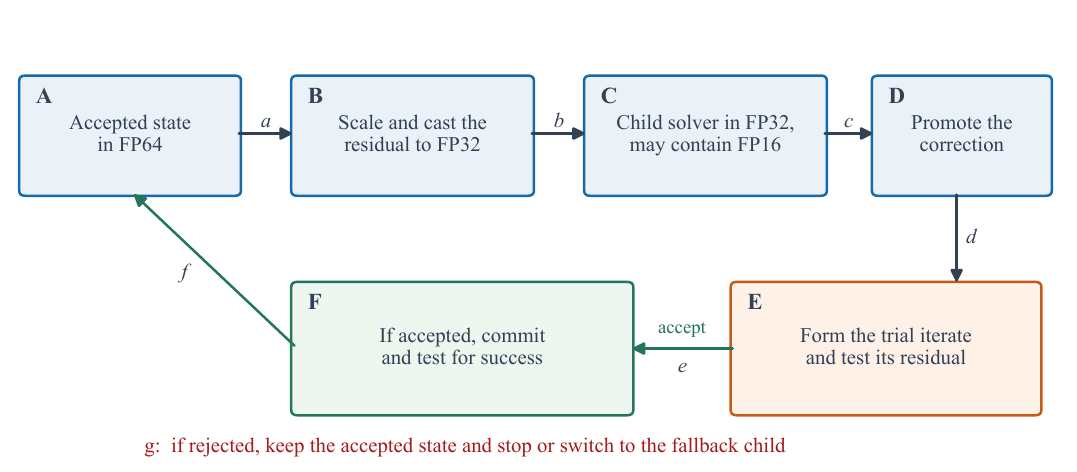}
\caption{One adapter call at the boundary between FP64 and FP32. Blocks A to F
are the steps of Algorithm~\ref{alg:adapter}, and edges a to g show what
passes between them. A: the accepted iterate and its residual, kept in FP64
(step~1). B: the residual is scaled by a power of two and cast to FP32
(step~3, edge a). C: the child solver computes a correction and may contain
further adapters (step~4, edge b, Section~\ref{zero-contract}). D: the
correction is promoted to FP64 and unscaled (step~5, edge c). E: the
adapter forms the trial iterate and recomputes its residual in FP64 (steps~6
and~7, edge d, Section~\ref{contract}). F: if the test accepts, the trial
becomes the accepted state, and the adapter checks for convergence (step~7,
edges e and f). Edge g: a rejected trial leaves the accepted state unchanged,
and the adapter stops or switches to its fallback child.}
\label{fig:adapter-call}
\end{figure}

An adapter solves a parent-level system $A_p y = g_p$. It starts from
$y_p=0$ and makes at most $J$ attempts, where an attempt is any correction
that reaches the acceptance test. Rejected attempts use up the budget just
as accepted ones do. Algorithm~\ref{alg:adapter} gives the full procedure, and
Figure~\ref{fig:adapter-call} shows one call as a diagram. The block letters in
the figure match the steps of the algorithm.

At attempt $t$, the adapter computes the residual
$r_p^{(t)}=g_p-A_p y_p^{(t)}$ in the parent precision~$p$. Our code can
also form it in a higher precision, but no experiment in this paper does.
If $\norm{r_p^{(t)}}_2\le\tau_p\norm{g_p}_2+\mathrm{atol}_p$, the adapter
returns $y_p^{(t)}$. Otherwise, it scales the residual, converts it to the
child precision~$q$, and calls the child solver $S_q$. It converts the
child's correction $c_q^{(t)}$ back to precision~$p$, undoes the scaling,
and forms the trial iterate
$y_p^{(t+1)}=\fl_p(y_p^{(t)}+s^{-1}Q_p(c_q^{(t)}))$.

The adapter then recomputes the residual at the trial iterate and applies
the acceptance test of Section~\ref{contract}. It keeps the trial iterate
only if the test passes. If the test fails, it throws the trial away and
either stops with a status that says why or switches to a fallback child.

When we write a correction equation such as $A_q c_q=r_q$, we mean the
equation the child aims to solve, not one it directly solves. An ILU child,
for example, or an iterative solver stopped at a loose tolerance, only returns an approximate solution.

The scale $s$ is a power of two, so in binary arithmetic, scaling adds no
rounding error, as long as the scaled values do not overflow or
underflow. This is very important before a conversion to FP16, whose narrow
range would overflow large entries or flush small ones to zero.

Scaling the right-hand side cannot fix problems in the matrix itself.
Converting $A$ to precision~$q$ can overflow, underflow entries that
matter for its structure, or create a zero pivot. Avoiding these
requires equilibrating the matrix, which is a separate step.

All experiments in Section~\ref{experiments} run the full algorithm,
including power-of-two scaling, the acceptance test with rollback, the
final convergence check, and stagnation detection. Damping, the energy
test, and fallback children are used only where we say so.

\begin{algorithmblock}{Algorithm 1. \textsc{ApplyAdapter}$(p,q,A_p,g_p,S_q,J,\tau)$}
\label{alg:adapter}
\begin{enumerate}[label=\arabic*.,leftmargin=2.2em,itemsep=1.5pt,topsep=3pt]
  \item \emph{(Block A.)} Set $y_p=0$. Form $r_p=g_p-A_py_p$ in parent precision~$p$ and set
    $\rho=\norm{r_p}_2$. Call $(y_p,r_p)$ the \emph{accepted state} and
    $(q,S_q)$ the \emph{active pair}. If
    $\rho\le\tau\norm{g_p}_2+\mathrm{atol}$, return $y_p$ with status
    \textsc{Success}.
  \item For each attempt $t=0,\dots,J-1$, carry out steps 3--8. Rejected
    attempts count toward $J$, and so does the attempt on which the
    adapter switches to its fallback child.
  \item \hspace{1em}\emph{Scale and cast (block B).} If $q=p$, set $s=1$ and skip
    the cast. Otherwise, choose a power-of-two scale $s$ that places
    $\norm{s\,r_p}_\infty$ in a fixed interval $(2^{\zeta-1},2^{\zeta}]$. If the converted residual is zero or not
    finite, return $y_p$ with status \textsc{Failure}.
  \item \hspace{1em}\emph{Child solve (block C).} Compute $c_q=S_q(Q_q(s\,r_p))$
    from the accepted residual. If $c_q$ is not finite, return $y_p$ with
    status \textsc{Failure}.
  \item \hspace{1em}\emph{Promote (block D).} Set $d=s^{-1}Q_p(c_q)$. If damping is
    on, let $w=A_pd$. Set $\alpha=0$ if $w=0$ or
    $\langle r_p,w\rangle\le0$, and
    $\alpha=\min\{\langle r_p,w\rangle/\norm{w}_2^2,\ \alpha_{\max}\}$
    otherwise. Then replace $d$ by $\alpha d$ (Section~\ref{contract}).
  \item \hspace{1em}\emph{Trial iterate (block E).} Form $y'=\fl_p(y_p+d)$, the
    displacement $h=y'-y_p$ that would actually be committed, and the
    recomputed residual $r'=g_p-A_py'$ in parent precision, with
    $\rho'=\norm{r'}_2$.
  \item \hspace{1em}\emph{Acceptance test (blocks E and F).} The default test requires
    $\rho'\le\rho\,(1+\sigma u_p)$, a slack of a few parent ulps
    (Section~\ref{contract}). With a progress threshold $\varrho<1$, it
    requires $\rho'\le\varrho\,\rho$ instead. For SPD $A_p$, the
    energy test requires $\widehat{\Delta\phi}+E<0$, where
    $\widehat{\Delta\phi}=\fl(\tfrac12h^{\mathsf T}\!A_ph-r_p^{\mathsf T}h)$
    and $E$ is the error estimate \eqref{eq:energy-bound}.
    If the test passes, set $y_p\leftarrow y'$, $r_p\leftarrow r'$, and
    $\rho\leftarrow\rho'$. If now $\rho\le\tau\norm{g_p}_2+\mathrm{atol}$,
    return $y_p$ with status \textsc{Success}.
    If the test fails, discard $y'$ and $r'$, so the accepted state does
    not change. Under \textsc{on-violation: reject}, return $y_p$ with
    status \textsc{Rejected}. Under \textsc{on-violation: fallback}, switch
    the active pair once to the fallback $(q_f,S_f)$, with
    $u_p\le u_{q_f}$, for the remaining attempts. The next attempt then
    builds the child's right-hand side from the accepted $r_p$ at the new
    precision. The fallback may run in the parent precision itself, with
    $q_f=p$, $s=1$, and no cast. If the fallback child also fails the
    test, return $y_p$ with status \textsc{Rejected}.
  \item \hspace{1em}\emph{Stagnation.} If $m_{\mathrm{stag}}$ consecutive
    accepted corrections each leave $\rho$ above $\rho_{\mathrm{stag}}$
    times its previous value, return $y_p$ with status
    \textsc{Stagnated}.
  \item If all $J$ attempts are used, return $y_p$ with status
    \textsc{Max-Steps}. Every exit returns the last accepted iterate. For
    every attempt, the diagnostics record the active pair,
    $(\rho,\rho',\alpha)$, $(\widehat{\Delta\phi},E)$ when the energy test
    is used, and whether the attempt was accepted.
\end{enumerate}
\end{algorithmblock}

\subsection{The residual-reducing correction contract}\label{zero-contract}
Our composition results rest on one requirement. We state it first in
exact arithmetic. It applies to any nonsingular $A$,
whether dense, sparse, or structured. Fix an iterate $x_k$ with residual
$r_k=b-Ax_k$. We call a map $C$ a residual-reducing correction if,
starting from the zero vector, it returns a correction $d_k=C(A,r_k)$ with
\begin{equation}\label{eq:clean}
  \norm{r_k - A d_k}_2 \le \norm{r_k}_2 .
\end{equation}
In words, the correction never makes the residual larger. Nothing else
about $C$ is assumed. Applying the correction, $x_{k+1}=x_k+d_k$, gives
\[
  r_{k+1}=b-Ax_{k+1}=(b-Ax_k)-Ad_k=r_k-Ad_k,
\]
so \eqref{eq:clean} means directly that
$\norm{r_{k+1}}_2\le\norm{r_k}_2$. If the inequality is strict, the
residual drops at every step. If it holds with equality, the residual may
stay the same.

Every node that meets this condition looks the same to the solver above
it. Given the matrix, a right-hand side, and a zero initial guess, it
returns a correction that does not increase the residual it was given:
\[
  \boxed{\;\texttt{solve}(A,\,r,\,0)\;\longrightarrow\;d
  \quad\text{such that}\quad \norm{r-Ad}_2\le\norm{r}_2\;}
\]
A solver can therefore use any such node as a correction step or as a
preconditioner. Two properties of this condition shape the rest of the
framework.

\paragraph{Why the child starts from zero.} When the child starts from
$d=0$, its initial residual is the parent's current residual $r_k$. When
it returns $d_k$, its final residual $r_k-Ad_k$ is the parent's next
residual $r_{k+1}$. Thus, the child's own stopping test measures the
parent's progress directly, with no change of variables between levels.

This makes \eqref{eq:clean} automatic for many children. A
residual-monotone method such as GMRES or MINRES never ends with a larger
residual than it started with, so it returns $d$ with
$\norm{r_k-Ad}_2\le\norm{r_k-Ad_0}_2$, where $d_0$ is its initial guess.
With $d_0=0$, the right-hand side of this inequality is $\norm{r_k}_2$, so
\eqref{eq:clean} holds without any extra check. At worst, the residual
stays the same.

A nonzero start can also work. It satisfies \eqref{eq:clean} whenever
$d_0$ itself does, that is, whenever $\norm{r_k-Ad_0}_2\le\norm{r_k}_2$,
and a good initial guess can reduce the residual further. But no fixed
nonzero start works for every right-hand side. For any $d_0\ne0$, take
$r=-\epsilon Ad_0$ with $\epsilon>0$. Then
$\norm{r-Ad_0}_2=(1+\epsilon)\norm{Ad_0}_2$, which is larger than
$\norm{r}_2=\epsilon\norm{Ad_0}_2$. For a residual-monotone child,
starting from zero is thus sufficient for \eqref{eq:clean}, as it is the only fixed start that works for every right-hand side, but it is not a necessary condition.

\begin{figure}[H]
\centering
\includegraphics[width=0.7\linewidth]{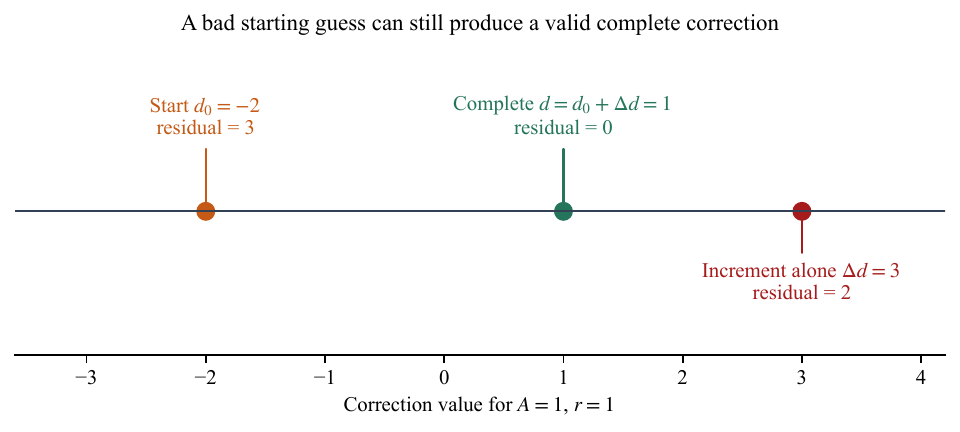}
\caption{A warm start with $A=1$ and $r=1$. The starting guess $d_0=-2$ has residual $3$, but the complete correction $d=d_0+\Delta d=1$ has residual zero. Returning the increment $\Delta d=3$ alone would give a residual of size $2$.}
\label{fig:warm-start}
\end{figure}

Children that reuse earlier work still fit the interface. The identity
$b-A(x+d)=r-Ad$ holds however the child produces $d$, so a child may
warm-start, keep a recycled or augmented Krylov
subspace~\cite{parks2006recycling}, or keep the history of a
variable-preconditioned conjugate gradient
method~\cite{notay2000fcg,knyazev2007variable}. It only has to return the
complete correction $d_0+\Delta d$, measured against the $r$ it was given.
The adapter tests that complete correction, so $d_0$ does not need to pass
the test on its own. Figure~\ref{fig:warm-start} gives a one-dimensional example.

All children in Section~\ref{experiments} start from zero, and none of
them reuses Krylov information between corrections. On the SPD problem of
Section~\ref{exp-sparse-spd}, this adds only about $14$ CG iterations
before the tolerance is reached at $64^{3}$.

\paragraph{The contract composes at any depth.} Condition
\eqref{eq:clean} involves $A$, $r$, and the returned $d$. It does not say anything
about where $C$ sits in the solver tree. Thus, any node that meets the
condition can supply corrections to a parent that uses this interface,
and that parent can meet the condition too.

For example, let $C_1$, $C_2$, and $C_3$ be an LU solve, a GMRES solve,
and a refinement adapter, each satisfying
$\norm{r-AC_i(A,r)}_2\le\norm{r}_2$. An outer solver can call any of them
in the same way. This gives a recursive definition,
\[
  \textsf{Solver} \;=\; \text{outer iteration} \;+\;
  \text{residual-reducing correction},
\]
where the correction can itself be a solver. The standard pairing of an outer
Krylov method with a fixed preconditioner is one case of this definition,
given that the preconditioner meets the contract. CAMPS also lets the
correction contain further solver layers.

\subsection{The mixed-precision contract with a finite-precision floor}\label{contract}
Converting between formats and rounding the residual and the update
introduce small errors, so \eqref{eq:clean} cannot be guaranteed in finite
precision. To account for these errors, we replace \eqref{eq:clean} with a
bound that requires contraction down to a floor. For a right-hand
side~$g$, suppose $S$ returns $y=S(g)$ with
\begin{equation}\label{eq:one-step}
  \norm{g-A_p S(g)}_2 \le \eta_p\norm{g}_2 + \omega_p,
  \qquad 0\le\eta_p<1,
\end{equation}
where $\omega_p$ is a floor term that accounts for rounding error. The
clean contract \eqref{eq:clean} corresponds to the boundary case
$\eta_p=1$ and $\omega_p=0$. Requiring $\eta_p<1$ tightens the
condition from non-increase to contraction.

The child does not receive the exact residual $r_t$. It receives the
computed residual $\hat r_t=r_t+e_t$, and \eqref{eq:one-step} bounds its
correction relative to that input. The parent's next exact residual also
depends on the evaluation error $e_t$ and on the rounding error
$\delta y_t$ in the update. Thus, a geometric bound for the parent needs
the recurrence
\begin{equation}\label{eq:recurrence-form}
  \norm{r_{t+1}}_2\le\eta_p\norm{r_t}_2+\omega_p
   +(1+\eta_p)\norm{e_t}_2+\norm{A_p\,\delta y_t}_2 .
\end{equation}
Lemma~\ref{lem:recurrence} in Section~\ref{guarantees} derives this
recurrence from the rounding model and gives its constants explicitly.
Lemma~\ref{lem:geom} below states the geometric bound that follows from
it.

\begin{lemma}[Per-edge geometric residual bound]\label{lem:geom}
Suppose the exact parent-level residuals $r_t=g-A_py_t$ of $J$ consecutive
accepted corrections satisfy
\[
  \norm{r_{t+1}}_2\le\eta\,\norm{r_t}_2+\omega,\qquad t=0,\dots,J-1,
\]
with constants $0\le\eta<1$ and $\omega\ge0$ that are uniform over these
steps, for instance $\eta=\hat\eta$ and $\omega=f(g)$ of
Lemma~\ref{lem:recurrence}. Then
\[
  \norm{r_J}_2
  \le \eta^{\,J}\,\norm{r_0}_2
    + \frac{1-\eta^{\,J}}{1-\eta}\,\omega .
\]
\end{lemma}

The proof appears in Appendix~\ref{app:contract-proofs}.

\begin{figure}[H]
\centering
\includegraphics[width=\linewidth]{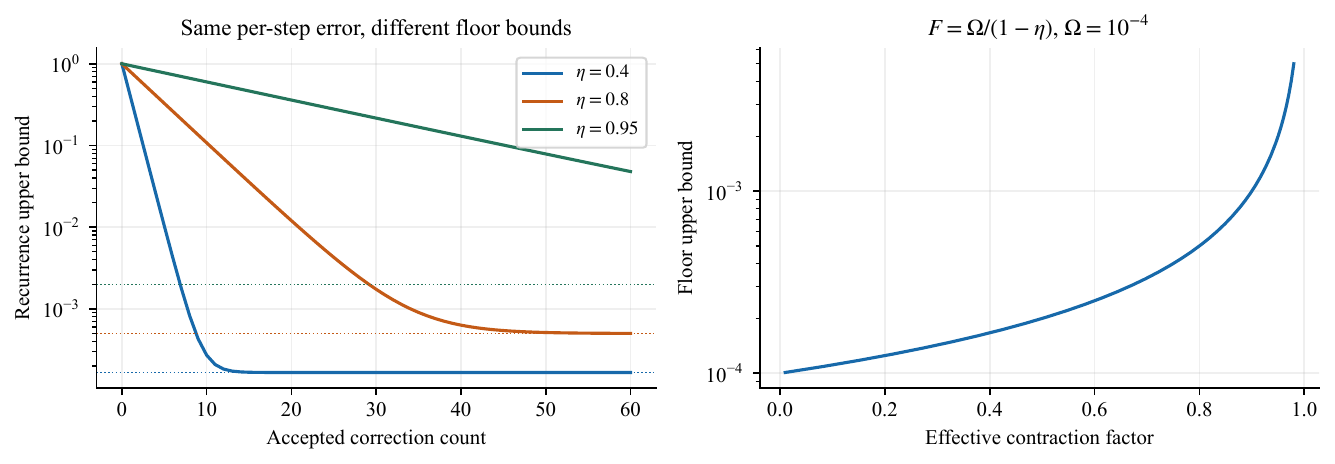}
\caption{Left: the recurrence $R_{t+1}=\eta R_t+\Omega$ with $\Omega=10^{-4}$ and $R_0=1$, for three contraction factors. Right: the floor $\Omega/(1-\eta)$ grows as $\eta$ approaches one. For a real solver these curves are upper bounds.}
\label{fig:geometric-floor}
\end{figure}

Lemma~\ref{lem:geom} bounds the residual whenever the recurrence holds, and
Figure~\ref{fig:geometric-floor} shows how its floor depends on the
contraction factor. It
is not a complete forward-error theorem, and it does not show that
$\eta_p<1$ holds for every child. The contract still applies to children
that are nonlinear or that change from one call to the next because it is
checked against the residual that each call achieves.

For a fixed linear child operator $B_p$, the condition
$\norm{I-A_pB_p}_2<1$ is sufficient for contraction in exact arithmetic.
In mixed precision, the analysis must also account for conversion,
residual evaluation, and rounding in the update. A cast to low precision
can also make $A_q$ singular or cause a factorization to fail. When that
happens, a refinement wrapper cannot recover the information lost in the
conversion.

Theorem~\ref{thm:propagation} in Section~\ref{propagation} derives
$\eta_p$ and $\omega_p$ from the child's own contract, the conversion
error, and the mismatch $A_p-A_q$ between the two stored matrices. The
statements below cover errors in the computed residual, the cases where the
contraction guarantee no longer holds, and two extensions of the
acceptance test.

\medskip
\noindent
\textbf{Errors in the computed residual.}
The floor $\omega_p$ must cover every source of rounding error in one
correction. These are the evaluation of the residual, the casts and
scaling, the mismatch term in \eqref{eq:propagation}, the rounding of the
update, and the evaluation of the norm. The floor is not a fixed constant.
Its size depends on the current iterate and is about
$u_p(\norm{g}_2+\norm{A_p}_2\norm{y}_2)$~\cite{higham2002asna}.
Lemma~\ref{lem:recurrence} gives it explicitly in \eqref{eq:fp-recurrence}.

Algorithm~\ref{alg:adapter} compares two computed norms and allows a slack
of $\sigma u_p$, a few units in the last place of the parent format. This
slack absorbs error in the comparison itself. The slack test measures progress in computed norms. It is a monitor, not a
proof that the exact residual decreased. It does not account for the
error that cancellation can cause when $g-A_py$ is computed. A
conservative test would use verified bounds $\epsilon$ and $\epsilon'$ on the
evaluation errors~\cite{rump2010verification,ogita2005accurate} and accept when
\[
  \norm{\hat r'}_2+\epsilon' \le \varrho\,\max\{0,\ \norm{\hat r}_2-\epsilon\},
  \qquad \varrho<1.
\]
When this test is inconclusive, the adapter could evaluate the residual in
higher precision~\cite{sleijpen1996reliable} or take another recovery
action. Our experiments use the simpler test with slack and report the
recomputed residuals it compared. Under the hypotheses of
Theorem~\ref{thm:recursive}, Proposition~\ref{prop:exits} shows that this
test does not reject a correction unless the residual is already within a
constant factor of the floor.

\medskip
\noindent
\textbf{Where the guarantee is lost.}
The classical sufficient condition for $\eta_p<1$ can fail. For a
backward-stable child, the standard bound is
$\norm{g-A_p\hat d}_2\lesssim c(n)\,u\,\norm{A_p}_2\norm{\hat d}_2$~\cite{higham2002asna},
where $u$ is the unit roundoff of the child's precision. Once
$\norm{A_p^{-1}g}_2\gtrsim\norm{g}_2/(c(n)\,u\,\norm{A_p}_2)$, the
right-hand side of this bound becomes larger than $\norm{g}_2$, so the
bound no longer guarantees contraction.

A failed sufficient condition does not make contraction impossible. For
example, take $A=\operatorname{diag}(1,2^{-100})$ and
$d=(g_1,\,2^{100}g_2)$. As long as these values are representable,
$g-Ad=0$ in FP64 with no rounding error, even though
$\kappa_2(A)u_{64}=2^{47}$.

Three different events can occur here. The sufficient condition can fail,
a particular computed correction can fail the acceptance test, or errors
in evaluating the residual can keep the test from recognizing a
correction that is actually good. On its own, none of these events shows
that every correction must fail.

Section~\ref{exp-dense-ic} measures the first two events for the children
we tested. It also shows that a child in the parent precision still
contracts in the same regime. We leave the third event as open for future work. A solver
node in this regime can still give an enclosing flexible Krylov node a
direction that has not passed the test (Section~\ref{outer-fgmres}). The
residual contract applies to refinement edges, not to every
preconditioner application, and \textsf{CAMPS-Ladder} relies on this
distinction.

\medskip
\noindent
\textbf{Damping and an energy contract for SPD systems.}
Two extensions of the acceptance test can avoid unnecessary rejections.
The first is damping: a direction $d$ that fails the test at unit step may
still reduce the residual at a different step size. For $w=A_pd\ne0$, the
step $\alpha_*=\langle r,w\rangle/\norm{w}_2^2$ minimizes
$\norm{r-\alpha w}_2$ over all real $\alpha$.

\begin{figure}[H]
\centering
\includegraphics[width=0.9\linewidth]{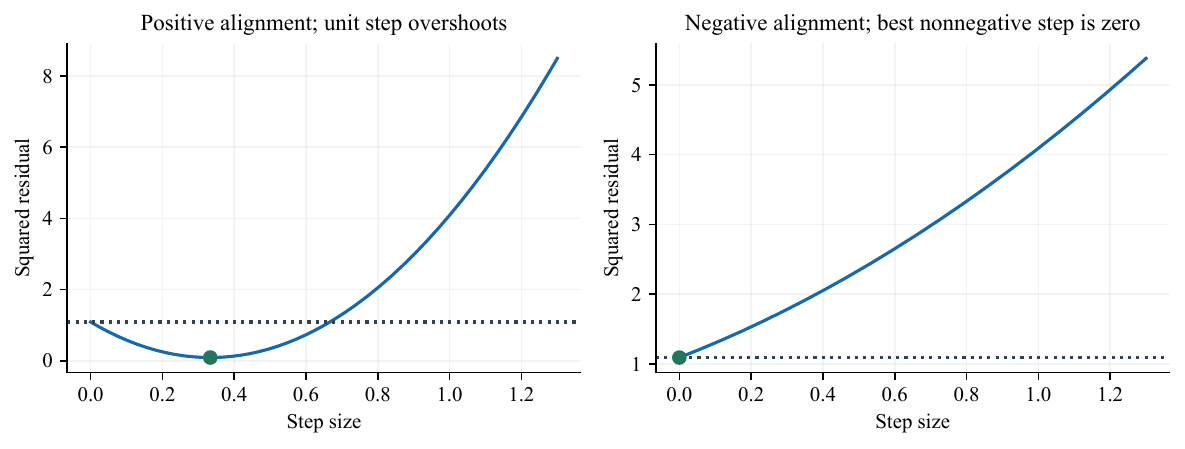}
\caption{Restricted damping for $r=(1,0.3)^{\mathsf T}$. Left: with $w=(3,0)^{\mathsf T}$, the best step is $\alpha=1/3$, and a unit step makes the residual larger. Right: with $w=(-1,0)^{\mathsf T}$, the best nonnegative step is zero.}
\label{fig:damping}
\end{figure}

Our implementation restricts the step to $[0,\alpha_{\max}]$. If $w=0$ or
$\langle r,w\rangle\le0$, it sets $\alpha=0$. This null step leaves the
residual unchanged, and the stagnation detector records the lack of
progress. Otherwise, it sets
$\alpha=\min\{\alpha_*,\alpha_{\max}\}\in(0,\alpha_*]$, which strictly
reduces the residual in exact arithmetic.
Appendix~\ref{app:contract-proofs} gives the derivation, and
Figure~\ref{fig:damping} shows both cases. Damping costs one
extra matrix-vector product in the parent precision, and the rounded
trial iterate must still pass the acceptance test in step~7 of
Algorithm~\ref{alg:adapter}.

\begin{figure}[H]
\centering
\includegraphics[width=0.9\linewidth]{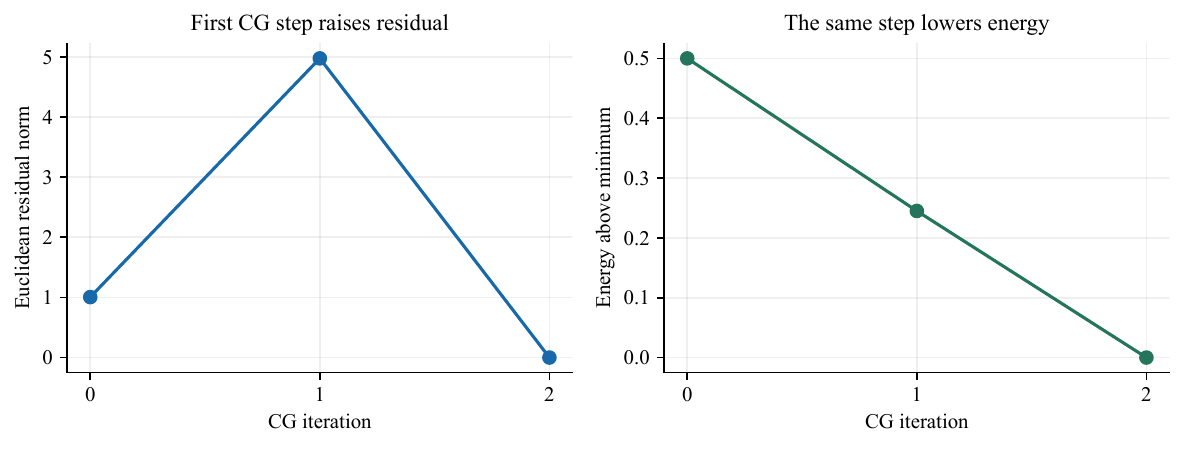}
\caption{CG in exact arithmetic on $A=\operatorname{diag}(1,100)$ with right-hand side $(1,0.1)^{\mathsf T}$ and a zero start. The first step raises the Euclidean residual (left) while it lowers the energy (right).}
\label{fig:cg-energy}
\end{figure}

The second extension is an energy test for SPD $A_p$. For these matrices,
a test on the Euclidean residual can reject useful corrections from a
conjugate gradient child. CG minimizes the $A$-norm of the error, but its
Euclidean residual does not have to decrease at every step. For example,
with $A=\operatorname{diag}(1,100)$ and $r=(1,0.1)^{\mathsf T}$, the first
CG step from zero raises $\norm{r}_2$ from about $1.005$ to $4.975$, even
though the energy decreases (Figure~\ref{fig:cg-energy}).

Thus, for SPD parents, the adapter also offers an energy contract based
on $\phi(y)=\tfrac12y^{\mathsf T}A_py-g^{\mathsf T}y$. The contract must
test the energy change of the committed iterate $y'=\fl_p(y+d)$, not of
the proposed correction $d$:
\[
  \phi(y')-\phi(y)=\tfrac12h^{\mathsf T}\!A_ph-r^{\mathsf T}h,
  \qquad h=y'-y ,
\]
where $r=g-A_py$ is the exact residual at $y$.

This distinction matters because of rounding. Take $A=1$, $y=2^{53}$,
$g=2^{53}+2$, and $d=\tfrac14$. Then $r=2$, and the expression for the
proposed step, $\tfrac12d^{\mathsf T}\!Ad-r^{\mathsf T}d=-0.46875$,
predicts a strict decrease. But $\fl(y+d)=y$ in binary64, so the committed
iterate does not move, and its energy change is zero.

Our implementation evaluates
$\widehat{\Delta\phi}=\fl(\tfrac12\hat h^{\mathsf T}\hat w-\hat r^{\mathsf T}\hat h)$,
where $\hat h=\fl(y'-y)$, $\hat w=\fl(A_p\hat h)$, and $\hat r$ is the
computed residual. It accepts the step when $\widehat{\Delta\phi}+E<0$ and
rejects it otherwise. Under the rounding model of
Section~\ref{guarantees}, with $\gamma_k=\gamma^{(p)}_k$, the quantity
\begin{equation}\label{eq:energy-bound}
  E=\Bigl[(2u_p+2\gamma_n)\,\abs{\hat h}^{\mathsf T}\abs{A_p}\abs{\hat h}
   +\gamma_{n+1}(1+u_p)\bigl(\abs{g}+\abs{A_p}\abs{y}\bigr)^{\mathsf T}\abs{\hat h}
   +(u_p+\gamma_n)\,\abs{\hat r}^{\mathsf T}\abs{\hat h}
   +u_p\abs{\widehat{\Delta\phi}}\Bigr]\,(1+\gamma_{n+3})
\end{equation}
is our estimate of $\abs{\widehat{\Delta\phi}-(\phi(y')-\phi(y))}$. We have
not yet proved that it encloses this error for every input, so the energy test
is a monitor and not a rigorous certificate.

The four terms in $E$ account for four sources of error. The first covers
rounding in the displacement, the matrix-vector product, and the first
dot product. The second covers the error in the computed residual, which
satisfies $\abs{\hat r-r}\le\gamma_{n+1}(\abs{g}+\abs{A_p}\abs{y})$. The
third covers the second dot product, and the fourth covers the final
subtraction. The factor $(1+\gamma_{n+3})$ accounts for rounding in
computing $E$ itself. For a sparse matrix, $n$ in the product and residual
terms can be replaced by the largest number of nonzeros in a row of $A_p$.

In the counterexample above, $\hat h=0$, so $\widehat{\Delta\phi}=E=0$, and
the test rejects the step. The test costs one product with $A_p$ and two
products with $\abs{A_p}$ per attempt. Appendix~\ref{app-energy} reports
this cost and the margins we measured. A strict decrease in
energy shows progress in the $A$-norm. A convergence guarantee would also
need a condition on how much the energy decreases.

\subsection{Arbitrary-layer composition}\label{arbitrary-layer}

An adapter returns a vector in its parent's format. Thus, it can be used
anywhere a parent of the same format expects an approximate inverse,
including as the child of another adapter. For example, the tree
\[
  \IR{64}{32}\!\left[\IR{32}{16}\!\left[\ILU(0)_{16}\right]\right]
\]
behaves like an FP64 approximate inverse to whatever calls it. Its outer
adapter computes residuals and updates in FP64, and its inner adapter
computes them in FP32. The innermost base preconditioner is the one part
of the tree that runs in FP16. Theorem~\ref{thm:closure} gives conditions under which a tree of nested adapters is a well-defined solver at any depth.

\begin{theorem}[Arbitrary-Layer Adapter Closure]\label{thm:closure}
Let $T$ be a finite solver tree whose internal Krylov nodes accept variable
right-preconditioner applications. At any subset $E$ of compatible edges,
replace the child application at edge $e=(p,q)$ by an adapter
$\IR{p}{q}^{\,J_e,\tau_e}[S_q]$. Assume each inserted adapter returns a vector in $F_p^{\,n}$. Then the
modified tree is well defined at every depth. Additionally, assume that at edge $e$
the exact parent residuals of the accepted corrections satisfy
\[
  \norm{r_{t+1}}_2\le\eta_e\norm{r_t}_2+\Omega_e,\qquad 0\le\eta_e<1,
\]
where $\Omega_e$ bounds the whole one-step error in
\eqref{eq:recurrence-form}, including the residual-evaluation and
update-rounding terms. Lemma~\ref{lem:recurrence} gives an example of such a
recurrence, with $\eta_e=\hat\eta$ and $\Omega_e=f(g)$. After the
$j_e\le J_e$ corrections that were accepted, the local residual satisfies
the geometric bound of Lemma~\ref{lem:geom} with $\eta=\eta_e$,
$\omega=\Omega_e$, and $J=j_e$. The bound has the same form at every
position in $T$, but its constants depend on the subtree below the edge.
Every enclosing node that admits variable correction applications, such as
another adapter or FGMRES, can use the adapter as its child.
\end{theorem}

The proof appears in Appendix~\ref{app:closure-proof}.

Theorem~\ref{thm:closure} states the interface, the assumptions, and the
local residual bounds needed to place adapters anywhere in a solver tree.
Suzuki and Iwashita informally note that Richardson iteration could appear
at several levels of a nested solver~\cite{suzuki2025nested}, but the theorem
makes this idea precise.

The theorem assumes contraction at each edge where an adapter is
inserted. It does not derive the parent's contraction from the child's
behavior. Theorem~\ref{thm:propagation} gives that relation, using child
constants that are measured or assumed, and Theorem~\ref{thm:recursive}
combines the two results from the leaves up to the root.

The local bound has the same form at every depth, with no factor that
grows with depth. Its constants can still change from edge to edge. The
values of $(\eta_e,\omega_e)$ depend on the whole subtree below the edge,
through \eqref{eq:propagation}.

``Well defined at every depth'' means that each node returns a vector of
the required type and an explicit status after a bounded amount of work. Adapters stop after $J$ attempts or at one of the earlier exits in Algorithm~\ref{alg:adapter}, and Krylov nodes stop at their iteration caps. The two degenerate cases are handled differently. If the Arnoldi process breaks down ($h_{j+1,j}=0$), the Krylov node ends the cycle, commits the minimizer over the directions it has, and recomputes the residual. If a child returns a zero direction, our implementation substitutes the identity direction and
records each substitution.

Stopping does not mean the residual is small. Contract \eqref{eq:clean}
allows a zero correction at every step, so non-increase does not
imply convergence. Progress requires a strict reduction in the residual,
as in \eqref{eq:one-step}, or a strict decrease in energy for SPD parents.
When progress stops, the stagnation check returns an explicit status.

Composition also does not guarantee better performance. It creates a space
of possible designs. Each layer adds casts, residual evaluations,
reductions for the stopping tests, and child calls. A deeper tree helps
when the cheaper work in the child outweighs the added overhead and the
parent still receives a sufficiently contractive correction.
Section~\ref{exp-placement} shows a case where an extra layer makes the
solver slower.

\subsection{Child-to-parent propagation}\label{propagation}

The constants of \eqref{eq:one-step} at each edge depend on quantities at
the child level. To compare the two levels, we view the child's matrix in
the parent's format as $A_q := Q_p(Q_q(A_p))$. Converting back to the
parent format adds no error because every number in format~$q$ is also a
number in format~$p$. Let $\Delta_{pq}=\norm{A_p-A_q}_2$ be the mismatch
between the two stored matrices. For a parent right-hand side $g$, let
$\tilde g$ be the right-hand side the child receives after unscaling, and
let $d$ be the correction after it is promoted and unscaled.

\begin{theorem}[Child-to-parent propagation]\label{thm:propagation}
Suppose the conversion, the child solve, and the child's output size satisfy
\[
  \norm{\tilde g-g}_2\le\epsilon_g\norm{g}_2,\qquad
  \norm{\tilde g-A_q d}_2\le\theta_q\norm{\tilde g}_2+\omega_q,\qquad
  \norm{d}_2\le M_q\norm{g}_2+\nu_q .
\]
Then the parent-level residual of the correction satisfies
\begin{equation}\label{eq:propagation}
  \norm{g-A_p d}_2
  \le \bigl[\epsilon_g+\theta_q(1+\epsilon_g)+\Delta_{pq}M_q\bigr]\norm{g}_2
      +\omega_q+\Delta_{pq}\nu_q .
\end{equation}
Consequently a single correction satisfies \eqref{eq:one-step} with
$\eta_p=\epsilon_g+\theta_q(1+\epsilon_g)+\Delta_{pq}M_q$ and
$\omega_p=\omega_q+\Delta_{pq}\nu_q$ whenever $\eta_p<1$.
\end{theorem}

The next proposition bounds the size of the vector that an adapter
returns.

\begin{proposition}[Output-size propagation]\label{prop:size}
Let $y_0=0$ and $R_t=\norm{r_t}_2$, with $R_0=\norm{g}_2$. Suppose
$R_{t+1}\le\eta R_t+\Omega$ with $0\le\eta<1$, and
$\norm{d_t}_2\le MR_t+\nu$. After $j$ accepted corrections, with
$S_j=(1-\eta^{\,j})/(1-\eta)$ and $D_j=\sum_{t<j}\norm{\delta y_t}_2$,
\[
  \norm{y_j}_2\le MS_j\norm{g}_2+\frac{M\Omega}{1-\eta}\,(j-S_j)+j\nu+D_j .
\]
In exact arithmetic with $\Omega=\nu=D_j=0$, this reduces to
$\norm{y_j}_2\le MS_j\norm{g}_2$.
\end{proposition}

The proofs of Theorem~\ref{thm:propagation} and
Proposition~\ref{prop:size} are in Appendix~\ref{app:propagation-proofs}.

\begin{figure}[H]
\centering
\includegraphics[width=0.85\linewidth]{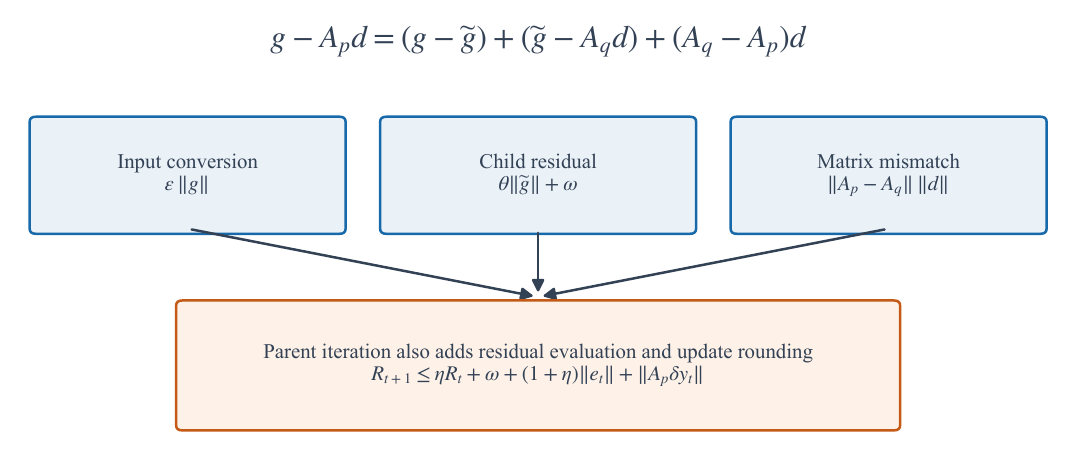}
\caption{The three terms of the propagation bound \eqref{eq:propagation}. The identity at the top holds without approximation, and the triangle inequality turns it into the bound. The parent's recurrence adds the residual-evaluation and update-rounding terms.}
\label{fig:error-budget}
\end{figure}

Equation~\eqref{eq:propagation} separates three sources of error (Figure~\ref{fig:error-budget}): the
conversion of the input, the quality of the child's solve, and the
mismatch between the stored matrices. The mismatch term is multiplied by
the correction size $M_q$, so a large correction amplifies $\Delta_{pq}$.
Thus, a correction can satisfy the child's residual contract and still
fail the parent's. A residual bound by itself is not enough, and the
child must also provide a bound on the size of its output.
Proposition~\ref{prop:size} shows how to carry that bound from one level
up to the next.

\subsection{Outer FGMRES and reliable residuals}\label{outer-fgmres}
The contract also gives a reason to use FGMRES as the outer solver. At the
start of a restart cycle, let $r_0=\beta v_1$ with $\norm{v_1}_2=1$.
Suppose the first adapter application $z_1=P(v_1)$ satisfies
$\norm{v_1-Az_1}_2\le\eta_P$. Then $x_0+\beta z_1$ lies in the FGMRES
search space, and its residual norm is at most $\eta_P\norm{r_0}_2$. In
exact arithmetic, FGMRES minimizes the residual over this space, so it
reduces the residual by at least this factor. This bound links the
measured quality of the adapter to progress in the outer solver. It does
not guarantee fast convergence in later iterations.

The preconditioner $P_j$ can differ between applications. Its output
depends on the input vector, on when the inner solvers stop, and on the
scale chosen for each call. Thus, CAMPS stores every vector $z_j$ that it
produces, which is what FGMRES requires.

In finite precision, the cheap residual estimate from the Hessenberg
matrix can drift away from the true residual. Thus, our implementation
recomputes $r=b-A_{64}x$ in FP64 at every restart and uses this recomputed
residual for the final convergence check. It reports both the residual at
which it stopped and a residual recomputed independently after the solve.

\subsection{Guarantees for arbitrary hierarchies: stability, convergence, and cost}\label{guarantees}
Theorem~\ref{thm:closure} is a local composition result, and
Theorem~\ref{thm:propagation} carries a contract across one edge. This
subsection combines them over a complete solver tree.

The results answer three questions. First, Theorem~\ref{thm:stability}
bounds the error of every output that the root reports as
\textsc{Success}, under the rounding model and the root's stopping test,
with no assumptions on the subtree. Second, Theorem~\ref{thm:recursive} gives sufficient conditions for convergence. It computes the required constants one level at a time, starting at the leaves of the tree and ending at the root. Third, Proposition~\ref{prop:work} gives a criterion for deciding whether a hierarchy costs less than an alternative.

Neither convergence nor a speedup holds for every hierarchy. In
Section~\ref{exp-dense-ic}, the trees that only rely on an FP32 LU child
fail to converge once $\kappa u_{32}$ exceeds about one. In
Section~\ref{exp-sparse-spd}, every tree we tested is slower than CG in
uniform precision. Thus, the results below state explicit conditions for
convergence and for lower cost. Their proofs are in
Appendix~\ref{app:hierarchy-proofs}.

\paragraph{Rounding model and notation.}
We use the standard model of floating-point
arithmetic~\cite{higham2002asna,ieee2019floating} in each format~$p$, and we assume that no
overflow or harmful underflow occurs. Let
$\gamma^{(p)}_k=ku_p/(1-ku_p)$ for $ku_p<1$. For vectors and for the
stored operator in format~$p$,
\begin{align}
  \fl_p(g-A_py)&=g-A_py+\delta r,
    &\abs{\delta r}&\le\gamma^{(p)}_{n+1}\bigl(\abs{g}+\abs{A_p}\abs{y}\bigr),
    \tag{M1}\\
  \fl_p(y+d)&=y+d+\delta y,
    &\abs{\delta y}&\le u_p\abs{y+d},
    \tag{M2}\\
  \fl_p\bigl(\norm{v}_2\bigr)&=(1+\nu)\norm{v}_2,
    &\abs{\nu}&\le\gamma^{(p)}_{n+1},
    \tag{M3}
\end{align}
where absolute values and inequalities hold entrywise. For an operator
with at most $s$ nonzeros per row, $n$ can be replaced by $s$ in
\textup{(M1)}.

For conversions, we assume \textup{(M4)}: $Q_q$ rounds to nearest,
promotion from $F_q$ to $F_p$ does not add any error because $F_q\subset F_p$, and
multiplication by the power-of-two scale~$s$ does not add any error. The
right-hand side that the child receives, $\tilde g=s^{-1}Q_q(sg)$, thus
satisfies $\norm{\tilde g-g}_2\le\epsilon_{pq}\norm{g}_2$, with
$\epsilon_{pq}\le u_q+\sqrt n\,\lambda_q2^{-\zeta}$. Here $\lambda_q$ is the
smallest positive subnormal number in $F_q$, and the scale places
$\norm{sg}_\infty$ in $(2^{\zeta-1},2^{\zeta}]$. For a unit-norm vector cast
without scaling, the same bound holds with $2^{-\zeta}$ replaced by $1/2$. We
also assume that the stored child operator $A_q=Q_q(A_p)$ is nonsingular
and satisfies $\abs{A_q-A_p}\le u_q\abs{A_p}$.

Finally, let
\[
  \kappa_p=\norm{A_p}_2\norm{A_p^{-1}}_2,\qquad
  \kappa'_p=\norm{\abs{A_p}}_2\norm{A_p^{-1}}_2\ \ge\ \kappa_p,\qquad
  \mu_p=\norm{\abs{A_p}}_2/\norm{A_p}_2\in[1,\sqrt n\,],
\]
and define the relative mismatch of an edge as
$\delta_{pq}=\norm{A_p-A_q}_2\norm{A_q^{-1}}_2\le
u_q\norm{\abs{A_p}}_2\norm{A_q^{-1}}_2$.

\begin{definition}[Corrector]\label{def:corrector}
A node $N$ at precision~$p$ is a \emph{$\theta$-corrector} if, for every
admissible right-hand side $g\in F_p^{\,n}$, its output $d=N(g)$ satisfies
$\norm{g-A_pd}_2\le\theta\norm{g}_2$, where the residual is the exact
residual with respect to the stored operator~$A_p$. A right-hand side is
admissible if it is finite and nonzero and its conversions satisfy
\textup{(M4)}.
\end{definition}

Using a relative bound here is not a restriction because each child
receives inputs of a fixed size. An adapter scales its child's right-hand
side so that its largest entry lies in a fixed interval
$(2^{\zeta-1},2^{\zeta}]$ (Algorithm~\ref{alg:adapter}), and a flexible Krylov
node calls its child on vectors of unit norm. With the input size fixed,
an absolute residual floor can be folded into the relative
factor~$\theta$.
If the child guarantees $\norm{h-A_qc}_2\le\theta'\norm{h}_2+\omega_q$ for
its scaled input $h$, then $\norm{h}_2>2^{\zeta-1}$ gives
$\theta=\theta'+\omega_q2^{1-\zeta}$. After the adapter divides by the scale
$s$, the floor on the unscaled residual is $\omega_q/s$.
A fallback child in the parent's own precision is the exception. It receives
the residual without scaling, so its $\theta$ has to hold as a relative bound
for inputs of every size.

\begin{lemma}[Size from residual]\label{lem:size}
If $N$ is a $\theta$-corrector at precision~$p$, then
$\norm{N(g)}_2\le(1+\theta)\norm{A_p^{-1}}_2\norm{g}_2$.
\end{lemma}

Lemma~\ref{lem:size} provides the bound on the output size that
Theorem~\ref{thm:propagation} needs. Apply it to a $\theta_q$-corrector
with input $\tilde g$, where
$\norm{\tilde g}_2\le(1+\epsilon_{pq})\norm{g}_2$. This gives
$M_q=(1+\theta_q)(1+\epsilon_{pq})\norm{A_q^{-1}}_2$ and $\nu_q=0$. Thus,
\eqref{eq:propagation} becomes
\begin{equation}\label{eq:T}
  \norm{g-A_pd}_2\le T_{pq}(\theta_q)\,\norm{g}_2,
  \qquad
  T_{pq}(\theta):=\epsilon_{pq}
     +(1+\epsilon_{pq})\bigl[\theta+(1+\theta)\,\delta_{pq}\bigr].
\end{equation}
If a sharper size bound has been measured, it can be used in place of
this one.

\begin{lemma}[Residual recurrence under rounding]\label{lem:recurrence}
Let $y_t$ be an accepted iterate with exact residual $r_t=g-A_py_t$, let
$\hat r_t=r_t+e_t$ be its evaluated residual, let $d_t$ be the promoted
correction computed from $\hat r_t$, and let $y_{t+1}=y_t+d_t+\delta y_t$ be
the committed iterate. If
$\norm{\hat r_t-A_pd_t}_2\le\eta\norm{\hat r_t}_2+\omega$, then
\begin{equation}\label{eq:recurrence-general}
  \norm{r_{t+1}}_2\le\eta\norm{r_t}_2+\omega+(1+\eta)\norm{e_t}_2
    +\norm{A_p\,\delta y_t}_2 ,
\end{equation}
and the next evaluated residual satisfies
$\norm{\hat r_{t+1}}_2\le\norm{r_{t+1}}_2+\norm{e_{t+1}}_2$. Under
\textup{(M1)}--\textup{(M2)}, with $\omega=0$, $\gamma=\gamma^{(p)}_{n+1}$,
$\gamma_1=\gamma^{(p)}_1$, $a_p=\gamma\kappa'_p$, $x_g=A_p^{-1}g$, and
$\gamma_1\kappa_p<1$, the evaluation error obeys
\begin{equation}\label{eq:dr}
  \norm{e_t}_2\le\gamma\bigl(\norm{g}_2+\norm{\abs{A_p}}_2\norm{x_g}_2\bigr)
   +a_p\norm{r_t}_2 ,
\end{equation}
and \eqref{eq:recurrence-general} becomes
\begin{equation}\label{eq:fp-recurrence}
\begin{gathered}
  \norm{r_{t+1}}_2\le\hat\eta\,\norm{r_t}_2+f(g),\qquad
  \hat\eta=\frac{\eta+(1+\eta)\,a_p}{1-\gamma_1\kappa_p},\\[3pt]
  f(g)=\frac{(1+\eta)\,\gamma\bigl(\norm{g}_2
        +\norm{\abs{A_p}}_2\norm{x_g}_2\bigr)
        +\gamma_1\norm{A_p}_2\norm{x_g}_2}{1-\gamma_1\kappa_p}
      \ \le\ \psi_p\norm{g}_2 ,
\end{gathered}
\end{equation}
with $\psi_p=\bigl[(1+\eta)\gamma(1+\kappa'_p)+\gamma_1\kappa_p\bigr]
/(1-\gamma_1\kappa_p)$.
\end{lemma}

\begin{theorem}[A posteriori accuracy of successful outputs]\label{thm:stability}
Let $b\ne0$, and let the root of a solver tree be a node at precision~$p_0$ that operates on
the given matrix $A=A_{p_0}$ and declares \textsc{Success} only when its
evaluated residual satisfies
$\fl\bigl(\norm{\fl(b-Ax)}_2\bigr)\le\tau\,\fl\bigl(\norm{b}_2\bigr)
+\mathrm{atol}$ with $\tau,\mathrm{atol}\ge0$, as the adapter of
Algorithm~\ref{alg:adapter} and the FGMRES node of
Section~\ref{outer-fgmres} both do. Under \textup{(M1)} and \textup{(M3)},
every $x$ returned with status \textsc{Success} satisfies, with
$\gamma=\gamma^{(p_0)}_{n+1}<1$, $\tau'=\tau(1+\gamma)/(1-\gamma)$, and
$a'=\mathrm{atol}/(1-\gamma)$,
\[
  \norm{b-Ax}_2\le\tau'\norm{b}_2+a'
     +\gamma\bigl(\norm{b}_2+\norm{\abs{A}}_2\norm{x}_2\bigr),
  \qquad
  \beta_2(x):=\frac{\norm{b-Ax}_2}{\norm{A}_2\norm{x}_2+\norm{b}_2}
     \le\tau'+\frac{a'}{\norm{A}_2\norm{x}_2+\norm{b}_2}+\mu_{p_0}\gamma ,
\]
and, when $\kappa_2(A)\beta_2(x)<1$, its relative forward error is at most
$2\kappa_2(A)\beta_2(x)/\bigl(1-\kappa_2(A)\beta_2(x)\bigr)$. No assumption is
made on the subtree: the bound holds whatever the precisions, linearity,
contraction, checks, or exit statuses of the nodes below the root.
\end{theorem}

This bound depends on the tolerance. The output is backward stable in the
working-precision sense when $\tau$ and
$\mathrm{atol}/(\norm{A}_2\norm{x}_2+\norm{b}_2)$ are of the order of the
root's unit roundoff. A small $\mathrm{atol}$ alone is not enough. With $A=1$
and $0<b\le\mathrm{atol}$, the zero vector passes the test and has
$\beta_2=1$. If the right-hand side of the stopping test is itself evaluated
in floating point, $\tau'$ and $a'$ gain a factor of at most
$(1+u_{p_0})^{2}$.

Theorem~\ref{thm:stability} describes what the root's check guarantees at
run time. Under the rounding model, every output reported as
\textsc{Success} satisfies the stated backward-error bound. The bound
holds no matter what the subtree contains. The subtree may include cast
edges that check nothing, children that fail, or poor choices of policy.
These can keep the solver from reaching \textsc{Success}, but they cannot
break the bound for an output that passes the root's stopping test.
Theorem~\ref{thm:recursive} gives sufficient conditions for reaching
\textsc{Success}.

The theorem covers outputs that pass the root's stopping test. An adapter
that exits with any other status also returns an accepted iterate, but that
iterate has not passed the stopping test, and the bound does not apply to
it. Table~\ref{tab:exits} lists what each exit returns.

\begin{table}[H]
\centering
\caption{The exits of Algorithm~\ref{alg:adapter}. Every exit returns the
last accepted iterate.}
\label{tab:exits}
\small
\begin{tabularx}{\linewidth}{@{}l>{\raggedright\arraybackslash}X>{\raggedright\arraybackslash}X@{}}
\toprule
Status & When it occurs & What is known about the returned iterate \\
\midrule
\textsc{Success} & The evaluated residual passes the stopping test. & At the
  root, the bound of Theorem~\ref{thm:stability}. \\
\textsc{Rejected} & A trial fails the acceptance test, and no fallback child
  remains. & Under rule \textup{(R3)} of Theorem~\ref{thm:recursive}, its
  residual is at the floor of the edge (Proposition~\ref{prop:exits}).
  Without \textup{(R3)}, no accuracy bound. \\
\textsc{Stagnated} & $m_{\mathrm{stag}}$ consecutive accepted corrections
  each leave the residual above $\rho_{\mathrm{stag}}$ times its previous
  value. & The same as for \textsc{Rejected}. \\
\textsc{Max-Steps} & All $J$ attempts are used. & Under \textup{(R3)}, the
  bound of Theorem~\ref{thm:recursive}. Without \textup{(R3)}, no accuracy
  bound. \\
\textsc{Failure} & The converted residual is zero or not finite, or the
  child's output is not finite. & No accuracy bound. \\
\bottomrule
\end{tabularx}
\end{table}

\begin{theorem}[Recursive convergence]\label{thm:recursive}
Assume \textup{(M1)--(M4)}, and that every adapter in the tree runs with
$\mathrm{atol}=0$ (an absolute tolerance adds
$\mathrm{atol}/((1-\gamma)(1-a_p))$ to the bound of \textup{(R3)}). Assign a
number $\theta(N)$ to the nodes of a finite solver tree, from the leaves
upward, by the following rules.
\begin{enumerate}[label=\textup{(R\arabic*)},leftmargin=3.0em,itemsep=3pt]
\item \emph{Base node.} $\theta(\operatorname{Base}_p)$ is any number for
  which $\operatorname{Base}_p$ is a $\theta$-corrector. It is a hypothesis,
  supplied by analysis or by measurement.
\item \emph{Cast edge.}
  $\theta\bigl(\operatorname{Cast}_{p\leftarrow q}[S]\bigr)
   =T_{pq}\bigl(\theta(S)\bigr)$.
\item \emph{Adapter edge.} The adapter uses the residual test of
  Algorithm~\ref{alg:adapter}, with damping and the energy test turned off.
  Let $\eta=T_{pq}(\theta(S))$. This bound must
  hold for every child the adapter may call, including a fallback child,
  and no attempt may end in \textsc{Failure}. Let $\hat\eta$,
  $f$, and $\psi_p$ be as in Lemma~\ref{lem:recurrence}, with
  $\gamma=\gamma^{(p)}_{n+1}$, $\gamma_1=\gamma^{(p)}_1$, and
  $a_p=\gamma\,\kappa'_p$, and let
  \[
    c_p(\varrho)=\frac{(1+\gamma)(1+a_p)+2}
         {\varrho(1-\gamma)(1-a_p)-(1+\gamma)(1+a_p)\,\hat\eta}.
  \]
  Let $\varrho_*$ be the smallest of $1$, the progress threshold $\varrho$
  when one is set, and the stagnation threshold $\rho_{\mathrm{stag}}$ when
  stagnation detection is enabled. If $\gamma_1\kappa_p<1$ and
  $c_p(\varrho_*)>0$, then with $c_p=c_p(\varrho_*)$,
  \[
    \theta\bigl(\IR{p}{q}^{\,J,\tau}[S]\bigr)
      =\max\bigl\{\hat\eta^{\,J},\,\tau_+\bigr\}+c_p\,\psi_p,
    \qquad
    \tau_+=\frac{\tau\,(1+\gamma)}{(1-\gamma)(1-a_p)} ,
  \]
  where $J$ is the attempt budget of Algorithm~\ref{alg:adapter}.
\item \emph{Flexible Krylov node.} If $\theta=\theta(S)<1$, then in exact
  arithmetic for the Arnoldi recurrence, the projected least-squares problem,
  and the update, with restart residuals evaluated under \textup{(M1)} and
  \textup{(M3)}, a node that exits on its evaluated restart residual at
  relative tolerance $\tau_K$, and that completes at least $k_{\min}$
  restart cycles before any other exit, satisfies
  \begin{gather*}
    \theta\bigl(\FGMRES_p^{\,k_{\min},\tau_K}[S]\bigr)
      =\max\bigl\{\hat\theta^{\,k_{\min}},\,\tau_{K+}\bigr\}
       +\frac{(1+\theta)\,\gamma^{(p)}_{n+1}(1+\kappa'_p)}
             {(1-\hat\theta)(1-a_p)},\\[3pt]
    \hat\theta=\theta+(1+\theta)\,a_p<1,\qquad
    \tau_{K+}=\frac{\tau_K(1+\gamma)}{(1-\gamma)(1-a_p)} .
  \end{gather*}
  An upper limit of $k$ cycles does not give $k_{\min}=k$ when an iteration
  cap can end the solve earlier. A truncated cycle counts toward $k_{\min}$
  if its minimizer is committed.
\end{enumerate}
Then the following hold.
\begin{enumerate}[label=\textup{(\roman*)},itemsep=3pt]
\item \emph{Closure.} Every node $N$ for which the rules yield $\theta(N)<1$
  is a $\theta(N)$-corrector for the operator stored at its precision. In
  particular, if $\theta$ is defined and less than one at the root, the
  returned vector satisfies
  $\norm{b-Ax}_2\le\theta(\mathrm{root})\norm{b}_2$.
\item \emph{Rate and termination.} Let the root be an adapter with
  effective factor $\hat\eta_0<1$ from \textup{(R3)}, and let
  $x_\star=A^{-1}b$. After $t$ accepted corrections,
  \begin{equation}\label{eq:root-rate}
    \norm{r_t}_2\le\hat\eta_0^{\,t}\norm{b}_2+\frac{f_0}{1-\hat\eta_0},
    \qquad
    f_0=\frac{(1+\eta_0)\,\gamma\bigl(\norm{b}_2
          +\norm{\abs{A}}_2\norm{x_\star}_2\bigr)
          +\gamma_1\norm{A}_2\norm{x_\star}_2}{1-\gamma_1\kappa_2(A)} .
  \end{equation}
  Put $\tau_-=\tau(1-\gamma)/(1+\gamma)$ and
  $\phi_0=\bigl[(1+a_{p_0})/(1-\hat\eta_0)+1\bigr]f_0/\norm{b}_2$, and let
  $K(\tau)$ be the smallest integer with
  $(1+a_{p_0})\,\hat\eta_0^{\,K}\le\tau_--\phi_0$. If the tolerance lies above
  the floor, $\tau_-\ge\bigl[(1+a_{p_0})c_{p_0}+1\bigr]f_0/\norm{b}_2$, then
  no attempt is rejected before \textsc{Success}, so attempted and accepted
  corrections coincide, and if the attempt budget satisfies $J\ge K(\tau)$
  the adapter returns \textsc{Success} after at most $K(\tau)$ attempts. If
  $J<K(\tau)$, it returns \textsc{Success} or \textsc{Max-Steps}, and in
  the second case the bound \eqref{eq:root-rate} holds at $t=J$. When recovery attempts occur, that is, when
  the floor hypothesis fails for the configured child and a rejected attempt
  switches the active pair, the budget must also cover them: the guarantee
  then applies to the fallback pair with $J$ reduced by the attempts already
  consumed.
\item \emph{Limiting accuracy.} If the root adapter is run with $\tau=0$ and
  $J$ large enough that $\hat\eta_0^{\,J}\norm{b}_2\le f_0$, the returned $x$
  satisfies $\norm{b-Ax}_2\le c_{p_0}f_0$, and hence
  $\beta_2(x)\le B/(1-\kappa_2(A)B)$ with
  $B=c_{p_0}\bigl[(1+\eta_0)\gamma\mu_{p_0}+\gamma_1\bigr]
     /\bigl(1-\gamma_1\kappa_2(A)\bigr)$
  whenever $\kappa_2(A)B<1$. Thus, the limiting backward error is $O(n\,u_{p_0})$ as long as $c_{p_0}$ and
  $1/(1-\gamma_1\kappa_2(A))$ stay bounded. The subtree enters
  the bound through the single number $\hat\eta_0$, which sets how fast the
  floor is reached and the constant $c_{p_0}$ in front of it, but not its
  order $n\,u_{p_0}$. The constant $c_{p_0}$ grows without bound as $\hat\eta_0$
  approaches one.
\end{enumerate}
\end{theorem}

Rule \textup{(R3)} requires $c_p(\varrho_*)>0$. This means the configured
threshold has to exceed the guaranteed factor $\hat\eta$ by a rounding
margin. A stricter progress threshold can reject corrections that contract.
For example, let $A=1$ and $g=1$, let the child return half of its input, and
set $\varrho=1/4$. The first trial leaves a residual of $1/2$, the test
rejects it, and the adapter returns zero, although the child is a
$1/2$-corrector. Damping and the energy test are excluded for a similar
reason. A damped step with $\alpha_{\max}<1$ can be shorter than the child's
correction. The energy test can reject a correction that reduces the
residual. For $A=\operatorname{diag}(1,128)$, $g=(0,1)^{\mathsf T}$, and
$d=(1/2,1/128)^{\mathsf T}$, the residual falls to $1/2$ and the energy rises
by $1/8-1/256$. With these options the adapter still satisfies
Theorems~\ref{thm:closure} and~\ref{thm:stability}, but rule
\textup{(R3)} does not cover it.

\begin{figure}[H]
\centering
\includegraphics[width=0.8\linewidth]{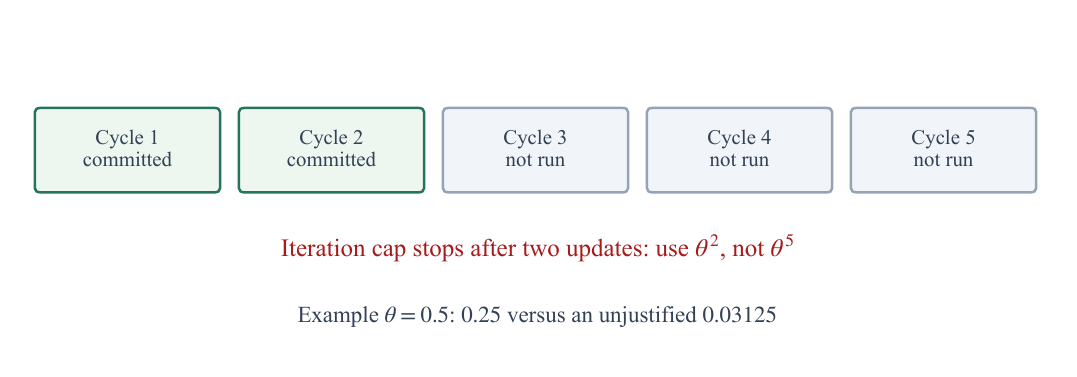}
\caption{Rule \textup{(R4)} counts completed cycles. With $\theta=0.5$ and an iteration cap that allows two committed cycles out of a budget of five, the guaranteed factor is $\theta^{2}=0.25$, not $\theta^{5}\approx0.03$.}
\label{fig:cycle-budget}
\end{figure}

Figure~\ref{fig:cycle-budget} illustrates the cycle count $k_{\min}$ in rule
\textup{(R4)}.

Under the assumptions of \textup{(R3)}, the next proposition bounds the
residual at the two early exits: rejection and stagnation.

\begin{proposition}[Early exits occur only at the floor]\label{prop:exits}
In the setting of \textup{(R3)}, suppose a trial correction from an accepted
iterate $y_t$ yields evaluated norms with $\rho'>\varrho\,\rho$ for some
$\varrho\in(0,1]$ with $c_p(\varrho)>0$. Then
$\norm{r_t}_2<c_p(\varrho)\,f(g)$. Consequently:
\textup{(a)} the acceptance test of Algorithm~\ref{alg:adapter}
($\varrho=1$) can reject a correction only when $\norm{r_t}_2<c_p(1)f(g)$,
that is, within a constant factor of the limiting residual
$f(g)/(1-\hat\eta)$, so a \textsc{Rejected} exit returns an iterate that is
already at the floor of its edge; and \textup{(b)} each event that counts
toward a \textsc{Stagnated} exit ($\varrho=\rho_{\mathrm{stag}}$) satisfies
the same bound, and the returned iterate satisfies
$\norm{r}_2\le\bigl(1+\hat\eta\,c_p(\rho_{\mathrm{stag}})\bigr)f(g)
\le c_p(\rho_{\mathrm{stag}})\,f(g)$.
\end{proposition}

Under the stated hypotheses, the acceptance test does not stop
convergence while the residual is above the floor. The proposition also
helps to explain failures. If a correction is rejected while
$\norm{r_t}_2\ge c_p(1)f(g)$, then at least one hypothesis fails. In
practice, the likely cause is that the child does not contract on that
right-hand side. Section~\ref{exp-dense-ic} uses this reading for the
rejections it reports at $\kappa\ge10^{8}$.

\begin{corollary}[What each edge of a cascade sees]\label{cor:cascade}
For the chain $\IR{p_0}{p_1}^{\,J_0}\bigl[\IR{p_1}{p_2}^{\,J_1}[S]\bigr]$
with $\tau=0$ at the inner adapter, the top-edge factor is
$\eta_0=T_{p_0p_1}\bigl(\hat\eta_1^{\,J_1}+c_{p_1}\psi_{p_1}\bigr)$. As $J_1$
grows it decreases geometrically, at the rate $\hat\eta_1$ of the lower edge,
until it reaches the plateau $T_{p_0p_1}(c_{p_1}\psi_{p_1})
=O\bigl(n\,u_{p_1}\kappa'_{p_1}\bigr)$. The order of this plateau contains
the intermediate precision $p_1$ but neither $p_2$ nor $\theta(S)$. For a particular
right-hand side both terms of this bound, the floor
$f(g)/\norm{g}_2$ of \eqref{eq:fp-recurrence} and the mismatch term
$\Delta_{p_0p_1}\norm{d}_2/\norm{g}_2$ of \eqref{eq:propagation}, grow
linearly with the amplification
$\norm{A}_2\norm{A^{-1}g}_2/\norm{g}_2\in[1,\kappa_2(A)]$. They are upper
bounds. The actual errors also depend on the directions of the correction and
of the rounding errors, and they can be much smaller. The bound reaches its
$\kappa$-dependent value in the worst case. The lowest precision and the leaf enter through the single number
$\hat\eta_1$. It sets the rate, the requirement $\hat\eta_1<1$, and the
constant $c_{p_1}$ in front of the plateau, but not the plateau's order. If the leaf is an LU factorization with partial
pivoting in precision~$q$ and $\omega<1$, then
$\theta(\mathrm{LU}_q)\le\omega/(1-\omega)$ with
$\omega=\gamma^{(q)}_{3n}\norm{\abs{\hat L}\abs{\hat U}}_2
\norm{A_q^{-1}}_2$~\cite{higham2002asna}, and the requirement $\hat\eta<1$
at its edge reduces to the classical sufficient condition
$\kappa(A)\,u_q\lesssim1$ of~\cite{moler1967,carson2018three}.
\end{corollary}

The next proposition uses the convergence factors and the cost of each
node to bound the total work of a tree and to compare alternatives.

\begin{proposition}[Guaranteed work and the lowering criterion]\label{prop:work}
Let $m_p$ be the cost of one residual evaluation at precision~$p$ (one
product with $A_p$ and the accompanying vector operations), $\chi_{pq}$ the
cost of one scaled cast and promotion, and $c(\operatorname{Base})$ the
application cost of a base node, in any additive unit (time, energy, bytes
moved). Define
\[
  c\bigl(\operatorname{Cast}_{p\leftarrow q}[S]\bigr)=\chi_{pq}+c(S),\qquad
    c\bigl(\IR{p}{q}^{\,J}[S]\bigr)\le c_{\mathrm{init}}+J\,C_{\max},
\]
\[
  c\bigl(\FGMRES_p^{\,m,k}[S]\bigr)=k\bigl(m\,(m_p+c(S))+o_p(m)+m_p\bigr),
\]
where $c_{\mathrm{init}}$ is the cost of the initial residual and $C_{\max}$
bounds one attempt, including the child call, the conversions, the trial
residual, and any damping, energy test, or fallback work. Let $o_p(m)$ be
the orthogonalization and update cost of one cycle, and let
$\sigma(N)$ be the total one-time setup cost of the nodes of~$N$.
\begin{enumerate}[label=\textup{(\alph*)},itemsep=3pt]
\item $c(N)$ bounds the cost of every application of $N$, since each early
  exit performs fewer operations. If the root is an adapter satisfying
  Theorem~\ref{thm:recursive}\,\textup{(ii)}, the total cost to return
  \textsc{Success} at tolerance~$\tau$ is at most
  $W(N,\tau)=\sigma(N)+c_{\mathrm{init}}+K(\tau)\,C_{\max}$, so $N$ is
  guaranteed to cost no more than any method whose cost to reach $\tau$ is
  at least $W(N,\tau)$.
\item Let two alternatives for the correction child of one parent loop
  realize per-correction residual factors
  $\varrho_1,\varrho_2\in(0,1)$ at per-correction costs $C_1,C_2$, each
  including the parent's residual evaluation and conversions, and let
  $\ell_i=\ln(1/\varrho_i)$. To reduce the residual by the factor $e^{-L}$,
  above both floors, alternative~$i$ costs $C_i\lceil L/\ell_i\rceil$.
  Alternative~1 is cheaper whenever $C_1/\ell_1+C_1/L\le C_2/\ell_2$, and is
  not cheaper whenever $C_1/\ell_1\ge C_2/\ell_2+C_2/L$. As $L\to\infty$, lowering the precision of a child $S$ from $p$ to $q$ behind an adapter edge is cheaper when
  \begin{equation}\label{eq:lowering}
    \frac{m_p+\chi_{pq}+c(S_q)}{m_p+c(S_p)}
      <\frac{\ln\varrho_q}{\ln\varrho_p},
  \end{equation}
  and more expensive when the inequality is reversed. At finite $L$,
  compare the costs $C_i\lceil L/\ell_i\rceil$ and the setup costs directly.
  Here $\varrho_p$ and $\varrho_q$ are the realized factors. By
  Lemma~\ref{lem:recurrence}, each satisfies
  $\varrho\le\hat\eta+f(g)/\norm{r_t}_2$, with $\hat\eta$ computed from
  $\eta=\theta(S_p)$ or $\eta=T_{pq}(\theta(S_q))$. In other words, the lower
  precision must keep a smaller fraction of the per-correction cost than
  of the convergence rate, measured in digits per correction.
\end{enumerate}
\end{proposition}

\begin{figure}[H]
\centering
\includegraphics[width=\linewidth]{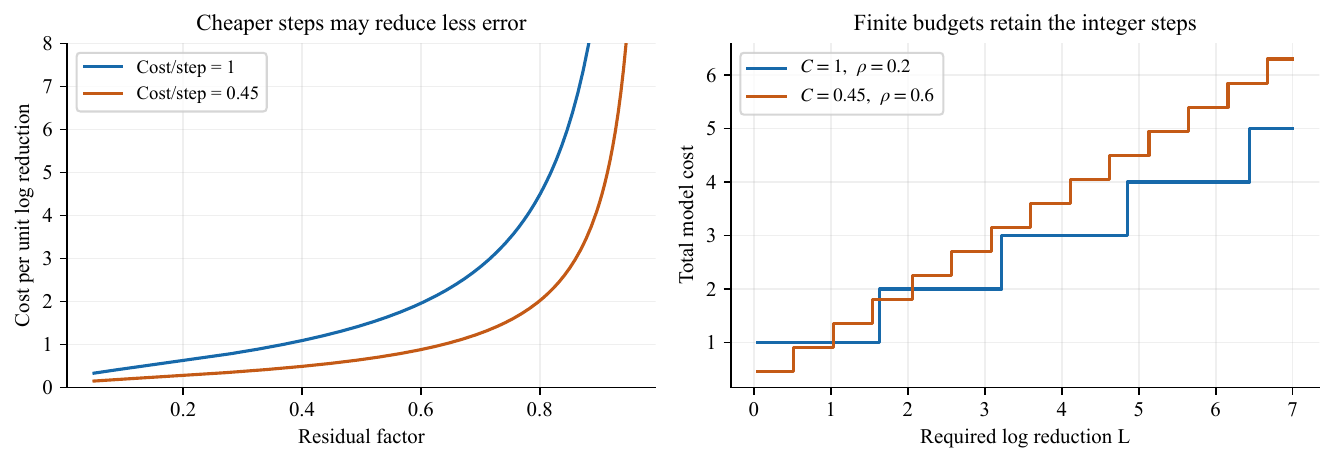}
\caption{A cost model with constant factors, no setup cost, and no rounding floor. Left: the cost per unit of logarithmic reduction depends on the residual factor as well as on the cost per step. Right: for a finite target, the total cost keeps the integer steps of $C\lceil L/\ln(1/\varrho)\rceil$, so the ordering for large $L$ need not hold for a short solve.}
\label{fig:cost-progress}
\end{figure}

Proposition~\ref{prop:work} makes the performance comparison depend on
both the hardware and the problem. The costs depend on the hardware, for
example on how much faster it runs FP32 than FP64. The residual factors
depend on the problem, through how much each correction reduces the
residual. Section~\ref{experiments} measures both. In the dense experiment
of Section~\ref{exp-dense-wc}, part~(a) favors the adapter because its
total cost includes the much cheaper FP32 factorization. In the SPD
experiment of Section~\ref{exp-sparse-spd}, criterion~\eqref{eq:lowering}
favors the uniform-precision method. Figure~\ref{fig:cost-progress}
illustrates the criterion.

\paragraph{Scope of these results.} These results have four limits.

First, the conditions in Theorem~\ref{thm:recursive} are sufficient, not
necessary. A tree may converge even when its computed $\theta$ is larger
than one. Theorem~\ref{thm:stability} still bounds any output that passes
the root's stopping test.

Second, the constants are worst-case bounds. The factors
$\gamma^{(p)}_{n+1}$ and $\kappa'_p$ can overestimate typical rounding
effects by orders of magnitude~\cite{higham2019probabilistic}. For a
half-precision parent, the rule gives no usable bound once
$n\,u_{16}\ge1$, unless the matrix is sparse enough to allow a smaller
factor. Section~\ref{exp-verify} measures the gap between the bounds and
the observed behavior.

Third, \textup{(R4)} bounds the progress of a restart cycle by its first
Arnoldi direction. Thus, it ignores the extra acceleration that later
directions provide. It also gives no bound when the child is not itself a
contraction, which can happen with a single $\ILU(0)$ application. In that case, the Krylov node has to be treated as a base node under \textup{(R1)}, with its factor taken from GMRES convergence theory or from measurement.

Fourth, \textup{(R4)} assumes exact arithmetic for the Arnoldi process,
the projected solve, and the update. Existing work analyzes the attainable
accuracy of flexible GMRES in finite
precision~\cite{arioli2009fgmres,buttari2026modular}, but a self-contained
analysis with a nonlinear adapter as the preconditioner is still open. A forward-error analysis of nested refinement is also open, though Carson and Higham give such analyses for refinement in two and three precisions~\cite{carson2017analysis,carson2018three}.

These limits apply to the recursive convergence result. They do not
affect Theorem~\ref{thm:stability}, which does not need anything beyond the root's
residual check and the rounding model.

\subsection{A construction strategy and the evaluated instances}\label{strategy}
CAMPS defines a family of methods. We build and evaluate each instance one
precision edge at a time.
\begin{enumerate}
  \item Choose a highest-precision parent. Its precision sets the accuracy
    of the final residual and of the updates.
  \item Insert one adapter at a candidate edge, and measure its
    contraction at the parent level, its cost, and its memory use.
  \item Recurse inside the child if the one-level adapter still contracts
    and the next lower precision reduces the total work or storage.
    Otherwise, stop at one level.
  \item Compare each instance against the parent without nesting, against
    a one-level alternative, and against a fixed nested baseline such as
    F3R~\cite{suzuki2025nested}.
\end{enumerate}
This strategy follows the local structure of the theory. Measurements at
each parent decide whether to keep an edge, and composition assembles the
full tree from the edges that are kept.

\subsubsection*{The named instance family}
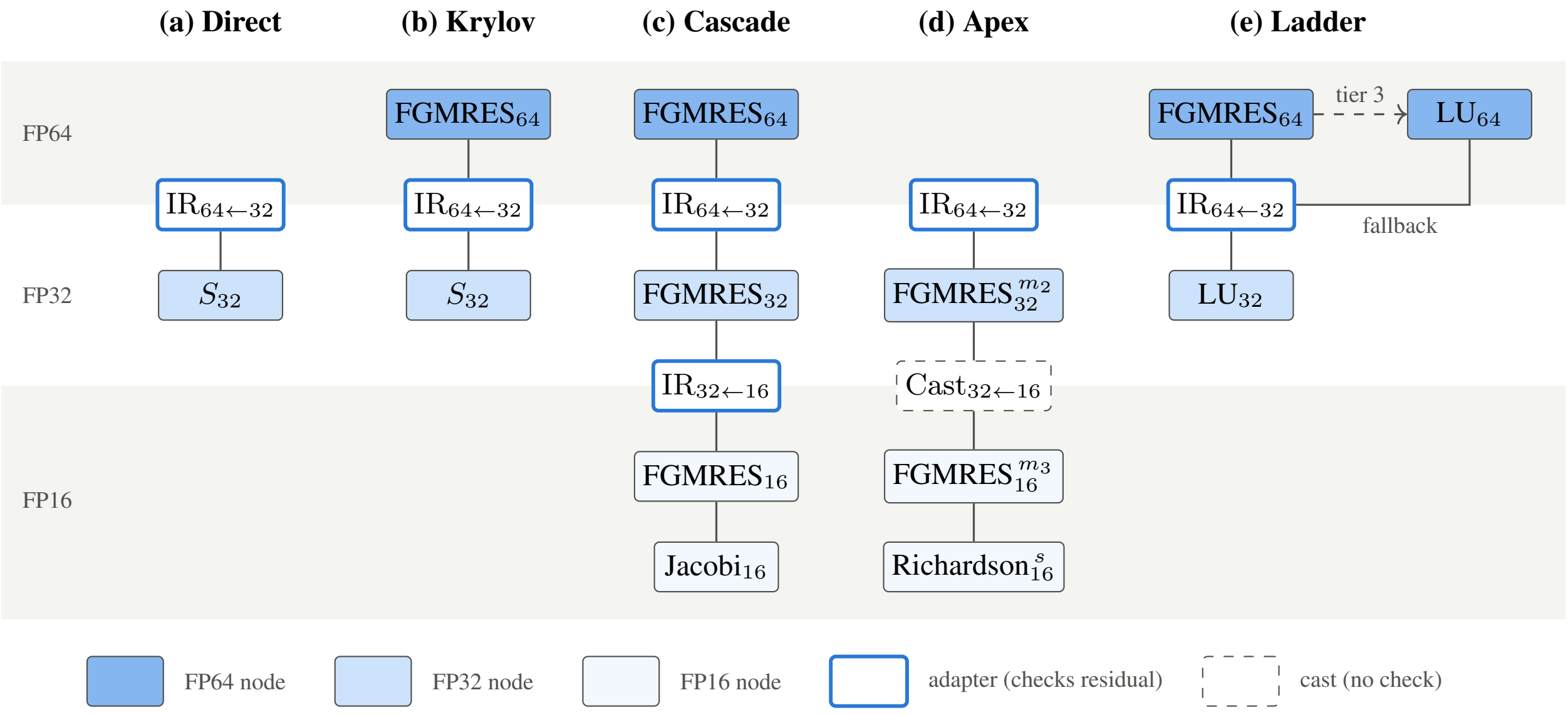
\begin{figure}[H]
\centering
\definecolor{campsfpA}{HTML}{86B6EF}
\definecolor{campsfpB}{HTML}{CDE2FB}
\definecolor{campsfpC}{HTML}{F3F8FE}
\definecolor{campsadapter}{HTML}{2A78D6}
\definecolor{campsink}{HTML}{52514E}
\definecolor{campsband}{HTML}{F4F4F1}
\resizebox{\linewidth}{!}{%
\begin{tikzpicture}[
  font=\footnotesize,
  nd/.style={draw=campsink, rounded corners=1.5pt, minimum height=5.2mm,
             minimum width=13mm, inner xsep=2.5pt, align=center},
  p64/.style={nd, fill=campsfpA},
  p32/.style={nd, fill=campsfpB},
  p16/.style={nd, fill=campsfpC},
  ad/.style={nd, draw=campsadapter, line width=1.1pt, fill=white},
  ca/.style={nd, dashed, fill=white},
  ed/.style={draw=campsink, line width=0.6pt},
  ttl/.style={font=\footnotesize\bfseries},
  lbl/.style={font=\scriptsize, text=campsink},
]
\fill[campsband] (0,0.55) rectangle (16.4,-0.95);
\fill[campsband] (0,-2.85) rectangle (16.4,-5.3);
\node[lbl, anchor=west] at (0.1,-0.2) {FP64};
\node[lbl, anchor=west] at (0.1,-1.9) {FP32};
\node[lbl, anchor=west] at (0.1,-4.05) {FP16};

\node[ttl] at (2.3,0.95) {(a) Direct};
\node[ad]  (a1) at (2.3,-0.95) {$\mathrm{IR}_{64\leftarrow32}$};
\node[p32] (a2) at (2.3,-1.9)  {$S_{32}$};
\draw[ed] (a1) -- (a2);

\node[ttl] at (4.9,0.95) {(b) Krylov};
\node[p64] (b0) at (4.9,0)     {FGMRES$_{64}$};
\node[ad]  (b1) at (4.9,-0.95) {$\mathrm{IR}_{64\leftarrow32}$};
\node[p32] (b2) at (4.9,-1.9)  {$S_{32}$};
\draw[ed] (b0) -- (b1) -- (b2);

\node[ttl] at (7.5,0.95) {(c) Cascade};
\node[p64] (c0) at (7.5,0)     {FGMRES$_{64}$};
\node[ad]  (c1) at (7.5,-0.95) {$\mathrm{IR}_{64\leftarrow32}$};
\node[p32] (c2) at (7.5,-1.9)  {FGMRES$_{32}$};
\node[ad]  (c3) at (7.5,-2.85) {$\mathrm{IR}_{32\leftarrow16}$};
\node[p16] (c4) at (7.5,-3.8)  {FGMRES$_{16}$};
\node[p16] (c5) at (7.5,-4.75) {Jacobi$_{16}$};
\draw[ed] (c0) -- (c1) -- (c2) -- (c3) -- (c4) -- (c5);

\node[ttl] at (10.2,0.95) {(d) Apex};
\node[ad]  (d1) at (10.2,-0.95) {$\mathrm{IR}_{64\leftarrow32}$};
\node[p32] (d2) at (10.2,-1.9)  {FGMRES$_{32}^{\,m_2}$};
\node[ca]  (d3) at (10.2,-2.85) {$\mathrm{Cast}_{32\leftarrow16}$};
\node[p16] (d4) at (10.2,-3.8)  {FGMRES$_{16}^{\,m_3}$};
\node[p16] (d5) at (10.2,-4.75) {Richardson$_{16}^{\,s}$};
\draw[ed] (d1) -- (d2) -- (d3) -- (d4) -- (d5);

\node[ttl] at (13.6,0.95) {(e) Ladder};
\node[p64] (e0) at (12.9,0)     {FGMRES$_{64}$};
\node[ad]  (e1) at (12.9,-0.95) {$\mathrm{IR}_{64\leftarrow32}$};
\node[p32] (e2) at (12.9,-1.9)  {LU$_{32}$};
\node[p64] (e3) at (15.4,0)     {LU$_{64}$};
\draw[ed] (e0) -- (e1) -- (e2);
\draw[ed] (e1.east) -| node[lbl, pos=0.3, below] {fallback} (e3.south);
\draw[ed, dashed, ->] (e0.east) -- node[lbl, above] {tier 3} (e3.west);

\node[p64, minimum width=8mm] at (1.3,-5.95) {};
\node[lbl, anchor=west] at (1.8,-5.95) {FP64 node};
\node[p32, minimum width=8mm] at (3.9,-5.95) {};
\node[lbl, anchor=west] at (4.4,-5.95) {FP32 node};
\node[p16, minimum width=8mm] at (6.5,-5.95) {};
\node[lbl, anchor=west] at (7.0,-5.95) {FP16 node};
\node[ad, minimum width=8mm] at (9.1,-5.95) {};
\node[lbl, anchor=west] at (9.6,-5.95) {adapter (checks residual)};
\node[ca, minimum width=8mm] at (13.0,-5.95) {};
\node[lbl, anchor=west] at (13.5,-5.95) {cast (no check)};
\end{tikzpicture}}
\caption{Solver trees of the five CAMPS instances evaluated in
Section~\ref{experiments}. Shaded bands mark the precision of each node.
Adapters and casts sit on the boundary between the two precisions they
join. An adapter checks the parent residual on every correction, and a cast
does not. $S_{32}$ is the FP32 child, which is LU, CG, or FGMRES depending
on the matrix class. The dense form of \textsf{CAMPS-Cascade} replaces the
FP16 FGMRES leaf with an approximate inverse $\mathrm{INV}_{16}$. In
\textsf{CAMPS-Ladder}, the adapter switches to the FP64 LU fallback after a
rejected correction, and the dashed arrow is the unchecked third tier.}
\label{fig:instances}
\end{figure}

This strategy produces a few tree shapes that recur across our
experiments. We name them so that Section~\ref{experiments} can refer to
each instance by its role. The name gives the shape, and the bracketed
argument gives the child family or configuration. The same shape can serve
different matrix classes by using different children.
Figure~\ref{fig:instances} draws the solver tree of each instance, and
Algorithms~\ref{alg:direct}--\ref{alg:ladder} state each one in full.

\begin{description}[leftmargin=1.6em,itemsep=3pt]
  \item[{\textsf{CAMPS-Direct}$[S_q]$}]
    (Figure~\ref{fig:instances}a, Algorithm~\ref{alg:direct}.)
    Uses the adapter itself as the solver. Algorithm~\ref{alg:adapter}
    runs to the final tolerance with no outer Krylov method. This shape
    suits children that contract strongly, with measured $\hat\eta\ll1$,
    such as an FP32 factorization on dense well-conditioned systems. For
    multiple right-hand sides, it applies corrections to whole blocks with
    vectorized kernels.

  \item[{\textsf{CAMPS-Krylov}$[S_q]$}]
    (Figure~\ref{fig:instances}b, Algorithm~\ref{alg:krylov}.)
    Uses the adapter as a variable right preconditioner for an outer
    flexible Krylov method, which combines the directions the adapter
    returns. This shape suits children whose correction quality varies.
    On moderately ill-conditioned dense systems, it needs fewer outer
    iterations than plain refinement, but each iteration costs more, and in
    our tests it was never faster than \textsf{CAMPS-Direct}.

  \item[{\textsf{CAMPS-Cascade}$[\operatorname{Base}_{q_2}]$}]
    (Figure~\ref{fig:instances}c, Algorithm~\ref{alg:cascade}.)
    Uses a recursive three-precision hierarchy with an adapter at both
    precision edges. This is the multi-edge setting of
    Theorem~\ref{thm:closure}. Sparse leaf children run their
    matrix-vector products natively in half precision on CSR data. Dense
    leaf children apply a stored half-precision operator through
    tensor-core matrix products.

  \item[{\textsf{CAMPS-Apex}$[m_2,m_3,s]$}]
    (Figure~\ref{fig:instances}d, Algorithm~\ref{alg:apex}.)
    Places a single adapter, at the topmost precision edge. Below it,
    Krylov levels with fixed iteration budgets are joined by plain casts,
    which carry no residual check (Section~\ref{grammar}). It checks
    one of its two precision edges. It uses the interior structure of
    F3R~\cite{suzuki2025nested} and directly competes with F3R.

  \item[\textsf{CAMPS-Ladder}]
    (Figure~\ref{fig:instances}e, Algorithm~\ref{alg:ladder}.)
    Uses three tiers for unknown or extreme conditioning. It first tries
    checked corrections in low precision. After a correction fails the
    acceptance test, it switches to checked corrections in the parent
    precision (\textsc{on-violation: fallback}). If no correction passes the test at unit step, it supplies raw directions in the parent precision, which is what a classical preconditioner would do. These last directions are unchecked (Section~\ref{contract}).
\end{description}

In Algorithms~\ref{alg:direct}--\ref{alg:ladder}, \textsc{ApplyAdapter} is Algorithm~\ref{alg:adapter}, $\tau$
is the tolerance on the final relative residual, and $\tau_{\mathrm{ad}}$
is a looser tolerance used inside each adapter application.

Two conventions apply when an adapter serves as an FGMRES preconditioner.
First, if an application rejects its first correction, it returns the zero
vector, and a guard replaces it with the identity direction $z_j=v_j$.
\textsf{CAMPS-Ladder} is the exception and uses its FP64 LU direction
instead. The implementation counts and reports every substitution. Second,
if an application returns a nonzero vector, FGMRES uses that vector
whatever the adapter's exit status is. FGMRES builds its search space from
the directions it actually receives.

Classical one-edge designs are also members of this family. For example,
$\FGMRES_{64}^{\,m,\tau_{\mathrm{out}}}
[\IR{64}{32}^{\,J,\tau_{\mathrm{ad}}}[\ILU(0)_{32}]]$ is
\textsf{CAMPS-Krylov}$[\ILU(0)_{32}]$ in our CAMPS grammar. In the same sense,
F3R is one instance of the broader nesting strategy of Suzuki and
Iwashita~\cite{suzuki2025nested}. The test matrices are inputs to these
instances, and they are not part of the definition of the framework.

\begin{algorithmblock}{Algorithm 2. \textsf{CAMPS-Direct}$[S_q]$ $(A_p, b, \tau)$}
\label{alg:direct}
\emph{Solver tree: Figure~\ref{fig:instances}a.}
\begin{enumerate}[label=\arabic*.,leftmargin=2.2em,itemsep=1.5pt,topsep=3pt]
  \item Return $\textsc{ApplyAdapter}(p,q,A_p,b,S_q,J_{\max},\tau)$: the
    adapter is the solver, run with a large correction cap
    $J_{\max}$, final tolerance $\tau$, and stagnation detection
    ($m_{\mathrm{stag}}$ consecutive checks with relative decrease worse than
    $\rho_{\mathrm{stag}}$).
  \item For a block right-hand side $B\in F_p^{\,n\times k}$: scaling,
    conversion, acceptance, and stopping are applied per column; passing
    columns accumulate, violating columns roll back individually.
\end{enumerate}
\end{algorithmblock}

\begin{algorithmblock}{Algorithm 3. \textsf{CAMPS-Krylov}$[S_q]$ $(A_p, b, \tau)$}
\label{alg:krylov}
\emph{Solver tree: Figure~\ref{fig:instances}b.}
\begin{enumerate}[label=\arabic*.,leftmargin=2.2em,itemsep=1.5pt,topsep=3pt]
  \item Define the variable right preconditioner
    $P(v):=\textsc{ApplyAdapter}(p,q,A_p,v,S_q,J,\tau_{\mathrm{ad}})$ with
    small $J$ (2--3).
  \item If $P(v)=0$ (first-correction rejection), substitute $z=v$
    [zero guard; counted].
  \item Return $x=\FGMRES_p(A_p,b,\tau,m)$ with preconditioner $P$, storing
    every realized $z_j$, recomputing the true residual $b-A_p x$ at each
    restart, and claiming convergence only on that recomputed value.
\end{enumerate}
\end{algorithmblock}

\begin{algorithmblock}{Algorithm 4. \textsf{CAMPS-Cascade}$[\operatorname{Base}_{q_2}]$ $(A_p, b, \tau)$ with $p=64$, $q_1=32$, $q_2=16$}
\label{alg:cascade}
\emph{Solver tree: Figure~\ref{fig:instances}c.}
\begin{enumerate}[label=\arabic*.,leftmargin=2.2em,itemsep=1.5pt,topsep=3pt]
  \item Leaf: $S_{q_2}:=\FGMRES_{q_2}$ on $A_{q_2}$, preconditioned by
    $\operatorname{Base}_{q_2}$ (e.g.\ Jacobi$_{16}$), tolerance
    $\tau_{\mathrm{leaf}}$, cap $k_{\mathrm{leaf}}$; norms and Krylov scalars
    accumulate in FP32 (a recorded promotion).
  \item Inner edge: $P_{q_1}(v):=\textsc{ApplyAdapter}
    (q_1,q_2,A_{q_1},v,S_{q_2},J_{\mathrm{in}},\tau_{\mathrm{in}})$, with the
    zero guard of Algorithm~\ref{alg:krylov}.
  \item Mid solver: $S_{q_1}:=\FGMRES_{q_1}$ on $A_{q_1}$ with variable
    preconditioner $P_{q_1}$, tolerance $\tau_{\mathrm{mid}}$, cap
    $k_{\mathrm{mid}}$.
  \item Outer edge: $P_{p}(v):=\textsc{ApplyAdapter}
    (p,q_1,A_{p},v,S_{q_1},J_{\mathrm{out}},\tau_{\mathrm{ad}})$, zero-guarded.
  \item Return $x=\FGMRES_p(A_p,b,\tau,m)$ with preconditioner $P_p$. Each
    precision edge stops independently; only the innermost base runs in
    $q_2$.
\end{enumerate}
\end{algorithmblock}

\begin{algorithmblock}{Algorithm 5. \textsf{CAMPS-Apex}$[m_2,m_3,s]$ $(A_{64}, b, \tau)$}
\label{alg:apex}
\emph{Solver tree: Figure~\ref{fig:instances}d.}
\begin{enumerate}[label=\arabic*.,leftmargin=2.2em,itemsep=1.5pt,topsep=3pt]
  \item Leaf: $S_{16}:=\FGMRES_{16}$ on $A_{16}$ run for exactly $m_3$
    iterations (no local stopping test), preconditioned by $s$ fixed
    Jacobi--Richardson sweeps in FP16.
  \item Interior cast: $S_{32}(w):=\FGMRES_{32}$ on $A_{32}$ run for exactly
    $m_2$ iterations, right-preconditioned by
    $Q_{32}(S_{16}(Q_{16}(\cdot)))$ [plain casts: no residual
    formation, no scaling, no acceptance at this edge].
  \item Return $x=\textsc{ApplyAdapter}(64,32,A_{64},b,S_{32},J_{\max},\tau)$:
    the adapter is the outermost solver. Per correction it costs one
    FP64 residual, one power-of-two scale, one cast, and one acceptance
    check---and no FP64 Krylov basis exists anywhere in the tree.
\end{enumerate}
\end{algorithmblock}

\begin{algorithmblock}{Algorithm 6. \textsf{CAMPS-Ladder} $(A_{64}, b, \tau)$}
\label{alg:ladder}
\emph{Solver tree: Figure~\ref{fig:instances}e.}
\begin{enumerate}[label=\arabic*.,leftmargin=2.2em,itemsep=1.5pt,topsep=3pt]
  \item Setup: factor $A_{32}$ (child $\mathrm{LU}_{32}$) and $A_{64}$
    (fallback child $\mathrm{LU}_{64}$); both factorizations are charged to
    setup. A lazy variant would defer the FP64 factorization to the first
    contract violation.
  \item Tier 1/2 preconditioner: $P(v):=\textsc{ApplyAdapter}
    (64,32,A_{64},v,\mathrm{LU}_{32},J{=}2,\tau_{\mathrm{ad}})$ configured
    with \textsc{on-violation: fallback}---a correction that violates the
    acceptance test is discarded, the active pair switches to
    $(64,\mathrm{LU}_{64})$, a same-precision recovery with scale $1$ and no
    cast, and the remaining attempts of that application rebuild their
    right-hand side from the retained accepted residual and use
    $\mathrm{LU}_{64}$.
  \item Tier 3: if $P(v)=0$ (every attempt of the application was rejected,
    including the fallback child's), substitute the raw
    direction $z=\mathrm{LU}_{64}^{-1}v$, unchecked; after the first
    substitution, commit to tier 3 for the remainder of the solve. Count and
    report all substitutions.
  \item Return $x=\FGMRES_{64}(A_{64},b,\tau,m)$ with preconditioner $P$.
    Degradation is monotone, per-application, and visible: the contract
    record documents every rejection that triggered a tier change.
\end{enumerate}
\end{algorithmblock}

\providecommand{\figplaceholder}[1]{%
  \fbox{\parbox{0.92\linewidth}{\centering\vspace{1.5em}%
  \textit{Figure placeholder:} #1\vspace{1.5em}}}}

\section{Experiments}\label{experiments}
The contract of Section~\ref{zero-contract} does not factor in sparsity,
symmetry, or conditioning. Thus, the experiments are not trying to show that
one CAMPS instance is fast on one matrix. We test a broader claim: that for each matrix class, the CAMPS grammar should have a tree that composes correctly and either beats a comparable uniform-precision solver or shows where it stops helping. We test this claim on four matrix classes: dense
well-conditioned, dense ill-conditioned, sparse symmetric positive definite
(SPD), and sparse nonsymmetric. We use the named instances of
Section~\ref{strategy}, and Figure~\ref{fig:instances}
draws the solver tree of each one.

We also test two placements that Theorem~\ref{thm:closure} permits beyond a
single edge. The first is \textsf{CAMPS-Cascade}, an FP64/FP32/FP16
hierarchy with refinement at two precision edges. The second is an
ablation that moves one child across different sets of edges.

Throughout this section, we denote $\hat\eta$ as the measured ratio
$\norm{r_{t+1}}_2/\norm{r_t}_2$ of parent residuals across one correction.
This is what contract \eqref{eq:one-step} bounds, so we check
the hypothesis of the theorem on every run instead of just assuming it.
The measured ratio comes from one run and one right-hand side. It is not the
uniform factor $\hat\eta$ of Lemma~\ref{lem:recurrence}, which has to hold
for every right-hand side.

All the raw results, operation counts per precision, environment records, seeds, configuration files, and scripts are located in the CAMPS repository at \url{https://github.com/CVC-Lab/CAMPS} (commit \texttt{b44dc2de32fb}). In the repository, the file \texttt{REPRODUCIBILITY.md} links every result file to the command, seeds, protocol, and environment that produced it.

\subsection{Implementation}\label{exp-impl}
CAMPS is written in Python on PyTorch~2.11 with CUDA~12.8. All timings in
this section come from an NVIDIA GeForce RTX~5070~Ti Laptop GPU. On this GPU,
FP64 arithmetic runs at $1/64$ of the FP32 rate, which strongly favors
lower-precision factorizations. However, on a datacenter GPU with full-rate FP64, the
dense speedups below would be expected to be smaller.

The core package is about $1{,}700$ lines. The precision conversions in it are explicit, and nodes that receive a vector in the wrong format raise an error instead of casting it. The adapter is based on Algorithm~\ref{alg:adapter} step by step and records every attempt and its outcome. We also turn off TF32 and reduced-precision reductions, so FP32 means IEEE FP32 multiplies. Half-precision matrix products and norms accumulate in FP32, and code records each of these precision promotions.

There is a hardware gap that influences the dense experiments. cuSOLVER does not expose a standalone half-precision factorization or triangular solve. Its mixed-precision \texttt{gesv} drivers can factorize in half precision internally~\cite{nvidia2026cusolver}, but they are complete solvers with their own refinement loop, and our code does not use them. Thus, our FP16 dense children are approximate inverses, computed in FP64 and applied in half precision through tensor-core matrix products. Our FP16 sparse children are
half-precision Krylov iterations with Jacobi or Richardson preconditioning.

On the same GPU, the reruns reproduce iteration counts, acceptance decisions, and residuals bit for bit. We also ran the full verification suite on an
NVIDIA A100 node of the Lonestar6 system at the Texas Advanced Computing
Center, and it passed every check. Appendix~\ref{app-impl} describes the
software, the verification suite, and the A100 results in full.

\subsection{Protocol}\label{exp-protocol}
\paragraph{Stopping rule.} 
We stop every method when $\norm{b-Ax}_2\le\tau\norm{b}_2$ with $\tau=10^{-10}$. We evaluate this
residual from the FP64 operator and never from the Hessenberg estimate alone
(Section~\ref{outer-fgmres}). After every solve, the test code recomputes $\norm{b-A_{64}x}_2$ independently and reports convergence on that
value only.

\paragraph{Accuracy measures.} We choose a manufactured vector $x_{\mathrm m}$ and form the right-hand side $b=\fl(Ax_{\mathrm m})$ in FP64. We report the forward error $\norm{x-x_{\mathrm m}}_2/\norm{x_{\mathrm m}}_2$ as a check, not a stopping criterion. On ill-conditioned systems a small residual does not
guarantee a small forward error~\cite{carson2017analysis,skeel1980}.
Thus, the
ill-conditioned sweep also reports the normwise and componentwise
backward errors~\cite{oettli1964compatibility,arioli1989,higham2002asna}
\[
\beta_{\mathrm{norm}}=\frac{\norm{b-Ax}_\infty}{\norm{A}_\infty\norm{x}_\infty+\norm{b}_\infty},
\qquad
\beta_{\mathrm{comp}}=\max_i\frac{|b-Ax|_i}{(|A||x|+|b|)_i}.
\]
These use the $\infty$-norm, and the $\beta_2$ of Theorem~\ref{thm:stability} uses the $2$-norm.
The conventions for these quantities and explanations for how rounding in $b$ limits the attainable forward error are placed in Appendix~\ref{app-measures}.

\paragraph{Timing and memory.} We time setup (casts, factorizations, and
inverse construction) separately from solves. Caches of converted operators are also cleared between setup runs, so each setup time clearly indicates the cost of operator conversion. Each configuration runs 3 untimed warmup solves, then 15 timed solves for dense problems or 10 for sparse problems. Before and after each timed solve, we wait for the GPU to finish all work, so the timer only measures the time of the solve. We report the medians of the timed runs. If two methods are close, we also record the interquartile range, and all the interquartile ranges are in the archived tables. We also report memory in two ways. We record the count of the bytes in every
object the solver owns, such as operator copies, factors, inverses, and
Krylov bases. We also record the CUDA allocator's own measurement, which
provides resident memory after setup and peak memory during a solve.

\paragraph{Convergence plots.} For each experiment we plot the relative
residual $\norm{b-Ax}_2/\norm{b}_2$ against the outer iterations of the root
solver. When the root is an adapter, one outer iteration is one accepted
correction, and the plotted value is the residual that the adapter recomputes
in FP64. When the root is GMRES or FGMRES, one outer iteration is one Krylov
iteration. The plotted value is the residual estimate from the Arnoldi
recurrence, replaced by the recomputed FP64 residual at each restart and at
the last iteration. One outer iteration of a CAMPS tree can contain a full
inner solve in lower precision. Thus, these plots show how the residual
decreases, and the tables give the cost.

\paragraph{Baselines.} For each matrix class, we compare the CAMPS instances against the uniform-FP64
direct solve and against FP64 Krylov with a matching FP64 preconditioner. This is $\FGMRES_{64}(\mathrm{LU}_{64})$ for dense problems and CG$_{64}$
for SPD problems. We also include GMRES$_{64}$ with a fixed preconditioner,
because a flexible method should not be assumed to be faster
(Section~\ref{discussion}). \textsf{CAMPS-Direct} is classical mixed-precision iterative refinement. \textsf{CAMPS-Krylov} is the same solver with an outer FGMRES added, and both share the same children, scaling, and contract. Thus, comparing them shows what the FGMRES layer contributes, with no differences in implementation mixed in.

\paragraph{Child families.} The same adapter accepts six different kinds of
child, with no change to its code and no special cases.
\begin{itemize}
\item Dense LU ($\mathrm{LU}_{32}$, with $\mathrm{LU}_{64}$ as a fallback
  child).
\item An approximate inverse applied in half precision through tensor cores
  ($\mathrm{INV}_{16}$, $\mathrm{INV}_{\mathrm{b}16}$).
\item Symmetric Krylov ($\mathrm{CG}_{32}$).
\item Nonsymmetric Krylov ($\FGMRES_{32}$, $\FGMRES_{16}$).
\item Stationary sweeps (Jacobi, and fixed-count Richardson inside
  \textsf{CAMPS-Apex}).
\item Other adapters.
\end{itemize}
The adapter only asks each of them for the contract stated in Section~\ref{zero-contract}, and the parent never knows which kind of child
it called.

\subsection{Dense, well-conditioned systems}\label{exp-dense-wc}
We build each test matrix as $A=U\Sigma V^{\mathsf T}$, where $U$ and $V$
are random orthogonal matrices and $\Sigma$ holds singular values spaced
evenly on a log scale. This allows us to precisely set the condition number, and
we set $\kappa_2(A)=10^{2}$. We test sizes $n=2048$, $4096$, and $8192$. On our
GPU, FP32 LU is $5\times$ faster than FP64 LU at $n=2048$ and $13\times$
faster at $n=8192$. This is the classic case where mixed precision is worth it, because most of the cost is a factorization that becomes much cheaper
in lower precision.
These matrices are synthetic and do not model a physical system.
Figure~\ref{fig:matrix-structure} in the appendix shows a small example next
to the two sparse test matrices.

\begin{figure}[H]
\centering
\includegraphics[width=\linewidth]{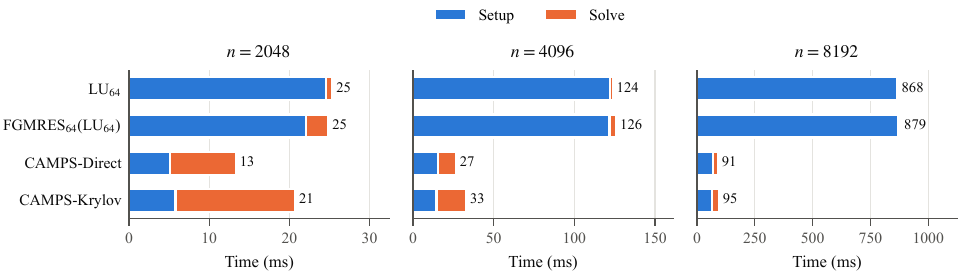}
\caption{Setup and solve time on dense well-conditioned systems with $\kappa=10^{2}$. The number after each bar is the total in ms. The FP64 factorization dominates both baselines, and the FP32 factorization removes most of that cost.}
\label{fig:dense-wc}
\end{figure}

\begin{table}[H]
\centering
\caption{Dense well-conditioned, $n=4096$, $\kappa=10^{2}$,
$\tau=10^{-10}$. Medians over 15 timed runs after 3 warmups (ms). ``Mem'' is
the analytic solver-owned total (operator copies, factors, inverses, bases).}
\label{tab:dense-wc}
\scriptsize
\resizebox{\textwidth}{!}{%
\begin{tabular}{@{}lrrrrlr@{}}
\toprule
Method & Setup & Solve & Total & Outer & $\norm{b-Ax}_2/\norm{b}_2$ & Mem (MiB) \\
\midrule
$\mathrm{LU}_{64}$ (direct) & 121.8 & 2.0 & 123.9 & 1 & $1.0\times10^{-13}$ & 256 \\
$\FGMRES_{64}(\mathrm{LU}_{64})$ & 121.8 & 4.3 & 126.0 & 1 & $1.1\times10^{-13}$ & 257 \\
\textsf{CAMPS-Direct}$[\mathrm{LU}_{32}]$ & 15.3 & 11.4 & \textbf{26.7} & 3 & $8.6\times10^{-13}$ & 256 \\
\textsf{CAMPS-Krylov}$[\mathrm{LU}_{32}]$ & 14.5 & 18.4 & 33.0 & 2 & $3.7\times10^{-16}$ & 257 \\
\textsf{CAMPS-Cascade}$[\mathrm{INV}_{16}]$ & 458.1 & 23.5 & 481.5 & 1 & $2.6\times10^{-11}$ & 258 \\
\textsf{CAMPS-Cascade}$[\mathrm{INV}_{\mathrm{b}16}]$ & 458.1 & 39.5 & 497.6 & 1 & $9.2\times10^{-13}$ & 258 \\
\bottomrule
\end{tabular}}
\end{table}
\noindent
Table~\ref{tab:dense-wc} and Figure~\ref{fig:dense-wc} show where the time
is being spent. At $n=4096$, both FP64 methods spend almost all of their time in the
FP64 factorization, about $122$\,ms, and the solve itself only takes a few
milliseconds. \textsf{CAMPS-Direct} replaces that factorization with an
FP32 one that takes $15.3$\,ms. Its solve is slower than the FP64 solve
because it needs three corrections instead of one, but it still finishes in
$26.7$\,ms in total. The FP64 direct solve takes $123.9$\,ms and
FP64-preconditioned FGMRES takes $126.0$\,ms, so \textsf{CAMPS-Direct} is
$4.6\times$ faster. All ratios here are computed from the medians in the
tables. The speedup grows with $n$, because the FP32 factorization gains
more over FP64 at larger sizes. It is $1.9\times$ at $n=2048$ ($13.3$
against $25.4$\,ms) and $9.6\times$ at $n=8192$ ($90.7$ against
$868.3$\,ms).

For this class, the adapter by itself with no outer Krylov solver is the
right tree. At $n=4096$, its FP32 LU child cuts the residual by a factor of
about $8\times10^{-5}$ on every correction. With a child this strong,
wrapping the adapter in FGMRES, which is what \textsf{CAMPS-Krylov} does, only saves
one correction, and each step costs more so the solve
is slower. \textsf{CAMPS-Krylov} does end at a smaller residual, but that
only helps if an application needs more accuracy than the stopping rule
asks for. Proposition~\ref{prop:work}\,(a) predicts the cost of
\textsf{CAMPS-Direct} from the measured factor. It gives
$K=\lceil\ln10^{-10}/\ln\hat\eta\rceil=3$ corrections, which matches what we
observe. The measured solve time of $11.4$\,ms is $3.8$\,ms per correction.

Mixed precision usually saves memory, but it does not here. The FP64
direct solve and \textsf{CAMPS-Direct} both cost about $256$\,MiB. The FP32
child keeps its own copy of the matrix so that it can check its own
residual, and $A_{32}$ plus $\mathrm{LU}_{32}$ ends up taking the same amount of space as
$\mathrm{LU}_{64}$. Without that extra copy, \textsf{CAMPS-Direct} would need about $192$\,MiB, a quarter less than the FP64 solve.

\textsf{CAMPS-Cascade}, with adapters at both precision boundaries, also
converges on this class in a single outer iteration. It ends at a residual
of $2.6\times10^{-11}$ with the FP16 child and $9.2\times10^{-13}$ with the
BF16 child, and its outer contraction factor is about $5\times10^{-6}$. The cascade puts one adapter inside another, and each one decides on its own when to stop. Theorem~\ref{thm:closure} says that nesting them this way still gives a working solver, and our run confirms it. The BF16 version is slower, at
$39.5$\,ms against $23.5$\,ms. BF16 stores fewer digits than FP16, so the inner level has to do more work
to reach the same accuracy. It uses $26$ FP32 and $8$ BF16 products
against $16$ and $4$ with FP16. That extra inner work also makes its single outer
correction more accurate, which is why it ends at a smaller residual.

The cascade's solve time is close to the solve time of \textsf{CAMPS-Direct}, but its
total time is about $480$\,ms. Almost all of that is setup, which computes the inverse that the half-precision child applies in FP64. We need this
step because cuSOLVER exposes no standalone half-precision factorization, so the cost
comes from the library interface, not from the framework. In
Section~\ref{exp-sparse-nsy} the children do not need factorization, and the
same cascade is faster than both FP64 baselines.

\begin{figure}[H]
\centering
\includegraphics[width=0.8\linewidth]{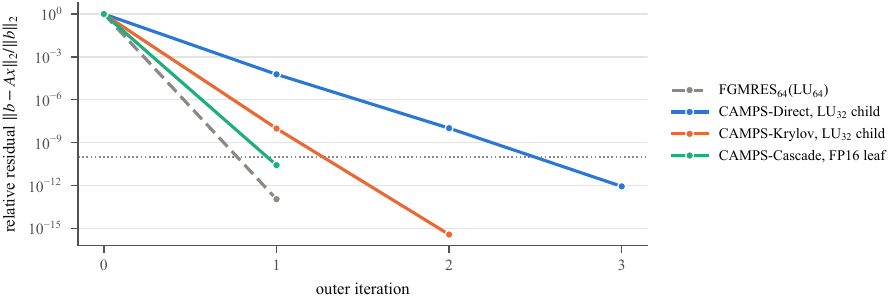}
\caption{Relative residual against outer iterations on the dense well-conditioned system with $n=4096$. The dotted line is the tolerance $\tau=10^{-10}$. LU$_{64}$ is a direct solve and has no history.}
\label{fig:conv-dense-wc}
\end{figure}

Figure~\ref{fig:conv-dense-wc} shows the residual after each outer iteration at
$n=4096$. Each correction of \textsf{CAMPS-Direct} reduces the residual by a
factor of about $10^{4}$, and three corrections reach the tolerance. With
FGMRES at the root, \textsf{CAMPS-Krylov} needs two iterations and the
cascade needs one.

\paragraph{Multiple right-hand sides.} When the same matrix is being solved for
many right-hand sides, the FP64 factorization is computed once and shared
across all of them. It might be expected that the setup savings of mixed precision
to matter less and less as the number of right-hand sides grows. However, 
Table~\ref{tab:multirhs} shows that this does not happen in the range we
tested. The adapter works on a block of right-hand sides directly, scaling
and checking each column separately (Algorithm~\ref{alg:direct}), so
\textsf{CAMPS-Direct} solves all the $k$ columns together in one batch.

\begin{table}[H]
\centering
\caption{Multi-RHS amortization, dense $n=4096$, $\kappa=10^{2}$: setup once
plus one $(n,k)$ block solve. Total ms (per-RHS ms).}
\label{tab:multirhs}
\scriptsize
\resizebox{\textwidth}{!}{%
\begin{tabular}{@{}lrrrr@{}}
\toprule
Method & $k=1$ & $k=4$ & $k=8$ & $k=16$ \\
\midrule
$\mathrm{LU}_{64}$ (direct) & 123.9 (123.9) & 126.7 (31.7) & 128.7 (16.1) & 130.9 (8.2) \\
\textsf{CAMPS-Direct}$[\mathrm{LU}_{32}]$ & \textbf{24.7} (24.7) & \textbf{39.6} (9.9) & \textbf{42.9} (5.4) & \textbf{46.2} (2.9) \\
\textsf{CAMPS-Krylov}$[\mathrm{LU}_{32}]$ & 27.1 (27.1) & 71.8 (18.0) & 119.9 (15.0) & 245.5 (15.3) \\
\textsf{CAMPS-Cascade}$[\mathrm{INV}_{16}]$ & 483.4 (483.4) & 551.4 (137.9) & 681.2 (85.2) & 888.8 (55.6) \\
\bottomrule
\end{tabular}}
\end{table}
\noindent
\textsf{CAMPS-Direct} is the fastest method at every $k$. At $k=1$ it wins
for the same reason as in Table~\ref{tab:dense-wc}. Setup dominates both
totals, and the FP32 factorization costs about an eighth of the FP64 one.
As $k$ grows, the FP64 direct solve spreads its factorization over more
columns, and its cost per column falls from $123.9$ to $8.2$\,ms.
\textsf{CAMPS-Direct} benefits in the same way, from $24.7$ to $2.9$\,ms per
column since it also solves all columns together. Each additional
column costs \textsf{CAMPS-Direct} a little more than it costs the direct
solve. Each column needs three corrections, each with an FP64 residual,
where the direct solve needs only one pair of triangular solves. Thus, if $k$
keeps growing, the direct solve eventually catches up. Extrapolating from
our measurements puts that point at $k\approx100$ or later depending on
which points we fit, which is well beyond the tested range of $k\le16$.

However, \textsf{CAMPS-Krylov} behaves differently. Its FGMRES wrapper solves one
column at a time, so it cannot share work across columns the way the batched
methods do. Its cost per column stops falling at about $15$\,ms and by
$k=16$ it is slower than the FP64 direct solve. \textsf{CAMPS-Cascade} gets
cheaper per column as $k$ grows, but building its inverse is so expensive
that it remains the slowest method at every $k$. We did not test larger
blocks, repeated solves, or a block Krylov method. For this nonsymmetric
matrix, the natural block baseline would be block GMRES or block
FGMRES~\cite{gutknecht2007block}. Block CG~\cite{oleary1980block} would not
apply since it requires a symmetric positive definite matrix.

The batched results do not carry over to right-hand sides that arrive one
at a time, where the factors are reused and each solve is separate. From
Table~\ref{tab:dense-wc}, \textsf{CAMPS-Direct} saves $121.8-15.3=106.5$\,ms
of setup and costs $11.4-2.0=9.4$\,ms more per solve. If every solve costs
the same as the first, the FP64 direct solve has the lower total time after
about $106.5/9.4\approx11$ solves. This is an estimate from the
single-solve timings, not a measurement.

\subsection{Dense, ill-conditioned systems}\label{exp-dense-ic}
We use the same kind of matrix at $n=2048$ but with condition numbers from
$10^{4}$ to $10^{12}$. Classical analysis says that refinement with an FP32
LU child converges as long as $\kappa u_{32}\lesssim1$, where
$u_{32}\approx6\times10^{-8}$ is the unit roundoff of FP32: that means up to
about $\kappa\approx1.7\times10^{7}$. All dense LU factorizations in this
section use standard partial pivoting, and no small pivots are replaced.

\begin{table}[H]
\centering
\caption{Ill-conditioned sweep, $n=2048$, $\tau=10^{-10}$. Cells: outer
iterations, or adapter attempts (accepted plus rejected corrections) to
convergence, or the failure mode (with the final relative residual). The first row is the median measured
per-correction parent-residual ratio of the $\mathrm{LU}_{32}$ child. At
$\kappa\ge10^{8}$ it is the midpoint of one accepted and one rejected ratio. Dense LU nodes run without pivot
boosting (see text).}
\label{tab:dense-ic}
\scriptsize
\resizebox{\textwidth}{!}{%
\begin{tabular}{@{}lcccccc@{}}
\toprule
 & $10^{4}$ & $10^{6}$ & $10^{7}$ & $10^{8}$ & $10^{10}$ & $10^{12}$ \\
\midrule
measured $\hat\eta$ ($\mathrm{LU}_{32}$ child) & $7.3\times10^{-4}$ & $3.8\times10^{-2}$ & $0.34$ & $3.4$ & $5.7$ & $3.1$ \\
\midrule
$\mathrm{LU}_{64}$ (direct) & ok & ok & ok & ok & ok & ok \\
$\FGMRES_{64}(\mathrm{LU}_{64})$ & ok\,1 & ok\,1 & ok\,1 & ok\,1 & ok\,1 & ok\,1 \\
\textsf{CAMPS-Direct}$[\mathrm{LU}_{32}]$ & ok\,3 & ok\,5 & ok\,13 & rejected & rejected & rejected \\
\textsf{CAMPS-Krylov}$[\mathrm{LU}_{32}]$ & ok\,1 & ok\,2 & ok\,5 & DNC & DNC & DNC \\
\quad same, loose acceptance & ok\,1 & ok\,2 & ok\,5 & DNC & DNC & DNC \\
\textsf{CAMPS-Direct}$[\mathrm{LU}_{32}\,|\,\mathrm{LU}_{64}]$ (fallback) & ok\,3 & ok\,5 & ok\,13 & ok\,3 & ok\,3 & ok\,3 \\
\quad same, damped, progress threshold $\varrho=0.9$ & ok\,3 & ok\,5 & ok\,11 & ok\,3 & ok\,3 & ok\,3 \\
\textsf{CAMPS-Ladder} & ok\,2 & ok\,3 & ok\,7 & ok\,2 & ok\,2 & ok\,2 \\
\textsf{CAMPS-Direct}$[\mathrm{LU}_{32}]$, damped & ok\,3 & ok\,5 & ok\,11 & stagnated ($1.6\times10^{-3}$) & stagnated ($2.9\times10^{-4}$) & stagnated ($8.4\times10^{-4}$) \\
\textsf{CAMPS-Krylov}$[\mathrm{LU}_{32}]$, damped & ok\,1 & ok\,2 & ok\,4 & DNC & DNC & DNC \\
\textsf{CAMPS-Ladder}, damped & ok\,2 & ok\,3 & ok\,6 & ok\,6 & ok\,4 & ok\,10 \\
\bottomrule
\end{tabular}}
\end{table}
\noindent
Table~\ref{tab:dense-ic} has two parts. The first row is the median
of the measured ratio $\hat\eta$ over the corrections of \textsf{CAMPS-Direct}.
In every run the first correction reduces the residual far more than the
later ones, by a factor between $10^{-5}$ and $2\times10^{-3}$. At
$\kappa\ge10^{8}$ there are two corrections. The first is accepted, and the
second is a rejected trial that would raise the residual by a factor of
$6.7$, $11$, and $6.2$ at $\kappa=10^{8}$, $10^{10}$, and $10^{12}$. The entry
in the table is the midpoint of the two. The
rest of the table shows how each method ends at each $\kappa$. The two FP64
methods converge everywhere, as expected, since FP64 LU is backward stable.
The instances whose only child is FP32 LU converge up to $\kappa=10^{7}$ and
fail beyond it. Past that point the second FP32 correction would make
the residual larger, and the acceptance test rejects it. The two instances that can switch
to an FP64 child, the fallback and the ladder, converge at every $\kappa$.
The table supports five findings.

\begin{figure}[H]
\centering
\includegraphics[width=0.62\linewidth]{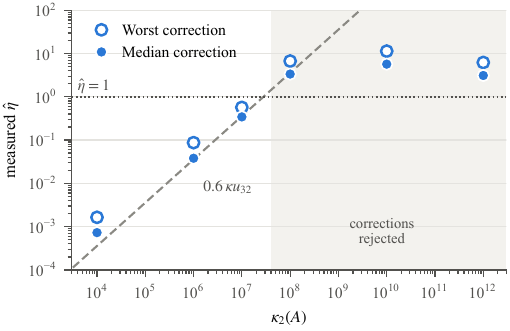}
\caption{Measured contraction factor $\hat\eta$ of the FP32 LU child against $\kappa$, $n=2048$. Filled points are the median correction and open points the worst. At $\kappa\ge10^{8}$ two corrections run, one accepted and one rejected, so the median is the midpoint of the two. Up to $\kappa=10^{7}$ the median follows $0.6\,\kappa u_{32}$, and the ratio crosses one between $\kappa=10^{7}$ and $10^{8}$.}
\label{fig:dense-ic-eta}
\end{figure}

\paragraph{The measured contraction follows the classical prediction.}
Figure~\ref{fig:dense-ic-eta} plots the measured $\hat\eta$ of the FP32 LU
child against $\kappa$. From $\kappa=10^{4}$ to $10^{7}$ the median stays within a factor of about two of
$0.6\,\kappa u_{32}$, and the ratio crosses $1$ between $\kappa=10^{7}$ and
$10^{8}$, which is where the classical analysis says it should. At $\kappa=10^{8}$ the
rejected trial has a ratio of $6.7$, about $1.1\,\kappa u_{32}$. Beyond $10^{8}$ the
rejected ratios stop growing with $\kappa$. The acceptance test computes $\hat\eta$ on every correction, and the theory
says refinement converges when this factor stays below one. Thus, the
solver checks the convergence condition directly while it runs
instead of just assuming it holds. The
condition that Lemma~\ref{lem:geom} relies on can be checked at every
edge on every correction while the solver runs.

\paragraph{Failure is fast and has a clear cause.} For $\kappa\ge10^{8}$,
the acceptance test stops \textsf{CAMPS-Direct} at the first correction that
increases the residual. This takes $4$--$8$\,ms, and the solver directly acknowledges the termination. On the other hand, the FGMRES-wrapped variants keep
iterating for $3.2$-$4.6$\,s until they hit the iteration limit, and they only report that they did not converge, not indicating why.

We also tried a fix suggested by GMRES-IR, which keeps using low-precision
LU factors as a preconditioner even when they no longer reduce the residual
on their own. We let the outer FGMRES accept these corrections as search
directions, allowing the residual to grow by up to $100\times$. The table
calls this loose acceptance, and it does not help. At $\kappa=10^{8}$, a single
FP32 LU correction increases the residual of a random unit right-hand side
by about $300\times$, and each further correction multiplies it again. Even
an unrestarted FGMRES with 400 search directions built this way stalls at a
relative residual of about $10^{-5}$.

This does not mean that GMRES-IR fails on these problems. Classical
GMRES-IR runs GMRES on the correction equation, with the low-precision LU
factors as its preconditioner. \textsf{CAMPS-Krylov} is built differently.
Its outer FGMRES takes the adapter's whole correction as one search
direction. Thus, the failure marks the limit of the configuration we
tested on a matrix class built to be hard for it. Going further in experimentation would mean
either applying the preconditioner in higher precision, as the
Carson--Higham analysis assumes~\cite{carson2017analysis}, or having a stronger
child.

\paragraph{A fallback child makes the solver robust.} The fallback version
of \textsf{CAMPS-Direct} has a second child, an FP64 LU factorization. The
adapter uses the cheap FP32 child as long as its corrections pass the
acceptance test. Up to $\kappa=10^{7}$, the fallback solver behaves the same as plain
iterative refinement. As soon as a correction fails, the adapter switches
to the FP64 child. With this switch, it converges at every $\kappa$ from
$10^{4}$ to $10^{12}$. For $\kappa\ge10^{8}$, where every FP32-only instance fails, it needs three corrections. The first FP32 correction is accepted and the second is
rejected. A single FP64 correction after the rejection reaches the
tolerance. The solve takes $6$-$7$\,ms, and the normwise backward error is $10^{-14}$ or smaller. The cost is in the setup, which has to build both factorizations and takes $28$\,ms, versus $5$\,ms for just the FP32 factorization. A lazy version could
only build the FP64 factors when the first correction fails.

\textsf{CAMPS-Ladder} (Algorithm~\ref{alg:ladder}) also converges at every
$\kappa$, in $2$-$7$ outer iterations. Its last resort, an unchecked FP64
LU direction, was never needed because the checked FP64 fallback always
worked. On this sweep the third tier is only a safety net and does not
affect any result.

At $\kappa=10^{12}$, the forward error is $1.2\times10^{-4}$ for the FP64
direct solve and $1.0\times10^{-2}$ for the fallback instance. The gap comes
from where each solver stops. The direct solve ends at a relative residual
of $9.9\times10^{-15}$. The fallback stops as soon as its residual falls
below $\tau=10^{-10}$, which happens at $1.9\times10^{-12}$. At this
condition number, a larger residual allows a larger forward error, and the
forward errors differ by a factor of $89$, which is close to the difference factor of $190$
between the residuals. Both solves have small normwise
backward errors, of $3.2\times10^{-16}$ and $4.0\times10^{-14}$. Thus, at this
conditioning, the stopping tolerance, not the working precision, controls
the forward error. Appendix~\ref{app-fwd} gives the bound behind this.

\begin{figure}[H]
\centering
\includegraphics[width=\linewidth]{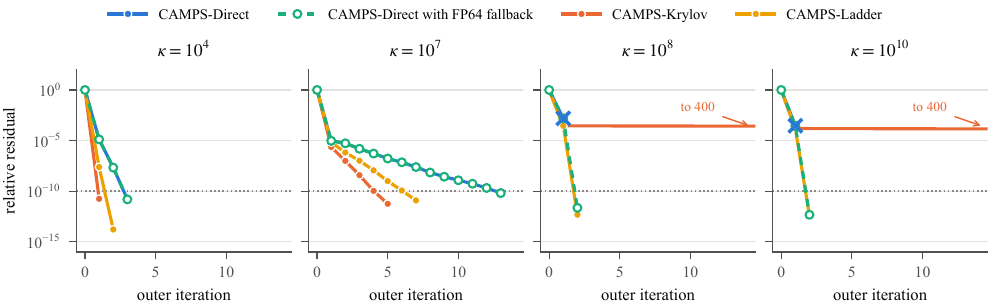}
\caption{Relative residual against outer iterations on dense ill-conditioned systems, $n=2048$. At $\kappa=10^{4}$ and $10^{7}$ the fallback is not used, so \textsf{CAMPS-Direct} follows the same curve with and without it. At $\kappa=10^{8}$ and $10^{10}$ the cross marks where \textsf{CAMPS-Direct} stops after the acceptance test rejects its second correction. \textsf{CAMPS-Krylov} stalls near $10^{-4}$ and runs to its limit of $400$ iterations. The fallback and ladder versions converge.}
\label{fig:conv-dense-ic}
\end{figure}

Figure~\ref{fig:conv-dense-ic} shows the residual histories at four condition
numbers.

\paragraph{Damping needs a progress threshold.} Damping scales each
correction by the factor that minimizes the new residual
(Section~\ref{contract}). Thus, a damped correction can never make the
residual grow, so it always passes the acceptance test. For
$\kappa\ge10^{8}$, the damped instances have no rejections at all, but the
corrections they accept barely do anything. The damped FP32 LU steps have
measured $\hat\eta$ between $0.99$ and $1.00$, so each step basically leaves the
residual unchanged. Damped \textsf{CAMPS-Direct} stalls after
$4$-$10$ corrections, at the same residual where the undamped version was
rejected. Damped \textsf{CAMPS-Krylov} accepts more than a thousand
corrections per solve and still runs out of iterations. Damping turns a
clear rejection into a clear stall, but it does not make the FP32 child
work. As Section~\ref{contract} notes, in principle, a damped child could still converge, but the children we tested do not.

Damping also blocks escalation. The adapter switches to a stronger child
only after a rejection, and a damped step that passes without making
progress never triggers one. At $\kappa=10^{8}$, the damped ladder
therefore needs $6$ outer iterations and $34$\,ms, against $2$ iterations
and $10$\,ms without damping. Its FP32 child keeps passing with
$\hat\eta\approx0.99$, so the fallback never starts. The fix is to require
real progress. With the stricter test \eqref{eq:one-step} and a progress
threshold of $\varrho=0.9$, any step that reduces the residual by less than
$10\%$ counts as a failure. For such a step, the damped fallback then switches to FP64. For $\kappa\ge10^{8}$ it converges in three
corrections, which is the same as the undamped fallback (the progress-threshold row
of Table~\ref{tab:dense-ic}). So when damping is on, the acceptance test
must check that the residual drops enough, not only that it does not grow.
This shows why contract \eqref{eq:clean} alone does not guarantee convergence.

\subsection{Sparse SPD systems}\label{exp-sparse-spd}
The problem we test is the 3D Poisson equation, discretized with the standard
seven-point stencil. Each row has $6$ on the diagonal and $-1$ for each
neighboring grid point. Figure~\ref{fig:poisson-problem} shows the equation and the stencil. We use grids of $32^{3}$, $48^{3}$, and $64^{3}$
points, so the largest system has $n=262{,}144$ unknowns. We compare
against CG$_{64}$, with and without Jacobi preconditioning, which is the
standard FP64 method for this class. We also compare against
$\FGMRES_{64}(\mathrm{Jacobi}_{64})$, the FP64 flexible method that
corresponds the most to the CAMPS instances.

\definecolor{campsblue}{HTML}{2A78D6}
\definecolor{campsorange}{HTML}{EB6834}
\definecolor{campsgray}{HTML}{52514E}
\tikzset{
  pdecube/.pic={
    \draw[campsgray,densely dashed,line width=0.4pt]
      (0,0)--(0.8,0.55)--(2.8,0.55) (0.8,0.55)--(0.8,2.55);
    \draw[campsgray,line width=0.6pt] (0,0) rectangle (2,2);
    \draw[campsgray,line width=0.6pt]
      (0,2)--(0.8,2.55)--(2.8,2.55)--(2,2) (2,0)--(2.8,0.55)--(2.8,2.55);
  },
  pdeaxes/.pic={
    \draw[-{Stealth[length=3pt]},campsgray,line width=0.4pt] (0,0)--(0.6,0)
      node[right,inner sep=1pt,font=\scriptsize]{$x_3$};
    \draw[-{Stealth[length=3pt]},campsgray,line width=0.4pt] (0,0)--(0,0.6)
      node[above,inner sep=1pt,font=\scriptsize]{$x_2$};
    \draw[-{Stealth[length=3pt]},campsgray,line width=0.4pt] (0,0)--(0.4,0.275)
      node[above right,inner sep=0.5pt,font=\scriptsize]{$x_1$};
  },
  stpt/.style={circle,draw=campsgray,line width=0.5pt,fill=white,inner sep=0pt,
    minimum size=18pt,font=\scriptsize},
}
\begin{figure}[H]
\centering
\begin{tikzpicture}
  \pic at (0,0) {pdecube};
  \shade[inner color=campsorange!85,outer color=campsorange!6] (1.4,1.28) circle (0.55);
  \node[font=\small] at (1.4,1.28) {$f$};
  \node[font=\scriptsize,anchor=west] (bc) at (3.2,2.3) {$u=0$ on $\partial\Omega$};
  \draw[-{Stealth[length=3pt]},campsgray,line width=0.4pt] (bc.west)--(2.45,1.8);
  \node[font=\small] at (1.4,-0.5) {$-\Delta u=f$ in $\Omega=(0,1)^{3}$};
  \pic at (-1.35,0) {pdeaxes};
  \begin{scope}[shift={(8.3,1.25)}]
    \foreach \p in {(-1.3,0),(1.3,0),(0,-1.3),(0,1.3),(-0.8,-0.55),(0.8,0.55)}
      \draw[campsgray,line width=0.5pt] (0,0)--\p;
    \node[stpt] at (0.8,0.55) {$-1$};
    \node[stpt] at (-1.3,0) {$-1$};
    \node[stpt] at (1.3,0) {$-1$};
    \node[stpt] at (0,1.3) {$-1$};
    \node[stpt] at (0,-1.3) {$-1$};
    \node[stpt,fill=campsblue!20] at (0,0) {$6$};
    \node[stpt] at (-0.8,-0.55) {$-1$};
    \pic at (-2.6,-1.25) {pdeaxes};
  \end{scope}
\end{tikzpicture}
\caption{The Poisson test problem. Left: the equation $-\Delta u=f$ on the
unit cube with $u=0$ on the boundary. It gives the steady temperature of a
body that is heated by a source $f$ while its surface is kept at a fixed
temperature. Right: the seven-point stencil on a grid with spacing $h$. After
multiplication by $h^{2}$, each grid point has weight $6$ and each of its six
neighbors has weight $-1$, which gives one row of $A$.}
\label{fig:poisson-problem}
\end{figure}
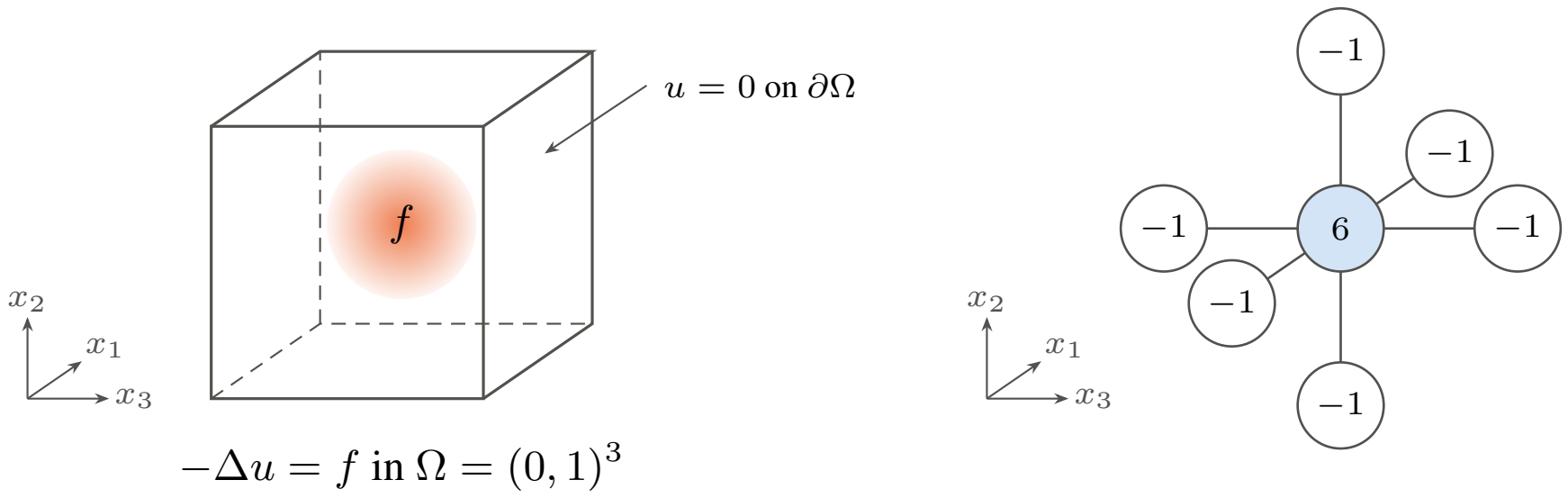

\begin{table}[H]
\centering
\caption{Poisson3D, $64^{3}$ ($n=262{,}144$), $\tau=10^{-10}$. Solve medians
over 10 runs. Matvec counts are by storage format, and peak is the measured
CUDA allocator peak during one solve.}
\label{tab:sparse-spd}
\scriptsize
\resizebox{\textwidth}{!}{%
\begin{tabular}{@{}lrrlrr@{}}
\toprule
Method & Solve (ms) & Outer & matvecs (fp64/fp32/fp16) & Peak (MiB) & fwd.\ error \\
\midrule
CG$_{64}$ & \textbf{97.5} & 236 & 237 / -- / -- & 14.7 & $4.3\times10^{-9}$ \\
CG$_{64}$(Jacobi$_{64}$) & 154.2 & 236 & 473 / -- / -- & 18.7 & $4.3\times10^{-9}$ \\
$\FGMRES_{64}$(Jacobi$_{64}$) & 1439.7 & 383 & 781 / -- / -- & 194.7 & $8.1\times10^{-8}$ \\
\textsf{CAMPS-Direct}$[\mathrm{CG}_{32}(\mathrm{J}_{32})]$ & 217.9 & 3 & 5 / 693 / -- & 20.0 & $5.0\times10^{-11}$ \\
\textsf{CAMPS-Krylov}$[\mathrm{CG}_{32}(\mathrm{J}_{32})]$ & 319.0 & 2 & 13 / 1012 / -- & 147.0 & $2.6\times10^{-15}$ \\
\textsf{CAMPS-Cascade}$[\mathrm{Jacobi}_{16}]$ & 1584.4 & 2 & 13 / 102 / 1518 & 210.5 & $8.8\times10^{-16}$ \\
\bottomrule
\end{tabular}}
\end{table}
\noindent
CG$_{64}$ is the fastest method in Table~\ref{tab:sparse-spd}, and no CAMPS instance beats it. The CG$_{64}$(Jacobi$_{64}$) row is slower than plain CG$_{64}$ and shows twice as many FP64 products. This is not because of Jacobi, since it only rescales the vector by a constant on this stencil. The extra $236$ products come from our code, where every base node checks its own result with one operator product that the timer includes. Thus, we compare against plain CG$_{64}$. We have not separated these products from the timed work yet, and Section~\ref{discussion} lists this as future work.

Every CAMPS instance converges and every correction is accepted. Each correction cuts the residual by a factor of about $10^{-4}$, so the last one lands well below the tolerance. The CAMPS instances end at relative residuals between about $2\times10^{-16}$ and $9\times10^{-13}$, while CG$_{64}$ stops right below $10^{-10}$. Their forward errors are smaller by factors between about $30$ and $5\times10^{6}$ across the three grids.

\begin{figure}[H]
\centering
\includegraphics[width=0.72\linewidth]{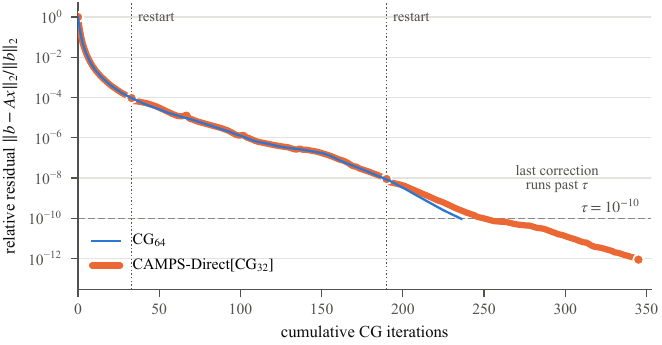}
\caption{Relative residual against cumulative CG iterations on Poisson at
$64^{3}$. The \textsf{CAMPS-Direct} curve is the inner FP32 CG residual of
each correction, converted to the units of the original system, and the dots
are the FP64 residuals after each correction. Each correction restarts CG
from zero. The two curves agree through the first two corrections.
\textsf{CAMPS-Direct} reaches $\tau$ after $250$ iterations, against $236$
for CG$_{64}$, but its last correction runs on to its own tolerance of
$10^{-4}$ and ends after $345$.}
\label{fig:sparse-spd}
\end{figure}

Figure~\ref{fig:sparse-spd} shows where \textsf{CAMPS-Direct} loses time at $64^{3}$. Its three corrections take $33$, $157$, and $155$ inner CG iterations, which is $345$ in total, versus $236$ for a single CG$_{64}$ solve. Although each correction restarts CG from zero, the two residual curves overlap through the first two corrections, so restarting costs almost nothing. The inner FP32 residual of \textsf{CAMPS-Direct} falls below the tolerance after $250$ iterations, $14$ more than CG$_{64}$. This is the child's own residual estimate, not a checked FP64 residual. Those few extra iterations come near the end. CG picks up speed the longer it runs~\cite{simoncini2005superlinear}, and the last correction loses that because it starts over from zero. The last correction runs for another $95$ iterations, until it reaches its own tolerance. A child that stops at the parent's tolerance could save part of this work, but we have not tested one. The table shows $693$ FP32 products, about twice the number of iterations, because the Jacobi node inside the child also does the extra check described earlier. On this problem, switching to FP32 also barely speeds up each iteration. Dividing solve time by iteration count, \textsf{CAMPS-Direct} takes $0.63$\,ms per iteration and CG$_{64}$(Jacobi$_{64}$), which does the same extra product, takes $0.65$\,ms. Plain CG$_{64}$ takes $0.41$\,ms. Proposition~\ref{prop:work}\,(b) compares solvers by their cost and
progress per unit of work. Taking one CG iteration as that unit, an FP32 iteration costs $1.53$ times as much as a CG$_{64}$ iteration and makes only $0.68$ times as much progress ($236/345$). Thus, the lowering criterion \eqref{eq:lowering} fails by a factor of $2.2$, which is exactly the ratio of the two solve times.

Three changes could each narrow this gap, and none of them requires
changing the interface (Section~\ref{zero-contract}). First, the last correction could stop as soon as the parent's tolerance is reached, which would save most of those $95$ iterations. Second, the Jacobi node could drop its extra product. Third, the child could keep its Krylov space from one correction to the next~\cite{parks2006recycling,notay2000fcg,knyazev2007variable}.
We have not tested any of these, so this result reflects the specific instances we ran, not a limit of the CAMPS framework.

\textsf{CAMPS-Krylov} is \textsf{CAMPS-Direct} with an FGMRES layer wrapped around it. That layer adds FP32 products ($1012$ against $693$) and has to store a Krylov basis, so \textsf{CAMPS-Krylov} is slower and uses much more memory ($147.0$ against $20.0$\,MiB). \textsf{CAMPS-Cascade} pushes most of the work down to an FP16 FGMRES solver with Jacobi preconditioning. It also needs $1518$ FP16 products and is the slowest of the three.

In principle, our acceptance test could throw away a good CG step. CG always reduces the error in the energy norm, but the residual can still go up along the way, and a residual-based test would reject such a step (Section~\ref{contract}). We looked for this at $32^{3}$ and $48^{3}$ and did not find it. On those runs, the residual test and the energy test accepted the same corrections and produced identical iterates. They agree because the child always runs to its own tolerance before returning, so its final residual is small. Appendix~\ref{app-energy} gives
the margins and the cost of the energy test.

For this class, we recommend CG$_{64}$. CAMPS pays off when the FP64 work it replaces is expensive, such as a dense factorization or the long orthogonalizations of GMRES. We find that a CG iteration on this problem is too cheap for lower precision to save much.

\subsection{Sparse nonsymmetric systems}\label{exp-sparse-nsy}
\begin{figure}[H]
\centering
\begin{tikzpicture}
  \foreach \y in {0.75,1.3,1.85}
    \draw[-{Stealth[length=5pt]},campsblue,line width=0.9pt] (-0.45,\y)--(3.45,\y);
  \pic at (0,0) {pdecube};
  \node[font=\scriptsize,campsblue,anchor=west] at (3.5,1.3) {flow, speed $c$};
  \node[font=\scriptsize,anchor=west] (bc) at (3.2,2.45) {$u=0$ on $\partial\Omega$};
  \draw[-{Stealth[length=3pt]},campsgray,line width=0.4pt] (bc.west)--(2.5,2.2);
  \node[font=\small] at (1.4,-0.5)
    {$-\Delta u+c\,\partial u/\partial x_{3}=f$ in $\Omega=(0,1)^{3}$};
  \pic at (-1.6,0) {pdeaxes};
  \begin{scope}[shift={(9.0,1.25)}]
    \foreach \p in {(-1.55,0),(1.4,0),(0,-1.3),(0,1.3),(-0.85,-0.6),(0.85,0.6)}
      \draw[campsgray,line width=0.5pt] (0,0)--\p;
    \node[stpt] at (0.85,0.6) {$-1$};
    \node[stpt,fill=campsorange!28,minimum size=25pt] (up) at (-1.55,0) {$-1{-}w$};
    \node[stpt] at (1.4,0) {$-1$};
    \node[stpt] at (0,1.3) {$-1$};
    \node[stpt] at (0,-1.3) {$-1$};
    \node[stpt,fill=campsblue!20,minimum size=25pt] at (0,0) {$6{+}w$};
    \node[stpt] at (-0.85,-0.6) {$-1$};
    \node[font=\scriptsize,campsorange!80!black,anchor=north] at (-1.75,-0.48) {upstream};
    \pic at (-3.1,-1.25) {pdeaxes};
  \end{scope}
\end{tikzpicture}
\caption{The convection--diffusion test problem. Left: the equation
$-\Delta u+c\,\partial u/\partial x_{3}=f$ on the unit cube with $u=0$ on the
boundary. It models heat that diffuses while a flow of speed $c$ carries it
in the $x_{3}$ direction. Right: the seven-point stencil with an upwind
difference for the flow term, after multiplication by $h^{2}$, where
$w=ch$. The upstream neighbor has weight $-1-w$ and the downstream neighbor
has weight $-1$, so $A$ is not symmetric.}
\label{fig:convdiff-problem}
\end{figure}
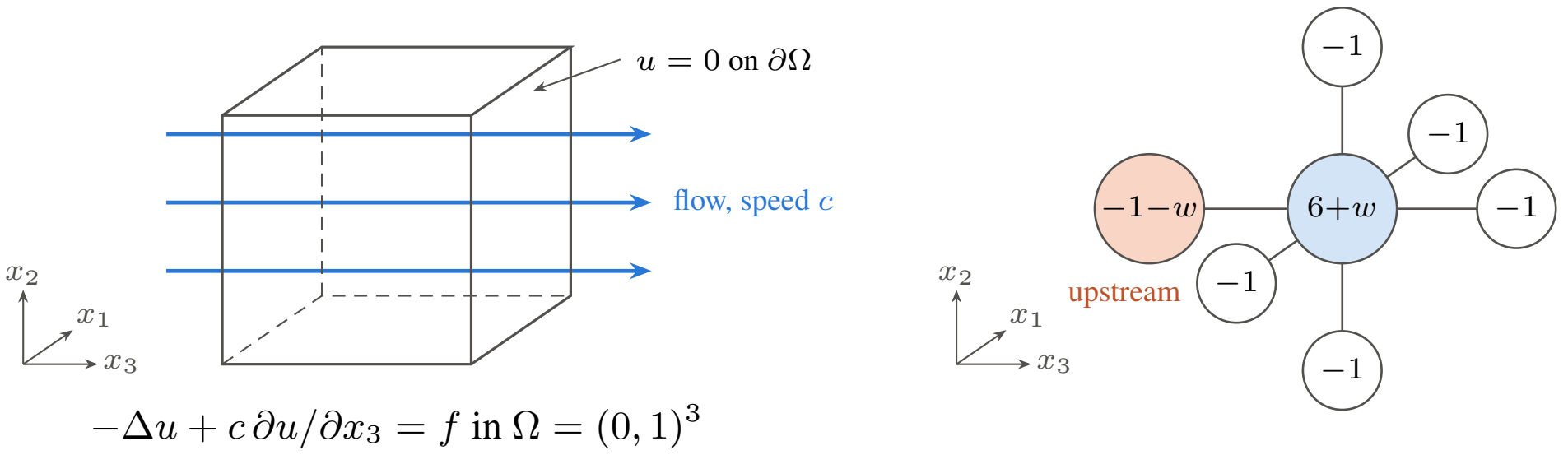

The test problem is 3D convection-diffusion, discretized with a
seven-point stencil. Each row has $6+w$ on the diagonal, $-1-w$ for the backward neighbor in the third direction, and $-1$ for the other five neighbors. Figure~\ref{fig:convdiff-problem} shows the equation and the stencil. The wind strength $w$ sets how strongly convection dominates, and a larger $w$ makes the matrix more nonsymmetric. We use $w=2$, $20$, and $100$ on a $32^{3}$ grid, and $w=20$ on a $48^{3}$ grid. We hold $w$ fixed when the grid changes, so the flow speed $c=w/h$ is larger on the finer grid. The main solver here is the sparse \textsf{CAMPS-Cascade} (Algorithm~\ref{alg:cascade}). It stores the matrix in three precisions, places an adapter at both precision
boundaries, and does its innermost matrix-vector products in FP16.

\begin{table}[H]
\centering
\caption{Convection--diffusion3D, $48^{3}$ ($n=110{,}592$), $w=20$,
$\tau=10^{-10}$. Solve medians [IQR] over 10 runs.}
\label{tab:sparse-nsy}
\scriptsize
\resizebox{\textwidth}{!}{%
\begin{tabular}{@{}lrrlr@{}}
\toprule
Method & Solve (ms) & Outer & matvecs (fp64/fp32/fp16) & Peak (MiB) \\
\midrule
GMRES$_{64}$(Jacobi$_{64}$) & 757.0 [749, 783] & 175 & 364 / -- / -- & 55.7 \\
$\FGMRES_{64}$(Jacobi$_{64}$) & 792.3 [782, 905] & 175 & 358 / -- / -- & 82.5 \\
\textsf{CAMPS-Krylov}$[\FGMRES_{32}(\mathrm{J}_{32})]$ & 1224.9 [1157, 1390] & 2 & 13 / 538 / -- & 100.8 \\
\textsf{CAMPS-Cascade}$[\mathrm{Jacobi}_{16}]$ & \textbf{535.9} [518, 595] & 1 & 8 / 31 / 403 & 90.1 \\
\bottomrule
\end{tabular}}
\end{table}
\noindent
Among the methods of this section, \textsf{CAMPS-Cascade} is the fastest in all four cases. Section~\ref{exp-f3r} tunes a wider set of trees on the same problems, and \textsf{CAMPS-Apex} is faster there. Against the best FP64 baseline, it is $1.48\times$ faster at $w=2$ ($357.9$ against $530.1$\,ms), $1.59\times$ faster at $w=20$ ($264.3$ against $419.3$\,ms), $1.15\times$ faster at $w=100$, and $1.41\times$ faster on the $48^{3}$ grid. It reaches the same tolerance as the baselines every time, and its forward errors are never larger than theirs.

\begin{figure}[H]
\centering
\includegraphics[width=\linewidth]{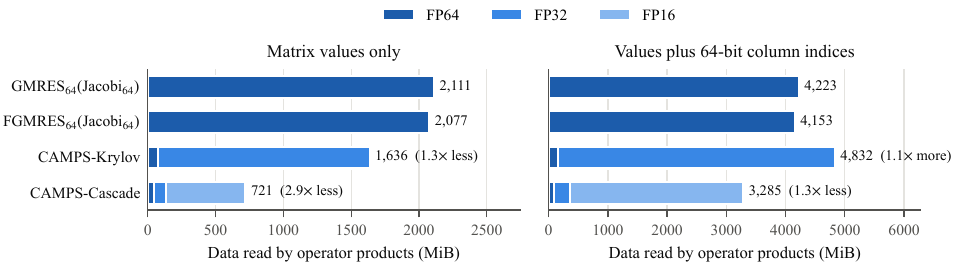}
\caption{Data read by sparse operator products in one solve of convection--diffusion at $48^{3}$ with $w=20$, split by storage precision. Ratios are relative to GMRES$_{64}$. The cascade performs more products than the FP64 baseline, but most of them read 16-bit values. Vector, row-pointer, and Krylov-basis traffic is not included.}
\label{fig:sparse-nsy-bytes}
\end{figure}

The product counts in Table~\ref{tab:sparse-nsy} and the data volumes in
Figure~\ref{fig:sparse-nsy-bytes} explain part of this. At $48^{3}$, the
FP64 baseline performs $364$ FP64 operator products. The cascade performs
$8$ in FP64, $31$ in FP32, and $403$ in FP16. Those are more products in
total, but most of them read 16-bit values, so the cascade reads
$2.9\times$ fewer bytes of matrix values. Every CSR product also reads a
column index for each nonzero, and our code stores these as 64-bit
integers. Counting the indices, each nonzero costs $16$, $12$, or $10$
bytes in FP64, FP32, or FP16, and the saving shrinks to
$364\cdot16/(8\cdot16+31\cdot12+403\cdot10)\approx1.3\times$. With 32-bit
indices it would be about $1.6\times$. Neither count includes the traffic
for vectors, row pointers, or Krylov bases. The measured speedup at
$48^{3}$ is $1.41\times$, which lies between the two byte ratios. The byte
counts and the solve time measure different things, so this does not show
that operator traffic explains the speedup. The FP64 GMRES baseline also
orthogonalizes against a basis that grows through each restart cycle, and
the cascade needs one outer FP64 iteration. We did not profile how the
time divides between operator products and orthogonalization.

Each precision boundary has its own adapter, and every correction was
accepted. The median contraction factor at the outer adapter was
$2.3\times10^{-5}$. Converting the matrix to FP16 does not lose anything, as
its entries are integers that are not larger than $106$ in absolute value, which FP16
stores exactly. The same tree with the same tuning converges for every
wind strength we tested, which was a $50\times$ range of $w$.

The two FP64 baselines take about the same time. FGMRES stores a second set
of basis vectors, so it needs more memory than GMRES ($82.5$ against
$55.7$\,MiB). The cascade's peak memory of $90.1$\,MiB is higher than both,
because it stores the matrix in three formats and keeps a Krylov basis at
each of its three levels.

\textsf{CAMPS-Krylov}, with a single adapter, is slower than both
baselines. Its inner FP32 solver needs $538$ FP32 products. The cascade
moves most of that work down to FP16, where each product is cheaper. On this
class, adding the second adapter and the FP16 level is what makes CAMPS
faster than the baselines. Both trees are built from the same parts, so the
framework is able to measure which one better suits the problem in question.

\begin{figure}[H]
\centering
\includegraphics[width=0.8\linewidth]{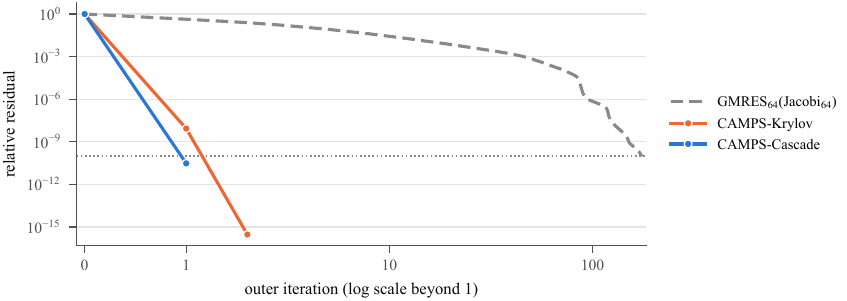}
\caption{Relative residual against outer iterations on convection--diffusion at $48^{3}$ with $w=20$. FGMRES$_{64}$ follows the same curve as GMRES$_{64}$ and is not drawn. The horizontal axis is linear up to one iteration and logarithmic beyond it.}
\label{fig:conv-sparse-nsy}
\end{figure}

Figure~\ref{fig:conv-sparse-nsy} shows the residual against outer iterations at
$48^{3}$. GMRES$_{64}$ needs $175$ iterations. \textsf{CAMPS-Krylov} needs two
outer iterations and the cascade needs one, because each of their outer
iterations contains a full inner solve in lower precision.

\subsection{Placing adapters at different edges}\label{exp-placement}
Theorem~\ref{thm:closure} says adapters can be placed at any precision
boundary in a solver tree and at more than one boundary in the same tree. To test this, we keep the same kind of child and move it to different places in the tree. The child is an approximate inverse of the matrix, computed in FP64 and then stored and
applied in FP32, FP16, or BF16. The test problems are dense, with $n=2048$
and $\kappa=10^{2}$, $10^{4}$, and $10^{6}$.

\begin{table}[H]
\centering
\caption{Placement ablation, $n=2048$, $\kappa=10^{4}$: the same INV child at
different edge subsets. Solve medians over 15 runs.}
\label{tab:placement}
\scriptsize
\resizebox{\textwidth}{!}{%
\begin{tabular}{@{}lrrl@{}}
\toprule
Instance (edge set) & Solve (ms) & Outer & Status \\
\midrule
$\FGMRES_{64}[\IR{64}{32}[\mathrm{INV}_{32}]]$ & 7.9 & 1 & converged \\
$\FGMRES_{64}[\IR{64}{16}[\mathrm{INV}_{16}]]$ (single deep edge) & 31.7 & 4 & converged \\
$\FGMRES_{64}[\IR{64}{\mathrm{b}16}[\mathrm{INV}_{\mathrm{b}16}]]$ & 582.0 & 60 & converged \\
$\FGMRES_{64}[\IR{64}{32}[\IR{32}{16}[\mathrm{INV}_{16}]]]$ (nested, no inner Krylov) & 42.5 & 3 & converged \\
$\FGMRES_{64}[\IR{64}{32}[\FGMRES_{32}[\IR{32}{16}[\mathrm{INV}_{16}]]]]$ (\textsf{Cascade}) & 41.7 & 1 & converged \\
$\FGMRES_{64}[\IR{64}{32}[\FGMRES_{32}[\IR{32}{\mathrm{b}16}[\mathrm{INV}_{\mathrm{b}16}]]]]$ & 857.6 & 3 & converged \\
\bottomrule
\end{tabular}}
\end{table}
\noindent
Every placement converges for $\kappa\le10^{4}$. That includes a single
deep edge from FP64 straight to FP16, two nested adapters with no Krylov
solver between them, and the full cascade with an adapter at both
boundaries. Nesting adapters in any of these ways gives a working solver,
as the closure part of Theorem~\ref{thm:closure} says.

Table~\ref{tab:placement} shows which placements are the best for each child. The FP32
inverse on a single edge is the fastest, at $7.9$\,ms with a single outer
iteration. FP16, with unit roundoff $u=2^{-11}$, also works on a direct
edge from FP64 and needs $4$ outer iterations. BF16, with $u=2^{-8}$,
struggles and needs $60$. Inserting an FP32 level between FP64 and
BF16 cuts the outer iterations to $3$ and the FP64 products from $365$ to
$18$, which is a $20\times$ drop in FP64 work. It does not save time though: the
new FP32 level performs $623$ products, and the total rises from $582.0$ to
$857.6$\,ms. Here, the new level adds more work than it removes. For FP16,
adding an inner FGMRES solver to the nested tree lowers the outer count
from three to one, but the time barely changes ($41.7$ against
$42.5$\,ms) because the inner Krylov work costs about as much as the outer
iterations it saves.

At $\kappa=10^{6}$, every FP16 and BF16 child fails. The zero guard
replaces each of their corrections with an identity direction, and the
solvers run until they hit their iteration limit. The FP32 inverse still
converges, and it does so in $15.2$\,ms. The FP16 child works at $\kappa=10^{4}$, where
$\kappa u_{16}\approx5$, but not at $10^{6}$, where
$\kappa u_{16}\approx490$. At $10^{6}$, the FP32 child has
$\kappa u_{32}\approx0.06$. Whether a child works or not depends on $\kappa$ times its own unit roundoff, just as we saw for the LU children in Section~\ref{exp-dense-ic}. The theorem says the same thing, since its bound for each edge depends only on the
child at that edge.

\begin{figure}[H]
\centering
\includegraphics[width=0.9\linewidth]{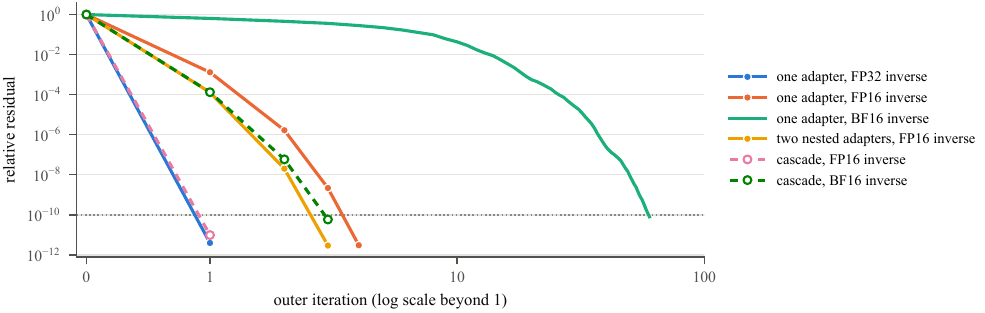}
\caption{Relative residual against outer iterations for the six placements at $\kappa=10^{4}$, $n=2048$. With one adapter, the FP32 inverse converges in one outer iteration, the FP16 inverse in four, and the BF16 inverse in $60$. An FP32 level above the BF16 inverse reduces its count to three.}
\label{fig:conv-placement}
\end{figure}

Figure~\ref{fig:conv-placement} shows the residual histories at
$\kappa=10^{4}$.

\subsection{Comparison with F3R}\label{exp-f3r}
F3R~\cite{suzuki2025nested} is the closest prior method to CAMPS. It nests three
FGMRES solvers, in FP64, FP32, and FP16, and ends in a fixed number of
Richardson sweeps. Each inner solver runs for a fixed number of iterations,
the precision boundaries are plain casts, and only the outermost solver
checks for convergence. The authors' code is
public\footnote{\url{https://github.com/suzuki-hpc/F3R}}, but it runs
hand-optimized C++ and CUDA kernels, so its timings would mostly measure
kernel speed. We instead rebuilt the F3R algorithm structure in our own code, using
the same kernels, matrices, and stopping rule as the CAMPS solvers. One
difference matters. The authors' released solver uses a block ILU(0)
preconditioner inside the Richardson sweeps, and we use Jacobi, which is the
preconditioner shared by every method in this section.

We compare F3R with three CAMPS trees. The first is the sparse
\textsf{CAMPS-Cascade}. The second is \textsf{CAMPS-Apex}
(Algorithm~\ref{alg:apex}), which keeps F3R's FP32 and FP16 levels but
replaces its outer FGMRES with an adapter. The third is a hybrid with an
adapter at the outer boundary and a cast at the inner one. All four trees go through the same tuning procedure. For each matrix class, we run a grid search over the inner iteration budgets on the $32^{3}$ problem, keep the fastest setting that converges, and reuse it at $48^{3}$. F3R gets the largest grid, with $12$ settings.

\begin{figure}[H]
\centering
\includegraphics[width=\linewidth]{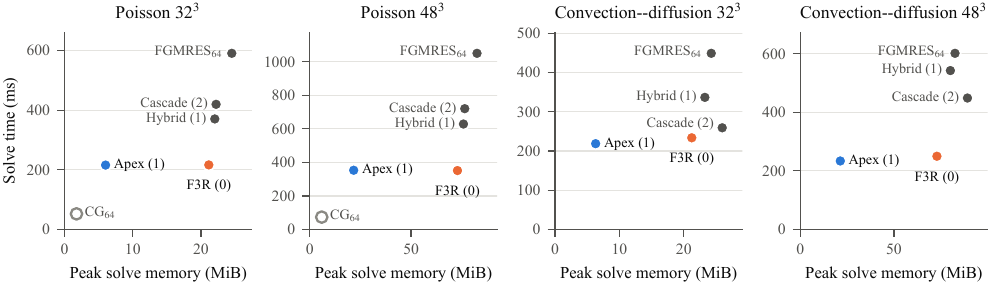}
\caption{Solve time against measured peak solve memory for the methods of Table~\ref{tab:f3r}. The number after F3R and each CAMPS tree counts its adapter edges. \textsf{CAMPS-Apex} matches our F3R implementation in time with about $3.4\times$ less memory. CG$_{64}$ is shown on Poisson for reference.}
\label{fig:f3r}
\end{figure}

\begin{table}[H]
\centering
\caption{Comparison with our F3R implementation. Solve times are medians over 10 runs in ms,
with $\tau=10^{-10}$ and Jacobi preconditioning throughout. The four nested
solvers are tuned the same way. ``Ad.'' is the number of adapter edges.
``Peak'' is the measured allocator peak during one $48^{3}$ solve, for
Poisson and convection--diffusion. For reference, CG$_{64}$ takes
$52.3$\,ms at $32^{3}$ and $73.6$\,ms at $48^{3}$ on Poisson.}
\label{tab:f3r}
\scriptsize
\resizebox{\textwidth}{!}{%
\begin{tabular}{@{}lcrrrrrr@{}}
\toprule
 & & \multicolumn{2}{c}{Poisson} & \multicolumn{2}{c}{conv--diff} & \multicolumn{2}{c}{Peak (MiB)} \\
\cmidrule(lr){3-4}\cmidrule(lr){5-6}\cmidrule(l){7-8}
Method & Ad. & $32^{3}$ & $48^{3}$ & $32^{3}$ & $48^{3}$ & Poisson & conv--diff \\
\midrule
$\FGMRES_{64}$(Jacobi$_{64}$) & -- & 590.5 & 1050.2 & 449.1 & 601.2 & 82.5 & 82.5 \\
F3R (reimplementation) & 0 & 216.4 & \textbf{351.2} & 233.7 & 249.9 & 72.9 & 72.9 \\
\textsf{CAMPS-Apex} & 1 & \textbf{215.8} & 353.0 & \textbf{218.7} & \textbf{233.7} & \textbf{21.7} & \textbf{21.4} \\
Hybrid (adapter top, cast inside) & 1 & 370.5 & 628.5 & 336.8 & 542.3 & 75.9 & 80.1 \\
\textsf{CAMPS-Cascade} & 2 & 419.5 & 720.1 & 259.2 & 448.6 & 76.5 & 89.2 \\
\bottomrule
\end{tabular}}
\end{table}
\noindent
We draw four conclusions from Table~\ref{tab:f3r} and
Figure~\ref{fig:f3r}.

\paragraph{The F3R algorithm works well in our implementation.} Our reimplementation is $1.9$-$3.0\times$
faster than the FP64 flexible baseline and $1.1$-$2.1\times$ faster than
\textsf{CAMPS-Cascade}, which checks both of its edges.

\paragraph{Checking every edge costs extra work.} On Poisson at $32^{3}$,
F3R performs $5$ FP64, $26$ FP32, and $360$ FP16 products, while the
cascade performs $19$, $148$, and $434$. The cascade does this extra work for two reasons. First, each adapter recomputes the residual in the parent precision before it accepts a correction. Second, the cascade's inner solvers keep iterating until they
reach a tolerance, while F3R's stop after a fixed, small number of
iterations. Our measurements do not show how much each reason contributes. The hybrid, which has a single adapter, is also much slower than \textsf{CAMPS-Apex}, which has one too. The two differ in three
ways. The hybrid keeps F3R's outer FP64 FGMRES, lets its FP32
solver run to a tolerance, and preconditions its FP16 solver with Jacobi
instead of Richardson sweeps. On Poisson at $32^{3}$ it performs $13$
FP64, $47$ FP32, and $455$ FP16 products, against $4$, $26$, and $360$ for
\textsf{CAMPS-Apex}.

\paragraph{\textsf{CAMPS-Apex} is as fast as our F3R implementation and uses less memory.}
\textsf{CAMPS-Apex} keeps F3R's FP32 and FP16 levels but replaces its outer
FP64 FGMRES with a plain CAMPS adapter. Each correction then costs one FP64
residual, a power-of-two scaling, a cast, and an acceptance check, with no
FP64 Arnoldi process, orthogonalization, or stored basis. On Poisson, the
two are within $1\%$ of each other, and their interquartile ranges overlap. On
convection-diffusion, \textsf{CAMPS-Apex} takes about $6\%$ less time,
and at $32^{3}$ the interquartile ranges do not overlap. It also uses about
$3.4\times$ less peak memory in every case, because it stores no FP64
Krylov basis. And unlike F3R, it checks every correction at its top edge,
so it measures the contraction, detects stagnation, and reports why a
correction was rejected.

On Poisson, \textsf{CAMPS-Direct}$[\mathrm{CG}_{32}]$ from
Section~\ref{exp-sparse-spd} is $2.1\times$ faster than F3R at $48^{3}$
($169.0$ against $351.2$\,ms) because its CG child takes advantage of the
symmetric matrix and F3R's general nesting does not. CG$_{64}$ is still
faster than every mixed-precision tree we tested on Poisson.

\paragraph{Every tree here depends on its inner settings.} F3R's inner
budgets matter a lot. In our tuning grid, doubling its FP16 budget
from $5$ to $10$ iterations made it $2.1$-$2.4\times$ slower, and the
worst setting was up to $7.6\times$ slower than the best setting. Every setting
still converged, so nothing in the solver signals a poor choice.
\textsf{CAMPS-Apex} keeps F3R's fixed budgets and is just as sensitive.
On Poisson, its worst setting was $3.5\times$ slower than its best. On the other hand,
\textsf{CAMPS-Cascade} has no fixed budgets. Its levels stop on their own
tolerances, and its default settings converged on every problem in
Sections~\ref{exp-sparse-spd} and~\ref{exp-sparse-nsy}. However, converging is not
the same as being fast: on Poisson at $32^{3}$, the cascade took
$923$\,ms with its default settings during tuning versus $419$\,ms with
its best setting.

A cast is a legal part of a CAMPS tree that skips the residual check
(Section~\ref{grammar}). Thus, F3R, the hybrid, \textsf{CAMPS-Apex}, and
\textsf{CAMPS-Cascade} are all trees in one design space, and a
user pays for checks only at the edges they choose. On these problems, the
best tree we found checks only the top edge.

\begin{figure}[H]
\centering
\includegraphics[width=\linewidth]{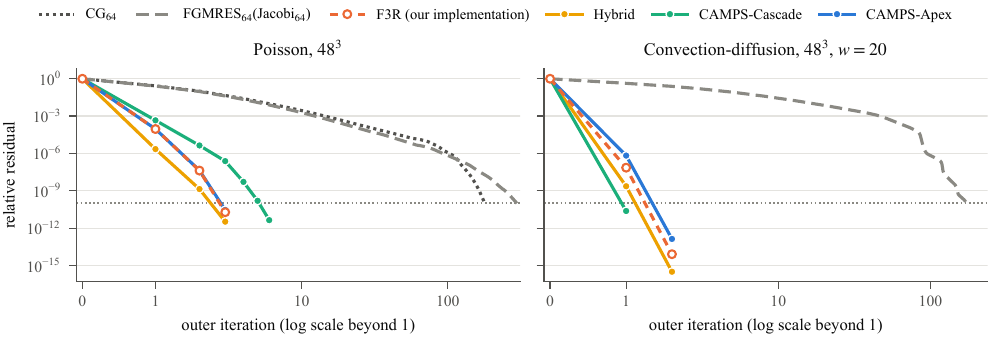}
\caption{Relative residual against outer iterations at $48^{3}$ for the methods of Table~\ref{tab:f3r}. On Poisson, F3R and \textsf{CAMPS-Apex} follow almost the same curve, and both need three outer iterations. On convection--diffusion both need two. The FP64 baselines need between $173$ and $297$ iterations.}
\label{fig:conv-f3r}
\end{figure}

Figure~\ref{fig:conv-f3r} shows the residual against outer iterations at
$48^{3}$.

\subsection{Checking the recursive bounds}\label{exp-verify}
The bounds of Theorem~\ref{thm:recursive} and Corollary~\ref{cor:cascade}
have three features for a chain of adapters. None of the experiments above
tests them directly.
\begin{enumerate}
\item The bound on the top-edge factor of a cascade falls like the lower
  edge's factor raised to the power $J_1$ until it levels off at a plateau.
\item The order of that plateau contains the middle precision. It does not
  contain the leaf solver or the lowest precision.
\item The order of the smallest backward error the root can reach contains
  the root's own precision.
\end{enumerate}
The bounds are inequalities, so they do not say that measured runs behave
the same way. We test whether they do on a chain made of adapters alone,
$\IR{64}{32}\bigl[\IR{32}{16}^{\,J_1}[B_{16}]\bigr]$. Because the chain
contains no Krylov solver, rule \textup{(R4)} and its exact-arithmetic assumptions are not needed. We also use the CAMPS framework's adapter code without changes.
The matrix comes from the dense class of Section~\ref{exp-dense-wc}, at
$n=512$ with $\kappa=10$, $10^{2}$, and $10^{3}$. The leaf $B_{16}$ is an
approximate inverse stored and applied in FP16. We use it either as
computed or deliberately weakened by scaling it by $1/2$, which makes its
contraction factor $\theta_{16}\approx0.5$.

We use two right-hand sides. The first is the manufactured $b$, whose
amplification $\mathrm{amp}(b)=\norm{A}_2\norm{A^{-1}b}_2/\norm{b}_2$ is
between $2.3$ and $4.1$. The second is the left singular vector for the
smallest singular value, which gives the worst case,
$\mathrm{amp}=\kappa$. Every quantity in the bounds is computed in FP64,
and no constant is fitted to the data. The leaf factor $\theta_{16}$ is
$\norm{I-A_{16}B_{16}}_2$ plus the measured rounding error of the FP16
product, and $\delta_{pq}$, $\kappa_p$, and $\kappa'_p$ come from singular
values. The inner adapters run with $\tau=0$, and the root runs until it
cannot make more progress, giving $57$ configurations in total. The constants in the bounds are computed in FP64 for these matrices and
right-hand sides. The comparison is an empirical consistency check, not a proof that the hypotheses hold for every input.

\begin{table}[!htb]
\centering
\caption{Recursive bounds against measurements on
$\IR{64}{32}[\IR{32}{16}^{\,J_1}[B_{16}]]$, $n=512$. ``Lower edge'' is the
worst pre-floor per-correction factor at the $32\leftarrow16$ edge and the
bound $\hat\eta_1$ of rule \textup{(R3)}, computed with the measured $\theta_{16}$. ``Top edge'' is the measured
first-correction factor $\norm{r_1}_2/\norm{r_0}_2$ at the $64\leftarrow32$
edge. The last columns divide the plateau by $u_{32}\,\mathrm{amp}(b)$ for
the manufactured and for the adversarial right-hand side.}
\label{tab:verify}
\scriptsize
\setlength{\tabcolsep}{3pt}
\resizebox{\textwidth}{!}{%
\begin{tabular}{@{}llllllllllll@{}}
\toprule
 & & \multicolumn{2}{c}{Lower edge} & \multicolumn{4}{c}{Top edge, leaf as computed} & & \multicolumn{2}{c}{Plateau$/(u_{32}\,\mathrm{amp})$} \\
\cmidrule(lr){3-4}\cmidrule(lr){5-8}\cmidrule(l){10-11}
$\kappa$ & $\theta_{16}$ & measured & bound & $J_1{=}1$ & $J_1{=}2$ & $J_1{=}3$ & $J_1{=}8$ & bound, $J_1{=}8$ & generic & adversarial \\
\midrule
$10^{1}$ & $2.0\times10^{-3}$ & $1.0\times10^{-3}$ & $7.0\times10^{-3}$ & $5.3\times10^{-4}$ & $3.2\times10^{-7}$ & $1.3\times10^{-7}$ & $1.3\times10^{-7}$ & $7.9\times10^{-3}$ & 0.93 & 0.75 \\
$10^{2}$ & $1.3\times10^{-2}$ & $7.3\times10^{-3}$ & $4.6\times10^{-2}$ & $2.3\times10^{-3}$ & $5.1\times10^{-6}$ & $1.3\times10^{-7}$ & $1.3\times10^{-7}$ & $6.1\times10^{-2}$ & 0.68 & 0.55 \\
$10^{3}$ & $8.8\times10^{-2}$ & $5.3\times10^{-2}$ & $3.9\times10^{-1}$ & $1.5\times10^{-2}$ & $2.2\times10^{-4}$ & $3.2\times10^{-6}$ & $1.2\times10^{-7}$ & vacuous & 0.51 & 0.44 \\
\midrule
 & & \multicolumn{2}{c}{Lower edge} & \multicolumn{4}{c}{Top edge, leaf degraded to $\theta_{16}\approx0.5$} & & \multicolumn{2}{c}{Plateau$/(u_{32}\,\mathrm{amp})$} \\
\cmidrule(lr){3-4}\cmidrule(lr){5-8}\cmidrule(l){10-11}
$\kappa$ & $\theta_{16}$ & measured & bound & $J_1{=}1$ & $J_1{=}2$ & $J_1{=}4$ & $J_1{=}8$ & plateau, $J_1{=}32$ & generic & \\
\midrule
$10^{1}$ & 0.501 & 0.500 & 0.508 & $5.00\times10^{-1}$ & $2.50\times10^{-1}$ & $6.25\times10^{-2}$ & $3.91\times10^{-3}$ & $8.9\times10^{-8}$ & 0.64 & \\
$10^{2}$ & 0.504 & 0.500 & 0.554 & $5.00\times10^{-1}$ & $2.50\times10^{-1}$ & $6.25\times10^{-2}$ & $3.90\times10^{-3}$ & $1.0\times10^{-7}$ & 0.56 & \\
$10^{3}$ & 0.530 & 0.501 & 0.962 & $5.00\times10^{-1}$ & $2.50\times10^{-1}$ & $6.25\times10^{-2}$ & $3.91\times10^{-3}$ & $9.3\times10^{-8}$ & 0.45 & \\
\bottomrule
\end{tabular}}
\end{table}
\paragraph{The bounds hold, but they are loose where rounding dominates.}
In every case where the bound gives a usable number, the measured factor
stays below it. At the lower edge, the bound comes very close to the
measurement when most of the error comes from the leaf. With the weakened leaf at $\kappa=10$, for example, we measure $0.500$ and the bound gives $0.508$. When most of the error comes from rounding, the bound is about ten times too high, and at $\kappa=10^{3}$ it stops being useful for the top edge at all. The FP32 adapter's constant comes out larger than one, so the bound allows no reduction, even though the measured factor is $1.2\times10^{-7}$. The problem is the worst-case rounding constants
$\gamma^{(32)}_{n+1}\kappa'_{32}$, which overestimate the real error by
orders of magnitude, which is what worst-case rounding bounds usually do~\cite{higham2019probabilistic}. Thus, the theorem's numbers are
pessimistic, but its predictions about how the factors behave can still be
right. We test those predictions in the next three paragraphs.

\begin{figure}[H]
\centering
\includegraphics[width=\linewidth]{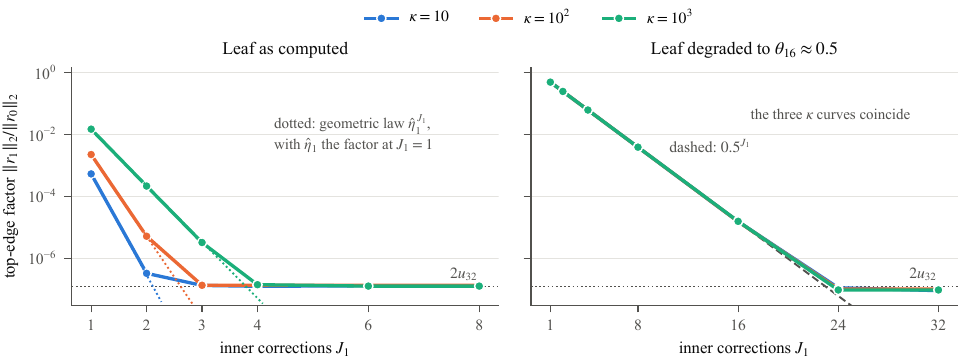}
\caption{Top-edge factor $\norm{r_1}_2/\norm{r_0}_2$ against the inner correction count $J_1$ on the chain $\IR{64}{32}[\IR{32}{16}^{\,J_1}[B_{16}]]$ with the manufactured right-hand side. The factor falls geometrically until it reaches a plateau near $2u_{32}$ that FP32 sets.}
\label{fig:verify}
\end{figure}

\paragraph{The top edge follows the geometric law.}
Figure~\ref{fig:verify} plots the top-edge factor against $J_1$. With the
weakened leaf, the factor for the first correction at the top edge is
$0.5^{J_1}$ to within $0.2\%$ for every $J_1\le8$ and every $\kappa$.
The weakened leaf has $I-A_{16}B_{16}\approx I/2$, so each inner correction
halves the residual, and $J_1$ corrections leave $0.5^{J_1}$ of it.
With the leaf as computed, it follows $\hat\eta_1^{\,J_1}$ closely until it
gets near the plateau. At $\kappa=10^{2}$, for example, it drops from
$2.3\times10^{-3}$ at $J_1=1$ to $5.1\times10^{-6}$ at $J_1=2$, which is
the square of the first value.

\paragraph{The plateau follows the middle precision.} With the leaf as
computed, the factor stops falling at about $1.3\times10^{-7}$, close to
$2u_{32}$, for every $\kappa$. The weakened leaf levels off at almost the
same value by $J_1=32$, between $8.9\times10^{-8}$ and $1.0\times10^{-7}$.
In these runs, neither FP16 nor the quality
of the leaf changes where the factor levels off. This agrees with the order of the bound in
Corollary~\ref{cor:cascade}, which contains the middle precision. Once the factor has leveled off, dividing it by
$u_{32}\,\mathrm{amp}(b)$ gives a value between $0.44$ and $0.96$ in every
case, even though the amplification itself ranges from $2.3$ to $1000$.

With the adversarial right-hand side, the plateau is $0.44$ to $0.75$
times $u_{32}\kappa$. In that case the floor $f(g)$ in
\eqref{eq:fp-recurrence} reaches its worst case, which grows with
$\kappa$. With an ordinary right-hand side it does not, which is why the
$\kappa$ in the bound does not show up in typical runs. The measured plateaus have the form that $f(g)$ predicts, but with a constant below one in place of the bound's $(n+1)\mu_p$.

\paragraph{The floor is set by the root.} In all $57$ configurations, the
root's final normwise backward error $\beta_2$ is between $0.43\,u_{64}$
and $0.88\,u_{64}$, no matter what the $\kappa$, leaf, $J_1$, or right-hand side is.
This is consistent with Theorem~\ref{thm:recursive}\,(iii). Since the root
keeps going until it cannot make more progress, every root run ended
with a rejected correction, and each rejection came at that floor, as
Proposition~\ref{prop:exits} requires.

\subsection{Summary}\label{exp-summary}
\begin{itemize}
\item \textbf{Dense, well-conditioned.} \textsf{CAMPS-Direct} is
  $9.6\times$ faster end to end than the FP64 direct solve at $n=8192$. It
  stays ahead at every tested block size up to $k=16$, and the
  extrapolated crossover is at $k\approx100$ or later.
  \textsf{CAMPS-Cascade} also converges, and it does so in a single outer iteration.
\item \textbf{Dense, ill-conditioned.} From $\kappa=10^{4}$ to $10^{7}$,
  the measured contraction of the FP32 child follows $0.6\,\kappa u_{32}$ within a factor of about two. Once
  the child stops reducing the residual, the adapter rejects it at once and
  reports why. The fallback converges at every $\kappa$ up to $10^{12}$
  with about $7$\,ms of solve time, and the ladder also converges
  everywhere. Damping lets every correction pass without making progress,
  so it needs a progress threshold to keep the fallback working.
\item \textbf{Sparse SPD.} Every instance converges, with forward errors
  $30$ to $5\times10^{6}$ times smaller than those of CG$_{64}$. However, plain
  CG$_{64}$ is still faster than every tested instance. This is because on this problem,
  FP32 barely saves any time per iteration, and the last correction runs past
  the tolerance. We recommend CG$_{64}$ for this class.
\item \textbf{Sparse nonsymmetric.} \textsf{CAMPS-Cascade}, with adapters
  at both precision boundaries and FP16 products at the leaf, is the
  fastest method of Section~\ref{exp-sparse-nsy} in all four cases. It is $1.15$-$1.59\times$ faster than
  the best FP64 baseline and reads $2.9\times$ fewer bytes of matrix
  values, or $1.3\times$ fewer when the 64-bit indices are counted. The
  same settings converge over a $50\times$ range of wind strengths.
\item \textbf{Placement.} Every tested placement converges when the
  children are accurate enough. Adding an FP32 level above the BF16 child
  cuts its FP64 work by $20\times$ but makes the whole solve slower.
  Whether a child works depends on $\kappa$ times its own unit roundoff,
  which is what the per-edge bound predicts.
\item \textbf{F3R.} Our reimplementation confirms that the F3R algorithm works well.
  \textsf{CAMPS-Apex} is within $1\%$ of it on Poisson and about $6\%$ faster on
  convection-diffusion, with about
  $3.4\times$ less peak memory, and \textsf{CAMPS-Direct}$[\mathrm{CG}_{32}]$
  beats it by $2.1\times$ on Poisson. \textsf{CAMPS-Cascade} is
  $1.1$-$2.1\times$ slower than F3R. It converges on every problem with
  default settings, but like every tree here, it is faster when tuned.
\item \textbf{Recursive bounds.} On a chain of FP64, FP32, and FP16
  adapters, the measured factors never exceed the bounds wherever the
  bounds give a usable number. Each extra inner correction multiplies the
  top-edge factor by the lower edge's factor, until it levels off at
  $0.44$-$0.96\,u_{32}\,\mathrm{amp}(b)$. Neither the leaf nor FP16
  affects where it levels off. In all $57$ configurations, the root's final
  backward error is between $0.43\,u_{64}$ and $0.88\,u_{64}$.
\end{itemize}

\noindent
These results support the claim we stated at the start of this
section. No single tree wins everywhere, and on SPD Poisson, none of ours
wins at all. However, for every class, our CAMPS grammar contains a tree that
either beats the FP64 baseline or shows clearly why it does not. The
measured contraction tells us in advance which children will converge and
which will not. The same interface accepted direct, approximate-inverse,
stationary, symmetric Krylov, nonsymmetric Krylov, and adapter children
without any changes. Every placement we tested gave a working solver, as
Theorem~\ref{thm:closure} predicts, and on the dense chain where we
checked them, the bounds of Theorem~\ref{thm:recursive} held.

\hypertarget{discussion}{%
\section{Conclusion \& Future Extensions}\label{discussion}}
This paper presented CAMPS, an interface for composing mixed-precision
correction solvers. An adapter computes residuals and updates in the
precision of its parent, and it checks every correction before it keeps it.
We showed that adapters can be placed at any compatible set of edges in a
solver tree. We gave a backward-error bound for every output that the root reports as
successful, and
sufficient conditions for convergence, with constants computed from the
leaves of the tree to its root. We tested five instances on four matrix
classes. The rest of this section discusses what the experiments show,
where the method falls short, and what remains to be done.

\paragraph{Where adapters help.} An adapter moves part of the computation to a lower precision. It does not make the correction solve easier because the child still has to reduce the residual. An adapter helps when the FP64 work it replaces costs much more than the extra work it adds, and our experiments show three cases of this. In the dense experiments, the adapter replaces an FP64 factorization with an FP32 factorization that is $5$--$13\times$ faster (Section~\ref{exp-dense-wc}). In the sparse nonsymmetric experiments, the cascade replaces most of the FP64 GMRES iterations and the orthogonalization that comes with them, with cheaper
iterations in FP32 and FP16 (Section~\ref{exp-sparse-nsy}). In \textsf{CAMPS-Apex}, the adapter replaces the outer FP64 FGMRES, so no FP64 Krylov basis is stored, and the peak memory is about $3.4\times$ lower than that of our F3R implementation (Section~\ref{exp-f3r}).

Adapters did not help on SPD Poisson, where no CAMPS instance was faster
than plain CG$_{64}$ (Section~\ref{exp-sparse-spd}). On that problem, FP32
saves almost no time in each CG iteration. The last correction also runs
until it reaches its own tolerance, well past the point where its inner
residual falls below the parent's tolerance. Restarting CG from zero at
each correction adds few iterations. A child that stops at the parent's
tolerance could save part of this work, but we have not tested one.

Every adapter also adds work in the parent precision. Each correction
needs one residual evaluation and two conversions. The cascade, which
checks both of its edges, is $1.1$--$2.1\times$ slower than our F3R
implementation (Section~\ref{exp-f3r}). It is slower for two reasons. Its
adapters perform these checks, and its inner solvers run until they reach
a tolerance instead of stopping after a fixed number of iterations.
Adapters can also use more memory than expected. When a child keeps its
own copy of the matrix to measure its residual, that copy can cancel the
storage savings of the lower precision (Section~\ref{exp-dense-wc}).

With many right-hand sides, the advantage of a cheaper setup becomes
smaller, because the FP64 factorization is shared across more columns.
\textsf{CAMPS-Direct} is still the fastest method at every block size we
tested, up to $k=16$. Extrapolating our measurements puts the crossover at
$k\approx100$ or later (Table~\ref{tab:multirhs}).

\paragraph{Outer refinement versus outer FGMRES.} An outer refinement loop
applies corrections one after another and converges when each one
contracts. An outer FGMRES uses the corrections as search directions, so
it can tolerate corrections of uneven quality, but it pays for basis
storage and orthogonalization. In our tests, FGMRES reduced the number of
outer iterations but never the time. On the ill-conditioned sweep at
$\kappa=10^{4}$, $10^{6}$, and $10^{7}$, \textsf{CAMPS-Krylov} needs $1$,
$2$, and $5$ outer iterations, against $3$, $5$, and $13$ corrections for
\textsf{CAMPS-Direct}. Its solves still take longer, $8.0$, $15.2$, and
$38.2$\,ms against $6.5$, $10.4$, and $26.0$\,ms. \textsf{CAMPS-Direct} is
also faster on dense well-conditioned systems, on SPD systems, and on
blocks of right-hand sides. In the F3R comparison, \textsf{CAMPS-Apex},
with an adapter on top, is as fast as our F3R implementation, which has
FGMRES on top. Once the child stops contracting, with $\hat\eta>1$,
neither outer loop converges, and a stronger child is needed, as in the
fallback and ladder instances. Flexibility also has a small cost when the
preconditioner is fixed. In the sparse nonsymmetric runs, GMRES was about
$4$--$5\%$ faster than FGMRES with the same Jacobi preconditioner and used
less memory.

\paragraph{What the acceptance test does and does not provide.} The test
measures progress. Every instance we report records $\hat\eta$ for each
correction, and on the ill-conditioned sweep $\hat\eta$ follows
$0.6\,\kappa u_{32}$ from $\kappa=10^{4}$ to $10^{7}$. The test also
detects failure early. $\hat\eta$ crosses one where convergence stops, and
the test replaces seconds of iterating without progress with a rejection
that arrives within milliseconds.

The test has limits. Beyond $\kappa u_{32}\approx1$, the classical bound no
longer implies contraction for the FP32 child (Section~\ref{contract}). In
our runs, the child's first correction is accepted and its second is
rejected. With damping, its corrections pass the test without making
progress. A rejection does not show that contraction is impossible, and
the FP64 fallback child converges at every tested $\kappa$. A rejection
also does not say which hypothesis failed. The cause can be an inaccurate
child, a lossy conversion, or cancellation in the residual. In the other
direction, a few accepted corrections do not prove a contraction bound for
all inputs. Checking each correction separately also does not control the
quality of the basis that an outer Krylov method builds from those
corrections. We did not measure the conditioning of that basis or its
loss of orthogonality, and we make no claim about either.

Theorems~\ref{thm:stability} and~\ref{thm:recursive} extend the per-edge
results to whole trees. Every output that the root reports as successful
satisfies a backward-error
bound set by the stopping tolerance and the root precision, with no
assumptions on the subtree. Convergence follows from one sufficient
condition per edge. Section~\ref{guarantees} lists the theoretical
questions that remain open.

\paragraph{Limitations and open experiments.} Several experiments are
still missing.
\begin{itemize}
\item \textbf{A control for the nested trees.} We did not run the
  cascade's tree entirely in FP64. That control would show how much of its
  speedup on nonsymmetric problems comes from lower precision and how much
  comes from nesting.
\item \textbf{F3R.} Our implementation uses Jacobi where the authors'
  code uses block ILU(0). We have not run the authors' public code on our
  matrices to verify our implementation, and we have not compared at their
  problem scales.
\item \textbf{Children.} We did not test a child that stops at the
  parent's tolerance, a child that keeps its Krylov space between
  corrections, or a short-budget CG child on which the residual and energy
  tests disagree.
\item \textbf{The energy test.} Its error estimate is not a proven
  enclosure (Section~\ref{contract}). A rigorous version needs verified
  bounds on each rounding error and new timings.
\item \textbf{Baselines.} The missing baselines include optimized
  algebraic multigrid, incomplete factorizations, recycled Krylov methods,
  FGMRES with longer restarts, and block FGMRES beyond $16$ right-hand
  sides.
\item \textbf{Problems and hardware.} Our sparse matrices are stencils
  whose entries FP16 stores without error. A broader study would use
  SuiteSparse matrices with irregular sparsity, anisotropy, and
  nonnormality~\cite{davis2011suitesparse,greenbaum1996gmres}, larger
  sparse problems, several architectures with different FP64 speeds, and
  performance profiles across matrix families~\cite{dolan2002profiles}.
\item \textbf{Protocol.} The diagnostic products should be separated from
  the timed work, the CSR indices could be stored in 32 bits, and timings
  should be reported as distributions over independent matrices.
\end{itemize}

\paragraph{Choosing a tree automatically.} We chose every tree in this
paper by hand, guided by measurements. Choosing the tree from features of
the problem and the hardware, and choosing the precision, budget, damping,
and fallback during a solve, is a solver-selection problem in the sense of
Rice~\cite{rice1976}. Learned comparators~\cite{jessup2016lighthouse} and
bandit-based precision selection~\cite{carson2026autotuning} are the
natural baselines. We plan to study a reinforcement-learning policy for
these choices. The acceptance test would check every step that such a
policy takes, independently of how the policy chose it.

\paragraph{Acknowledgements}
This research was supported in part by grants and gifts from the Peter O'Donnell Foundation, The Michael J. Fox Foundation, the Jim Holland-Backcountry Foundation and the Michael-Connie Rasor Foundation towards curing Parkinson's disease. Computational analyses used resources of the Texas Advanced Computing Center (TACC) at The University of Texas at Austin.

\paragraph{LLM usage disclaimer}
We used large language models to improve the wording and grammar of this
manuscript. AI tools also assisted in preparing some of the code, figures,
and table formatting. The main technical ideas, the theoretical results, and
the design of the experiments come from the authors.

\bibliographystyle{plainnat}
\bibliography{reference}
\appendix
\providecommand{\figplaceholder}[1]{%
  \fbox{\parbox{0.92\linewidth}{\centering\vspace{1.5em}%
  \textit{Figure placeholder:} #1\vspace{1.5em}}}}
\section{Supplementary Details}\label{supp}
\subsection{Comparison with prior designs}\label{app:positioning}
\begin{table}[H]
\centering
\caption{CAMPS compared with the closest prior designs. Rows 1 to 4 are
designs used mainly for dense systems, rows 5 to 9 are designs for sparse
systems, and rows 10 to 13 apply to both. The columns are as follows.
\emph{Precision boundary} says where the precision changes: ``implicit''
means inside one refinement loop, with no separate object for the boundary,
and ``per level'' means each level of the method has its own precision.
\emph{Levels} is the number of solver nodes on the path from the root to the
leaf. A refinement loop and the solve it wraps count as two nodes, and a
Krylov method with its fixed preconditioner counts as one. For multigrid it
is the number of grid levels $L$. \emph{Correction solver} is the solver that
computes each correction, marked as direct or iterative. \emph{Check} is how
a correction is judged. ``A priori'' means convergence is guaranteed in
advance by a bound and nothing is tested during the solve. The $\kappa u_f$
bound is the classical condition $\kappa(A)\,u_f<1$, where $\kappa(A)$ is the
condition number and $u_f$ is the unit roundoff of the factorization
precision. \emph{On failure} is what the design does when a correction is
not good enough, and a dash means it has no response. \emph{Depth closure}
says whether the design comes with a proof that its interface is preserved
when it is nested to any depth.}
\label{tab:positioning}
\footnotesize
\setlength{\tabcolsep}{3pt}
\renewcommand{\arraystretch}{1.15}
\begin{tabularx}{\textwidth}{@{}>{\raggedright\arraybackslash}p{1.3in}>{\raggedright\arraybackslash}p{0.68in}>{\centering\arraybackslash}p{0.36in}>{\raggedright\arraybackslash}X>{\raggedright\arraybackslash}X>{\raggedright\arraybackslash}X>{\raggedright\arraybackslash}p{0.72in}@{}}
\toprule
Design & Precision boundary & Levels & Correction solver & Check & On failure & Depth closure \\
\midrule
\multicolumn{7}{@{}l}{\emph{Dense systems}} \\
LU-IR~\cite{langou2006exploiting,buttari2007mixed} & implicit & 2 & direct (LU) & a priori ($\kappa u_f$ bound) & iteration cap, refactorize & --- \\
GMRES-IR family~\cite{arioli2009fgmres,carson2017analysis,carson2018three,amestoy2024five} & implicit & 2 & iterative (LU-preconditioned GMRES) & a priori (three- to five-precision bounds) & --- & --- \\
Multistage IR~\cite{oktay2022multistage} & implicit & 2 & direct and iterative (LU-IR, SGMRES-IR, GMRES-IR) & convergence-rate monitor & escalate to next variant & --- \\
cuSOLVER IRS~\cite{nvidia2026cusolver} & implicit & 2 & direct (LU) & iteration cap & global fallback to FP64 factorization & --- \\
\midrule
\multicolumn{7}{@{}l}{\emph{Sparse systems}} \\
HSL\_MA79~\cite{hogg2010mixed} & implicit & 2 & direct (sparse factorization) & iteration cap & global fallback to FP64 factorization & --- \\
Inner--outer Krylov~\cite{buttari2008sparse,anzt2011errorcorrection,lindquist2022gmres,loe2021experimental} & implicit & 2 & iterative (Krylov) & fixed inner tolerance & --- & --- \\
Inexact, relaxed Krylov~\cite{simoncini2003inexact,vandeneshof2005relaxation} & --- & 2 & iterative (any inner solve) & a priori accuracy schedule & --- & --- \\
Progressive-precision multigrid~\cite{mccormick2021algebraic,tamstorf2021discretization} & per level & $L$ & iterative (multigrid) & a priori (discretization) & --- & multigrid analysis \\
F3R~\cite{suzuki2025nested} & per level & 4 & iterative (FGMRES and Richardson) & outermost residual & --- & --- \\
\midrule
\multicolumn{7}{@{}l}{\emph{Dense and sparse systems}} \\
PETSc~\cite{petsc} & none (one scalar type per build) & any & direct or iterative & convergence tests (user-defined) & user-coded & --- \\
Trilinos~\cite{heroux2005trilinos,loe2021experimental} & per object, joined by user-written wrappers & any & direct or iterative & status tests (user-defined) & user-coded & --- \\
Ginkgo~\cite{anzt2022ginkgo,ginkgo2026mixed} & per object, converted at each apply & any & direct or iterative, including refinement around a lower-precision solver & stopping criteria on the outer residual & none documented & --- \\
\textbf{CAMPS} & typed edge $\IR{p}{q}$ & any & direct or iterative, including adapters & residual or energy test on every correction & rollback, then fallback edge & Theorems~\ref{thm:closure}, \ref{thm:propagation}, \ref{thm:recursive} \\
\bottomrule
\end{tabularx}
\end{table}

\subsection{Proofs for the finite-precision contract}\label{app:contract-proofs}

\begin{proof}[Proof of Lemma~\ref{lem:geom}]
Unrolling the recurrence gives
\[
  \norm{r_J}_2
  \le \eta^J\norm{r_0}_2
     +\omega\sum_{j=0}^{J-1}\eta^j.
\]
Since $0\le\eta<1$, summing the geometric series yields
\[
  \norm{r_J}_2
  \le \eta^J\norm{r_0}_2
     +\frac{1-\eta^J}{1-\eta}\,\omega.
\]
\end{proof}

\paragraph{Residual decrease under restricted damping.}
Let $w=A_pd\ne0$. Expanding the squared residual gives
\[
  \norm{r-\alpha w}_2^2
  =\norm{r}_2^2-2\alpha\langle r,w\rangle
   +\alpha^2\norm{w}_2^2.
\]
The minimizer over all real $\alpha$ is
\[
  \alpha_*=\frac{\langle r,w\rangle}{\norm{w}_2^2},
\]
with
\[
  \norm{r-\alpha_*w}_2^2
  =\norm{r}_2^2
   -\frac{\langle r,w\rangle^2}{\norm{w}_2^2}.
\]

The implementation restricts $\alpha$ to $[0,\alpha_{\max}]$,
where $\alpha_{\max}>0$.
If $w=0$ or $\langle r,w\rangle\le0$, it chooses $\alpha=0$,
which leaves the residual unchanged.
Otherwise,
$\alpha=\min\{\alpha_*,\alpha_{\max}\}\in(0,\alpha_*]$.
Hence
$\alpha\norm{w}_2^2\le\langle r,w\rangle$, and
\[
  \norm{r-\alpha w}_2^2
   =\norm{r}_2^2-2\alpha\langle r,w\rangle+\alpha^2\norm{w}_2^2
   \le\norm{r}_2^2-\alpha\langle r,w\rangle
   <\norm{r}_2^2.
\]
The restricted step attains the unrestricted minimum when
$\alpha_*\le\alpha_{\max}$.
This argument holds in exact arithmetic.
The implementation separately tests the rounded trial iterate.

\subsection{Proof of Arbitrary-Layer Adapter Closure}\label{app:closure-proof}

\begin{proof}[Proof of Theorem~\ref{thm:closure}]
The adapter type rule maps a child result in $F_q^{\,n}$ back into $F_p^{\,n}$,
so replacing one compatible edge preserves the input and output types required
by its parent. Applying Lemma~\ref{lem:geom} at that edge, with the constants $\eta_e$ and
$\Omega_e$ and the number $j_e$ of accepted corrections, gives the
geometric bound. The bound refers to the edge's own parent residuals and
constants, so its form does not depend on the edge's depth in $T$. Repeating the argument from the leaves toward the root establishes
interface closure for any finite subset $E$: each replacement is type-correct
in isolation, and type-correct children compose into a type-correct parent.
Finally, admissibility at the consuming node: an enclosing adapter consumes
its child solely through the zero-start interface and judges every correction
by the measured parent residual, so it assumes neither fixedness nor
linearity of the child; and a flexible Krylov node---one that stores each
realized preconditioned vector and never re-applies the preconditioner, of
which FGMRES is the canonical instance---likewise requires neither, so a
variable adapter is an admissible correction operation for both. The
statement is deliberately local: it does not assert that $\eta_e<1$ holds for
every matrix or hierarchy, nor that insertion reduces runtime.
\end{proof}

\subsection{Proofs for child-to-parent propagation}\label{app:propagation-proofs}

\begin{proof}[Proof of Theorem~\ref{thm:propagation}]
Write $g-A_pd=(g-\tilde g)+(\tilde g-A_qd)+(A_q-A_p)d$ and apply the
triangle inequality to the three terms using the hypotheses and
$\norm{\tilde g}_2\le(1+\epsilon_g)\norm{g}_2$.
\end{proof}

\begin{proof}[Proof of Proposition~\ref{prop:size}]
Since $y_j=\sum_{t<j}(d_t+\delta y_t)$, we have
$\norm{y_j}_2\le\sum_{t<j}(MR_t+\nu)+D_j$. Lemma~\ref{lem:geom} gives
$R_t\le\eta^{\,t}\norm{g}_2+\Omega(1-\eta^{\,t})/(1-\eta)$. Summing over
$t<j$ gives the bound.
\end{proof}

\subsection{Proofs for hierarchy stability, convergence, and cost}\label{app:hierarchy-proofs}

\begin{proof}[Proof of Lemma~\ref{lem:size}]
$d=A_p^{-1}\bigl(g-(g-A_pd)\bigr)$ and
$\norm{g-A_pd}_2\le\theta\norm{g}_2$.
\end{proof}

\begin{proof}[Proof of Lemma~\ref{lem:recurrence}]
$r_{t+1}=g-A_p(y_t+d_t+\delta y_t)=(\hat r_t-A_pd_t)-e_t-A_p\,\delta y_t$
and $\norm{\hat r_t}_2\le\norm{r_t}_2+\norm{e_t}_2$ give
\eqref{eq:recurrence-general}. For \eqref{eq:dr}, \textup{(M1)} with
$y_t=x_g-A_p^{-1}r_t$ gives
$\norm{e_t}_2\le\gamma\bigl(\norm{g}_2+\norm{\abs{A_p}}_2(\norm{x_g}_2
+\norm{A_p^{-1}}_2\norm{r_t}_2)\bigr)$. For \eqref{eq:fp-recurrence},
\textup{(M2)} gives $\norm{\delta y_t}_2\le\gamma_1\norm{y_{t+1}}_2
\le\gamma_1\bigl(\norm{x_g}_2+\norm{A_p^{-1}}_2\norm{r_{t+1}}_2\bigr)$;
insert this and \eqref{eq:dr} into \eqref{eq:recurrence-general} and solve
for $\norm{r_{t+1}}_2$. The bound by $\psi_p$ uses
$\norm{x_g}_2\le\norm{A_p^{-1}}_2\norm{g}_2$.
\end{proof}

\begin{proof}[Proof of Theorem~\ref{thm:stability}]
Let $r=b-Ax$ and $\hat r=\fl(b-Ax)=r+\delta r$. By \textup{(M3)},
$\fl(\norm{\hat r}_2)\ge(1-\gamma)\norm{\hat r}_2$ and
$\fl(\norm{b}_2)\le(1+\gamma)\norm{b}_2$, so the stopping test gives
$\norm{\hat r}_2\le\tau'\norm{b}_2+a'$. By \textup{(M1)},
$\norm{\delta r}_2\le\gamma(\norm{b}_2+\norm{\abs{A}}_2\norm{x}_2)$, and
$\norm{r}_2\le\norm{\hat r}_2+\norm{\delta r}_2$. Dividing by
$\norm{A}_2\norm{x}_2+\norm{b}_2$ and using $\mu_{p_0}\ge1$ gives the bound on
$\beta_2$. For the forward error, $x-x_\star=-A^{-1}r$ with
$x_\star=A^{-1}b$, and
$\norm{r}_2\le\beta_2\norm{A}_2(\norm{x}_2+\norm{x_\star}_2)$, so
$\norm{x-x_\star}_2\le\kappa_2(A)\beta_2
(2\norm{x_\star}_2+\norm{x-x_\star}_2)$.
\end{proof}

\begin{proof}[Proof of Theorem~\ref{thm:recursive}]
\textup{(R2)} is \eqref{eq:T}; a cast edge performs no parent arithmetic,
because promotion is exact.

\textup{(R3)}. The child is applied to the scaled cast of the evaluated
residual $\hat r_t$, so by \eqref{eq:T} the promoted correction satisfies
$\norm{\hat r_t-A_pd_t}_2\le\eta\norm{\hat r_t}_2$, and
Lemma~\ref{lem:recurrence} with $\omega=0$ gives \eqref{eq:fp-recurrence}
for every accepted correction. Lemma~\ref{lem:geom} then gives
$\norm{r_t}_2\le\hat\eta^{\,t}\norm{g}_2+f(g)/(1-\hat\eta)$ after $t$ accepted
corrections. A \textsc{Failure} exit is excluded by hypothesis. On \textsc{Success},
\textup{(M3)} and \eqref{eq:dr} give
$\norm{r_t}_2\le\tau_+\norm{g}_2
+\gamma(\norm{g}_2+\norm{\abs{A_p}}_2\norm{x_g}_2)/(1-a_p)$. If no attempt
is rejected, a \textsc{Max-Steps} exit follows $J$ accepted corrections,
and the bound just stated applies with $t=J$. A rejection occurs when
$\rho'>\varrho\,\rho$ for the configured progress threshold, or when
$\rho'>\rho$ under the default test. The function $c_p(\varrho)$ decreases as
$\varrho$ grows, so in both cases Proposition~\ref{prop:exits} gives
$\norm{r_t}_2<c_pf(g)$ for the accepted iterate at the time of the
rejection, and it gives the same bound for the iterate returned by a
\textsc{Stagnated} exit. After a rejection the accepted iterate stays in
the set $\norm{r}_2\le c_pf(g)$. A rollback does not change it, and an
accepted correction of the fallback child, which satisfies the same
recurrence by hypothesis, gives
$\norm{r_{t+1}}_2\le\hat\eta\,c_pf(g)+f(g)\le c_pf(g)$ because
$c_p\ge3/(1-\hat\eta)$. Thus, every exit that follows a rejection,
including \textsc{Max-Steps} with fewer than $J$ accepted corrections,
returns $\norm{r}_2\le c_pf(g)$. Because $c_p\ge3/(1-\hat\eta)$ and
$c_p\ge3/(1-a_p)$, all cases are covered by
$\max\{\hat\eta^{\,J},\tau_+\}\norm{g}_2+c_pf(g)$, which is \textup{(R3)}.

\textup{(R4)}. A cycle starts from an iterate $x$ with evaluated residual
$\hat r=r+\delta r=\beta v_1$. The first application returns $z_1=S(v_1)$
with $\norm{v_1-A_pz_1}_2\le\theta$, so the candidate $x+\beta z_1$ lies in
the search space with $\norm{\hat r-\beta A_pz_1}_2\le\theta\norm{\hat r}_2$
(Section~\ref{outer-fgmres}). The update minimizes
$\norm{\hat r-A_pZ_my}_2$ over that space, so the new exact residual
$r-A_pZ_my$ has norm at most
$\theta\norm{\hat r}_2+\norm{\delta r}_2\le\hat\theta\norm{r}_2
+(1+\theta)\gamma(\norm{g}_2+\norm{\abs{A_p}}_2\norm{x_g}_2)$ by
\eqref{eq:dr}, and Lemma~\ref{lem:geom} applies over the $k_{\min}$ completed cycles. Because
$\theta<1$ forces $A_pz_j\ne0$, a breakdown $h_{j+1,j}=0$ ends the cycle at
the minimizer over the directions already formed and does not affect the
bound. A tolerance exit returns an iterate whose evaluated restart residual
satisfies $\fl(\norm{\hat r}_2)\le\tau_K\,\fl(\norm{g}_2)$; by \textup{(M3)}
and \eqref{eq:dr},
$\norm{r}_2\le\tau_{K+}\norm{g}_2+\gamma(\norm{g}_2+\norm{\abs{A_p}}_2
\norm{x_g}_2)/(1-a_p)$, which the stated bound covers. An iteration cap that
truncates a cycle leaves the bound of the completed cycles in force, because
the truncated cycle minimizes over a space that contains its starting
iterate.

\textup{(i)} follows by induction on the height of $N$: the rules state
exactly that each production maps a corrector for the child operator to a
corrector for the parent operator. \textup{(ii)} is
\eqref{eq:fp-recurrence} at the root with $g=b$; the evaluated norm of an
iterate with $\norm{r_t}_2\le\hat\eta_0^{\,t}\norm{b}_2+f_0/(1-\hat\eta_0)$
is at most $(1+\gamma)\bigl((1+a_{p_0})\norm{r_t}_2+f_0\bigr)$, which is below
$\tau\,\fl(\norm{b}_2)$ once $(1+a_{p_0})\hat\eta_0^{\,t}\le\tau_--\phi_0$;
and by Proposition~\ref{prop:exits} a rejection while the stopping test still
fails would force $\tau_-\norm{b}_2<[(1+a_{p_0})c_{p_0}+1]f_0$, contrary to
the hypothesis. Since every attempt is then accepted, the $K(\tau)$-th
attempt exists when $J\ge K(\tau)$, and Algorithm~\ref{alg:adapter} tests
success immediately after each accepted update, so \textsc{Success} is
declared at or before attempt $K(\tau)$; with $J<K(\tau)$ the loop ends after
$J$ accepted corrections, to which \eqref{eq:root-rate} applies.
\textup{(iii)}: every exit then returns
$\norm{r}_2\le c_{p_0}f_0$, and
$f_0\le[(1+\eta_0)\gamma\mu_{p_0}+\gamma_1]
(\norm{b}_2+\norm{A}_2\norm{x_\star}_2)/(1-\gamma_1\kappa_2(A))$ because
$\mu_{p_0}\ge1$; replace $\norm{x_\star}_2$ by
$\norm{x}_2+\norm{A^{-1}}_2\norm{r}_2$ and solve for $\norm{r}_2$.
\end{proof}

\begin{proof}[Proof of Proposition~\ref{prop:exits}]
By \textup{(M3)} and \eqref{eq:dr}, applied to $y'$ and to $y_t$,
$\rho'\le(1+\gamma)\bigl((1+a_p)\norm{r'}_2+\gamma G\bigr)$ and
$\rho\ge(1-\gamma)\bigl((1-a_p)\norm{r_t}_2-\gamma G\bigr)$, where
$G=\norm{g}_2+\norm{\abs{A_p}}_2\norm{x_g}_2$. Insert
$\norm{r'}_2\le\hat\eta\norm{r_t}_2+f(g)$ into $\rho'>\varrho\rho$, collect
the terms in $\norm{r_t}_2$, and use $\varrho\le1$ and $\gamma G\le f(g)$.
\end{proof}

\begin{proof}[Proof of Proposition~\ref{prop:work}]
\textup{(a)} is a count of the operations of Algorithm~\ref{alg:adapter} and
of one FGMRES cycle, combined with the bound $K(\tau)$ of
Theorem~\ref{thm:recursive}. \textup{(b)}:
$C_iL/\ell_i\le C_i\lceil L/\ell_i\rceil<C_i(L/\ell_i+1)$; compare the upper
bound of one alternative with the lower bound of the other.
\end{proof}

\subsection{Software and verification}\label{app-impl}
\paragraph{Core package.} The core package has about $1{,}700$ lines and
separates five concerns.
\begin{itemize}
\item \emph{Precision specifications} record the unit roundoff, range,
  subnormal threshold, and accumulation format of each storage format.
  Every conversion is explicit, and an operator or solver node rejects an
  input of the wrong format instead of casting it.
\item \emph{Operators} carry their format and symmetry. For the
  energy test they also provide the entrywise absolute product
  $\abs{A}\abs{x}$.
\item \emph{The adapter} implements Algorithm~\ref{alg:adapter} step by
  step. This covers the stored accepted state, power-of-two scaling and
  casting, the zero-start child call, promotion, optional damping, the
  recomputed residual of the committed iterate, the residual
  energy acceptance test, rollback, the fallback pair, stagnation
  detection, and the attempt budget. For every attempt it writes
  $(\rho,\rho',\alpha,\widehat{\Delta\phi},E)$ and the outcome to a contract
  record.
\item \emph{The residual contract and scaling} carry the comparison
  tolerance and report every conversion.
\item \emph{The solver nodes} are FGMRES with stored directions and FP64
  restart residuals, GMRES, CG, dense LU, stored approximate inverses, and
  Jacobi.
\end{itemize}

\paragraph{Backend settings.} A diagnostics module fixes the backend
settings. TF32 and reduced-precision reductions are turned off, so FP32
means IEEE FP32 multiplies. FP16 and BF16 matrix products accumulate in
FP32, as the platform specifies. Norms and Krylov scalars over
half-precision vectors also accumulate in FP32. We record each of these as
an explicit promotion, since the storage precision of a vector is not the
same as the precision of the arithmetic on it. Dense operators are stored
matrices. Sparse operators are in CSR format, with native matrix--vector
products in all four storage formats. Krylov scalars are computed on the
host, and this overhead is the same for every method in every comparison.

\paragraph{Experiment layer.} A further $1{,}800$ lines contain the matrix
generators, the counting operators, the named instances and baselines, and
the measurement harness. The generators build dense $U\Sigma V^{\mathsf T}$
matrices with a chosen $\kappa_2$ and the 3D Poisson and
convection--diffusion stencils. The counting operators record every product
by storage format. The harness handles warmups, medians with interquartile
ranges, separate setup timing, residuals recomputed in FP64, normwise and
componentwise backward errors, analytic and measured memory, and an
environment record in every result file. The verification scripts add
another $850$ lines. Dense LU nodes use partial pivoting without pivot
modification. A zero-pivot boost is kept only for incomplete
factorizations, for the reason given in Section~\ref{exp-dense-ic}.

\paragraph{Determinism.} Every result file records the GPU, driver, CUDA
and PyTorch versions, and backend flags under which it was produced, and
every problem is generated from a fixed seed. On the same GPU, reruns
reproduce iteration counts, acceptance decisions, and residuals bit for
bit. We confirmed this for the fallback, damped-fallback, and ladder rows
of Section~\ref{exp-dense-ic} and for the sweep of
Section~\ref{exp-verify} after every revision of the adapter.

\paragraph{Verification suite.} The repository has a one-command
verification suite. It reruns the check of Theorem~\ref{thm:recursive}, the
energy contract with its counterexample, the fallback and ladder
rows, and a reduced-size pass over all seven experiments. It then writes a
report that compares the outcome with the archived results. On a different
GPU, the report requires three things. Exit statuses, attempt counts, and
the split between accepted and rejected corrections must agree. Every
rigorous bound must hold. Residuals and contraction factors must agree
within a factor of ten. Timings are reported but not checked, because they
depend on the hardware. On a datacenter GPU with full-rate FP64, for
example, the dense speedups of Section~\ref{exp-dense-wc} should be much
smaller than on our laptop GPU, whose FP64 rate is $1/64$ of its FP32 rate.

\paragraph{Second architecture.} We also ran the suite on a Lonestar6 node
at the Texas Advanced Computing Center with an NVIDIA A100 (40\,GB, driver
570.195.03), under Linux with the same PyTorch and CUDA versions. It passed
every check.
\begin{itemize}
\item All $57$ recursive-bound configurations respect their bounds, with
  root backward errors of $0.39$--$0.83\,u_{64}$.
\item The counterexample is rejected.
\item Both SPD contracts converge in the same three accepted corrections.
\item Every fallback and ladder row reproduces its status and attempt
  counts.
\item The reduced-size sweep converges on the same $52$ of $57$ runs. The
  five exceptions are the pure-FP32 instances at $\kappa=10^{10}$, on both
  machines.
\end{itemize}
Residuals and contraction factors agree with the laptop run within a factor
of six. Differences of this size come from the different order in which the
two GPUs sum floating-point values. The report and its result files are in
the repository.

\subsection{Structure of the test matrices}\label{app-matrices}
\begin{figure}[H]
\centering
\includegraphics[width=\linewidth]{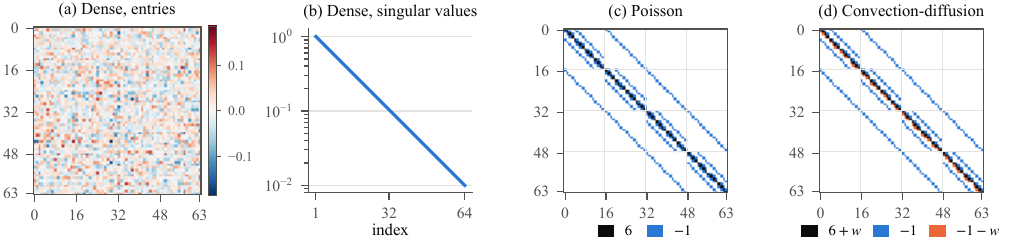}
\caption{The three kinds of test matrix at $n=64$. (a) Entries of a dense matrix $A=U\Sigma V^{\mathsf T}$ with $\kappa=10^{2}$. (b) Its singular values, which are spaced evenly on a log scale. (c) The Poisson matrix on a $4^{3}$ grid. (d) The convection--diffusion matrix on the same grid. In (c) and (d) the color shows the value of each nonzero entry, and the grid lines separate the planes of the grid.}
\label{fig:matrix-structure}
\end{figure}

Figure~\ref{fig:matrix-structure} shows small versions of the three kinds of
test matrix in Section~\ref{experiments}. Each kind has a different structure,
and the CAMPS instances we run on it are chosen to use that structure.

\paragraph{Dense matrices.} A dense test matrix has no zero entries and no
symmetry, and its entries are of order $1/\sqrt n$. The one property we
control is the condition number, which we set through the singular values.
The standard method for a dense system is LU factorization. It costs
$O(n^{3})$ operations, and each later triangular solve or matrix-vector
product costs $O(n^{2})$. \textsf{CAMPS-Direct} and \textsf{CAMPS-Krylov} use
this difference. They compute the factorization in FP32 and keep the cheaper
residual in FP64. The condition number decides whether the FP32 factors are
accurate enough (Section~\ref{exp-dense-ic}). When they are not, the fallback
child and \textsf{CAMPS-Ladder} switch to an FP64 factorization.

\paragraph{Poisson.} The Poisson matrix is sparse, symmetric, and positive
definite. For a grid of $g^{3}$ points, each row has at most seven nonzero
entries, which lie on the diagonal and at distances $1$, $g$, and $g^{2}$
from it. A matrix-vector product costs $O(n)$ operations, so no
factorization is needed. Symmetry and positive definiteness allow CG, which
stores a fixed number of vectors. \textsf{CAMPS-Direct} with a CG$_{32}$ child
uses this structure (Section~\ref{exp-sparse-spd}). The entries are $6$ and
$-1$, and FP32 and FP16 store both without error.

\paragraph{Convection--diffusion.} The convection--diffusion matrix has the
same sparsity pattern as the Poisson matrix. The entry next to the diagonal
is $-1-w$ on one side and $-1$ on the other, so the matrix is not symmetric
and CG does not apply. GMRES and FGMRES apply, but they store a Krylov basis
and orthogonalize against it, and that work grows with the number of
iterations. \textsf{CAMPS-Cascade} and \textsf{CAMPS-Apex} use the sparsity
to move most of the matrix-vector products and most of the Krylov iterations
to FP32 and FP16 (Sections~\ref{exp-sparse-nsy} and~\ref{exp-f3r}). For the
wind strengths we test, the entries are small integers, and FP16 stores them
without error.

\subsection{Error measures}\label{app-measures}
The backward errors of Section~\ref{exp-protocol} use the convention that
$\xi/0$ is zero when $\xi=0$ and infinite otherwise. Every denominator is
positive on the dense sweep, so the convention never applies to the
reported values.

The forward error is measured against the manufactured vector
$x_{\mathrm m}$. The right-hand side is
$b=\fl(Ax_{\mathrm m})=Ax_{\mathrm m}+e_b$, so $x_{\mathrm m}$ is not the
exact solution $x_\star=A^{-1}b$ of the stored system. For a computed $x$
with residual $r=b-Ax$,
\[
  x-x_{\mathrm m}=A^{-1}(e_b-r),\qquad x-x_\star=-A^{-1}r ,
\]
and thus
$\norm{x-x_{\mathrm m}}_2\le\norm{A^{-1}}_2(\norm{e_b}_2+\norm{r}_2)$. A
bound that uses the residual alone applies to $x_\star$ and not to
$x_{\mathrm m}$.

We measured $e_b$ for the stored operands of the dense sweep at
$\kappa=10^{4}$, $10^{8}$, $10^{10}$, and $10^{12}$, using error-free
products and exact summation. In every case $\norm{e_b}_2/\norm{b}_2$ is
between $1.8\,u_{64}$ and $2.0\,u_{64}$, far below the worst-case bound
$\gamma^{(64)}_{n}\norm{\abs{A}\abs{x_{\mathrm m}}}_2/\norm{b}_2
\approx6\times10^{-12}$. The two reference vectors differ by
$\norm{x_\star-x_{\mathrm m}}_2/\norm{x_{\mathrm m}}_2=1.2\times10^{-13}$ at
$\kappa=10^{4}$, $6.0\times10^{-10}$ at $10^{8}$, $4.9\times10^{-8}$ at
$10^{10}$, and $3.8\times10^{-6}$ at $10^{12}$, where we computed
$x_\star-x_{\mathrm m}=A^{-1}e_b$ with an FP64 solve. For the FP64 direct
solve, the forward error is between $24$ and $51$ times larger than this
difference at these four condition numbers. The script
\texttt{experiments/rhs\_error\_check.py} in the repository reproduces
these numbers.

\subsection{Forward error at $\kappa=10^{12}$}\label{app-fwd}
At $\kappa=10^{12}$ the forward errors relative to $x_{\mathrm m}$ are
$1.2\times10^{-4}$ for the FP64 direct solve and $1.0\times10^{-2}$ for the
fallback instance. Rounding in the right-hand side does not explain either
value, because it moves the reference by $3.8\times10^{-6}$
(Appendix~\ref{app-measures}). Measured against $x_\star$, the two errors
are the same to two digits. Both come from the attained residuals. The two
solves stop at relative residuals of $9.9\times10^{-15}$ and
$1.9\times10^{-12}$, with normwise backward errors of $3.2\times10^{-16}$
and $4.0\times10^{-14}$. The bound
\[
\frac{\norm{x-x_\star}_2}{\norm{x_{\mathrm m}}_2}
\le\frac{\norm{A^{-1}}_2\norm{r}_2}{\norm{x_{\mathrm m}}_2}
\]
allows forward errors of $1.4\times10^{-3}$ and $0.27$, and the observed
errors are about $12$ and $26$ times smaller. The residuals differ by a
factor of about $190$, and the forward errors by a factor of $89$.

The fallback instance stops as soon as its residual falls below
$\tau=10^{-10}$, two orders of magnitude above where the direct solve lands.
At this conditioning it pays for that with two orders of magnitude more
forward error. The direct solve's $1.2\times10^{-4}$ is close to
$\kappa_2(A)u_{64}\approx1.1\times10^{-4}$, the size expected of a backward
stable solve. Small residuals show backward accuracy, not forward
accuracy~\cite{carson2017analysis,higham2002asna}.

\subsection{Residual and energy contracts on SPD problems}\label{app-energy}
The Euclidean acceptance test can reject valid CG steps that lower the
energy norm of the error while the residual grows (Section~\ref{contract}).
On the reported runs at $32^{3}$ and $48^{3}$, every correction was
accepted under both the residual contract and the energy contract
on the committed displacement. Both contracts gave identical iterates and
residuals, with three corrections each, because the child ran to its own
tolerance before returning. The measured energy decrease exceeded its
error estimate \eqref{eq:energy-bound} by factors between $10^{6}$
and $5\times10^{10}$.

The energy test needs three extra products per correction. In paired runs
these cost $12$--$27\%$ of the solve time on the laptop GPU and $2\%$ on the
A100. The energy contract therefore did not screen out any CG progress
here. A short-budget CG child on which the two contracts disagree would
show whether the energy contract has a real advantage.
Section~\ref{discussion} lists that test as open.

\end{document}